\documentclass[CJK,CCT,openany,twoside]{cctbook}  
\usepackage{psfig,epsfig,amssymb,vatola} 
\usepackage{multicol}  
\usepackage{fancyhdr}
\pdfoutput=1
\input psfig.sty
\input cyracc.def
\input amssymb.sty
\input vatola.sty
\usepackage{color}
\usepackage{bm}                       

\def\le{\leqslant}

\def\zd{{\rm d}}

\def\zi{{\rm i}}

\newcommand{\qqbar}{q\bar{q}}
\newcommand{\QQbar}{Q\bar{Q}}

\newcommand{\rt}{\rightarrow}
\newcommand{\lm}{\Lambda}
\newcommand{\yones}{\Upsilon(1S)}
\newcommand{\ytwos}{\Upsilon(2S)}
\newcommand{\bbbar}{b\bar{b}}
\newcommand{\uubar}{u\bar{u}}
\newcommand{\ddbar}{d\bar{d}}
\newcommand{\ssbar}{s\bar{s}}
\newcommand{\ccbar}{c\bar{c}}
\newcommand{\pim}{\pi^{-}}
\newcommand{\pip}{\pi^{+}}
\newcommand{\piz}{\pi^{0}}
\newcommand{\leplep}{\ell^{+}\ell^{-}}
\newcommand{\psip}{\psi^{\prime}}
\newcommand{\chicone}{\chi_{c1}}

\newcommand{\mumu}{\mu^{+}\mu^{-}}
\newcommand{\pipi}{\pi^{+}\pi^{-}}

\newcommand{\ppbar}{p\bar{p}}
\newcommand{\DDbar}{D\bar{D}}

\newcommand{\chip}{\chi^{\prime}}

\newcommand{\jpsi}{J/\psi}

\newcommand{\lmb}{\bar{\Lambda}}

\newcommand{\twog}{\gamma\gamma}
\newcommand{\babar}{BaBar}
\newcommand{\DZero}{D0}
\newcommand{\ee}{e^{+}e^{-}}
\newcommand{\jp}{J/\psi}

\newfont{\yihao}{cmb10 at 18pt}

\usepackage{calc}                     

\newenvironment{OOExercises}[1][10..]{
\begin{list}
{\hfill\arabic{enumi}.}{
\settowidth{\labelwidth}{\bfseries #1}%
\setlength{\labelsep}{8pt}%
\setlength{\topsep}{10bp}%
\setlength{\partopsep}{5pt}%
\setlength{\parskip}{0pt}%
\setlength{\itemsep}{0bp}%
\setlength{\leftmargin}{\labelwidth+\labelsep}%
\usecounter{enumi}}}{\end{list}
}

\DeclareSymbolFont{lettersA}{U}{txmia}{m}{it}
\DeclareMathSymbol{\piup}{\mathord}{lettersA}{25}
\DeclareMathSymbol{\muup}{\mathord}{lettersA}{22}      
\DeclareMathSymbol{\gammaup}{\mathord}{lettersA}{13}      

\renewcommand{\baselinestretch}{1.06} \renewcommand{\arraystretch}{1.2}
 \catcode`@=11

{%
\fancyhead{} 

}

\definecolor{orangec}{cmyk}{.24,.91,.96,.18}

\definecolor{orangecc}{cmyk}{.24,.94,.96,.18}
\definecolor{oorangec}{cmyk}{.8,.2,.5,.4}
\definecolor{ooorangec}{cmyk}{1,.9,0.08,.04}      

\renewcommand\footnoterule{\vspace*{3pt}
\hrule height 0pt \vspace*{3pt}}

\newfont{\xbt}{cmb10 at 12pt}

\usepackage{graphicx}                 

\graphicspath{{figs/}}      

\begin{document}
\thispagestyle{empty}

\vspace*{-20.5mm} {\small Front.~Phys.
\ \\
\small DOI 10.1007/s11467-014-0449-6\ }\\[-3mm]
{\color{orangec}\def\temptablewidth{\textwidth}{\rule{\temptablewidth}{.5pt}}}
\\[2mm]
\vspace*{8mm}


\begin{center}

{\usefont{T1}{fradmcn}{m}{n}\yihao A new hadron spectroscopy} \footnotetext{~\\[-3mm]
\centerline{\hspace*{-10mm}\copyright\ Higher Education Press and
Springer-Verlag Berlin Heidelberg 2014}\vspace*{-8mm}}
\vspace*{5mm}

{\bf\small Stephen Lars Olsen}\vspace*{4mm}

{\footnotesize\it
 Center for Underground Physics, Institute for Basic Science, Daejeon 305-811, Korea\\[1.5mm]
    E-mail: solensnu@gmail.com\\[1mm]
    Received August 13, 2014; accepted September 16, 2014

 }\vspace{5mm}

\baselineskip 10pt
\renewcommand{\baselinestretch}{0.8}
\parbox[c]{152mm}
{\noindent{{\color{oorangec}QCD-motivated models for hadrons predict
an assortment of ``exotic'' hadrons that have structures that are
more complex than the quark-antiquark mesons and three-quark baryons
of the original quark-parton model.~These include pentaquark
baryons, the six-quark $H$-dibaryon, and tetraquark, hybrid and
glueball mesons.~Despite extensive experimental searches, no
unambiguous candidates for any of these exotic configurations have
been identified.~On the other hand, a number of meson states, one
that seems to be a proton-antiproton bound state, and others that
contain either charmed-anticharmed quark pairs or bottom-antibottom
quark pairs, have been recently discovered that neither fit into the
quark-antiquark meson picture nor match the expected properties of
the QCD-inspired exotics.~Here I briefly review results from a
recent search for the $H$-dibaryon, and discuss some properties of
the newly discovered states --the proton-antiproton state and the
so-called $XYZ$ mesons-- and compare them with expectations for
conventional quark-antiquark mesons and the predicted QCD-exotic
states.}\vspace{2mm}

{\color{ooorangec}\bf Keywords}~~quarks, charmonium, pentaquarks,
di-baryons, baryonium, tetraquarks\vspace{2mm}

{\color{ooorangec}\bf PACS numbers}~~14.40.Gx, 12.39.Mk, 13.20.He }}
\end{center}
\normalsize

\baselineskip 12pt \renewcommand{\baselinestretch}{1.06}
\parindent=10.8pt  \parskip=0mm \rm\vspace{2mm}

\begin{multicols}{2}
\setlength{\parindent}{1em}    

\centerline{Contents}\vspace{2.5mm}

\noindent 1\quad  Introduction\hfill 1\vspace{0.0mm}

\noindent 2\quad Pentaquarks and $H$-dibaryons\hfill 2\vspace{0.0mm}

\hspace{1.5mm}2.1\quad Belle $H$-dibaryon search\hfill
2\vspace{0.0mm}

\noindent 3\quad What we do see\hfill 3\vspace{0.0mm}

\hspace{1.5mm}3.1\quad Baryonium in radiative $\jp\rt\gamma\ppbar$
decays?\hfill 4\vspace{0.0mm}

\hspace{1.5mm}3.2\quad The $XYZ$ mesons\hfill 4\vspace{0.0mm}

\hspace{9.5mm}3.2.1\quad Charmoniumlike mesons\hfill 4\vspace{0.0mm}

\hspace{9.5mm}3.2.2\quad Bottomoniumlike mesons\hfill
22\vspace{0.0mm}

\hspace{1.5mm}3.3\quad Comments\hfill 24\vspace{0.0mm}

\hspace{9.5mm}3.3.1\quad Molecules?\hfill 25\vspace{0.0mm}

\hspace{9.5mm}3.3.2\quad Tetraquarks?\hfill 28\vspace{0.0mm}

\hspace{9.5mm}3.3.3\quad QCD-hybrids?\hfill 28\vspace{0.0mm}

\hspace{9.5mm}3.3.4\quad Hadrocharmonium?\hfill 28\vspace{0.0mm}

\hspace{9.5mm}3.3.5\quad A unified model?\hfill 29\vspace{0.0mm}

\noindent 4\quad Summary\hfill 29\vspace{0.0mm}

\hspace{1.5mm}Acknowledgements\hfill 30\vspace{0.0mm}

\hspace{1.5mm}References and notes\hfill 30\vspace{0.0mm}

\vspace*{7mm} {\color{ooorangec}\hrule}\vspace{2mm} \noindent
{\color{ooorangec}\large
\usefont{T1}{fradmcn}{m}{n}\xbt 1\quad Introduction}\vspace{3.5mm}

\noindent The strongly interacting particles of the Standard Model
are colored quarks and gluons.~In contrast, the strongly interacting
particles in nature are color-singlet ({i.e.}, white) mesons and
baryons.~In the theory, quarks and gluons are related to mesons and
baryons by the long-distance regime of Quantum Chromodynamics (QCD),
which remains the least understood aspect of the theory.~Since
first-principle lattice-QCD (LQCD) calculations are still not
practical for most long-distance phenomena, a number of models
motivated by the color structure of QCD have been proposed.~However,
so far at least, predictions of these QCD-motivated models that
pertain to the spectrum of hadrons have not had great success.

For example, it is well known that combining a $q=u,d,s$ light-quark
triplet with a $\bar{q}=\bar{u},\bar{d},\bar{s}$ antiquark
antitriplet gives the familiar meson octet of flavor-$SU(3)$.~Using
similar considerations based on QCD,  two quark triplets can be
combined to form a ``diquark'' antitriplet of antisymmetric $qq$
states and a sextet of symmetric states as illustrated in
Fig.~1(a).~In QCD, these diquarks have color: combining a red
triplet with a blue triplet -- as shown in the figure -- produces a
magenta (anti-green) diquark and, for the antisymmetric triplet
configurations, the color force between the two quarks is expected
to be attractive.~Likewise, green-red and blue-green diquarks form
yellow (anti-blue) and cyan (anti-red) antitriplets as shown in
Fig.~1(b).

Since these diquarks are not color-singlets, they cannot exist as
free particles but, on the other hand, the anticolored diquark
antitriplets should be able to combine with other colored objects in
a manner similar to antiquark antitriplets, thereby forming
multiquark color-singlet states with a more complex substructure
than the $\qqbar$ mesons and $qqq$ baryons of the original quark
model~[1].~These so-called ``exotic'' states include pentaquark
baryons, six-quark $H$-dibaryons and tetraquark mesons, as
illustrated in Fig.~1(c).~Other proposed exotic states are
glueballs, which are mesons made only from gluons, hybrids formed
from a $q$, $\bar{q}$ and a gluon, and deuteron-like bound states of
color-singlet ``normal'' hadrons, commonly referred to as
molecules.~These are illustrated in Fig.~1(d).~Glueball and hybrid
mesons are motivated by QCD; molecules are a generalization of
classical nuclear physics to systems of subatomic partics.

\vspace{3mm} \centerline{\psfig{figure=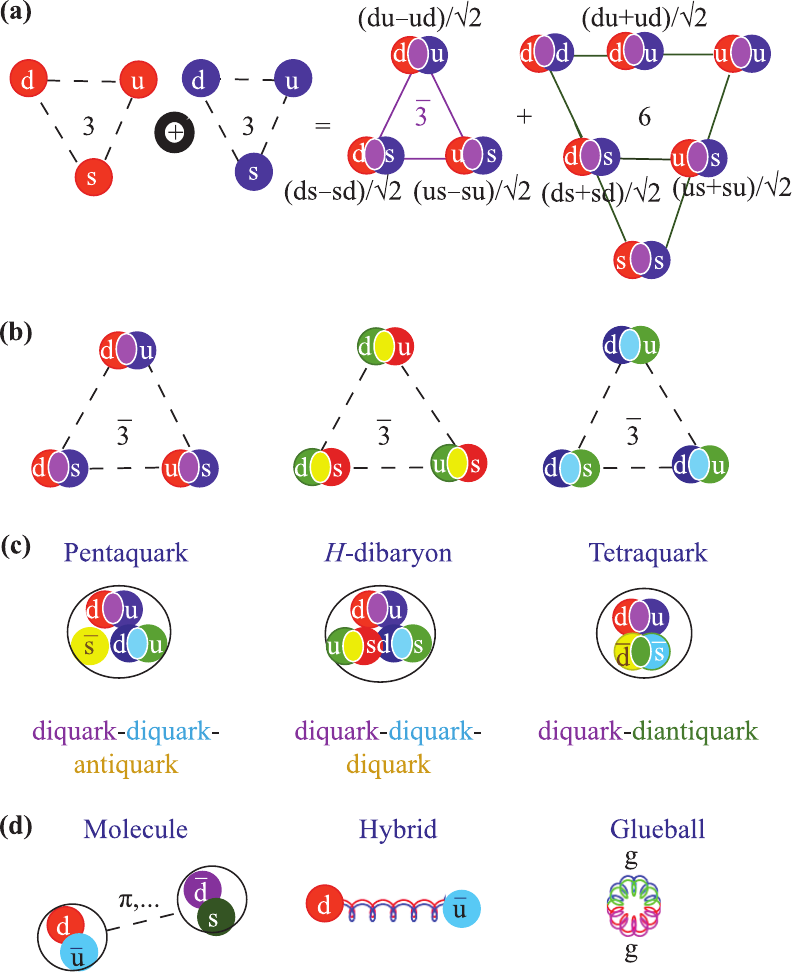}\vspace{1mm}}
 {\baselineskip
10.5pt\renewcommand{\baselinestretch}{1.05}\footnotesize \noindent
{\color{ooorangec}\bf Fig.~1}\quad {\bf(a)} Combining a red and blue
quark triplet produces a magenta (anti-green) antitriplet and
sextet.~{\bf(b)}  The three anticolored diquark
antitriplets.~{\bf(c)}  Some of the multiquark, color-singlet states
that can be formed from quarks, antiquarks, diquarks and
diantiquarks.~{\bf(d)} Other possible multiquark/gluon systems.

}\vspace{0mm}

\renewcommand\footnoterule{\vspace*{3pt}
\noindent
\def\temptablewidth{\textwidth}{\rule{0.3\temptablewidth}{0.5pt}}\vspace*{3pt}}

\vspace*{7mm} {\color{ooorangec}\hrule}\vspace{2mm} \noindent
{\color{ooorangec}\large
\usefont{T1}{fradmcn}{m}{n}\xbt 2\quad Pentaquarks and $\bm H$-dibaryons }\vspace{3.5mm}

\noindent All of the above-mentioned candidates for exotic states
have been the subject of numerous theoretical and experimental
investigations during the four decades that have elapsed since QCD
was first formulated.~This activity peaked in 2003 when the LEPS
experiment at the SPring-8 electron ring in Japan reported the
observation of a peak in the $K^+ n$ invariant mass distribution in
$\gamma n\rt K^+K^- n$ reactions on a carbon target~[2], with
properties close to those that had been predicted for the $S=+1$
${\it\Theta}_5^+$ pentaquark~[3].~This created a lot of excitement
at the time~[4] but subsequent, high-statistics experiments~[5, 6]
did not confirm the LEPS result and, instead, gave negative
results.~The current ``conventional wisdom'' is that pentaquarks do
not exist~[7], or at least have yet to be found.

The six-quark $H$-dibaryon, subsequently referred to as $H$, was
predicted by Jaffee in 1977 to be a doubly strange, tightly bound
six-quark structure $(uuddss)$ with isospin zero and
$J^P=0^+$~[8].~An $S=-2$ state with baryon number $B=2$ and mass
below $2m_{\lm}$ could only decay via weak interactions and, thus,
would be long-lived.~Although Jaffe's original prediction that the
$H$ would be $\sim 80$~MeV below the $2m_{\lm}$ threshold was ruled
out by the observation of double-$\lm$ hypernuclei, most notably the
famous ``Nagara'' event~[9] that limited the allowed $H$ region to
masses above $2m_{\lm}-7.7$~MeV, the theoretical case for an
$H$-dibaryon with mass near $2m_{\lm}$ continues to be strong, and
has been recently strengthened by two independent LQCD calculations,
both of which find an $H$-dibaryon state with mass near
$2m_{\lm}$~[10, 11].

\vspace*{4mm} \noindent {\color{ooorangec}2.1\quad Belle
$H$-dibaryon search }\vspace{3.1mm}

\noindent The Belle experiment recently reported results of a search
for production of an $H$-dibaryon with mass near $2m_{\lm}$ in
inclusive $\yones$ and $\ytwos$ decays~[12].~Decays of narrow
$\Upsilon(nS)$ $(n=1,2,3)$ bottomonium $(\bbbar)$ resonances are
particularly well suited for searches for multiquark states with
non-zero strangeness.~The $\Upsilon(nS)$ states are flavor-$SU(3)$
singlets that primarily decay via annihilation into three
gluons.~The gluons materialize into $\uubar$, $\ddbar$ and $\ssbar$
pairs with nearly equal probabilities, plus additional gluons that
also subsequently materialize as $\qqbar$ pairs.~This creates final
states with a high density of quarks and antiquarks in a limited
volume of phase space.~A benchmark rate for multiquark-state
production in these decays is set by the measured inclusive decay
branching fractions to antideuterons
($\bar{D}$)$^{1)}$\footnote{\hspace*{-5mm}$^{1)}$~The rates for
deuteron and antideuteron production in $\Upsilon(nS)$ decays are
almost certainly equal.~However, in $\ee$ experiments,
$\frac{}{}~~~$background deuterons are copiously produced by
particle interactions in the beam pipe and other material in the
inner part of the $\frac{}{}~~~$detector, and these make
experimental measurements of their true rate quite
difficult.~Therefore the rate is quoted for antideuterons,
$\frac{}{}~~~$which, because they do not suffer from these
backgrounds, are easier to measure.}: ${\mathcal B}(\yones\rt
\bar{D} + X)=(2.9\pm 0.3) \times 10^{-5}$ and ${\mathcal
B}(\ytwos\rt \bar{D} + X)=(3.4\pm 0.6) \times 10^{-5}$~[13].~If the
six-quark $H$-dibaryon is produced at a rate that is similar to that
for six-quark antideuterons, there should be many thousands of them
in the 102~million $\yones$ and 158~million $\ytwos$ event samples
collected by Belle.

For $H$ masses below $2m_{\lm}$, Belle searched for $H\rt\lm p\pim$
(\& $\bar{H}\rt \bar{\lm}\bar{p}\pip$) signals in the inclusive $\lm
p\pim$ invariant mass distribution.~For masses above $2m_{\lm}$, the
$H\rt \lm\lm$ (\& $\bar{H}\rt\bar{\lm}\bar{\lm})$ mode was
used.~Figure~2 shows the measured $\lm p\pim$ (a) \& $\lmb
\bar{p}\pip$ (b) invariant mass spectra for masses below $2m_{\lm}$
and the $\lm\lm$ (c) and $\lmb\lmb$ (d) mass spectra for masses
above $2m_{\lm}$.~Here results from the $\yones$ and $\ytwos$ data
samples are combined.~No signal is observed.~The solid red curves
show results of a background-only fit to the data; the dashed curve
shows the MC expectations for an $H$-dibaryon produced at $1/20$th
of the antideuteron rate.~Upper limits on the inclusive branching
ratios that are at least a factor of twenty below that for
antideuterons are set over the entire $|M_{H}-2m_{\lm}|<30$~MeV mass
interval.

Neither pentaquarks nor the $H$-dibaryon are seen in spite of the
stong theoretical motivation for their existence.~The absence of
pentaquarks led Wilczek to remark ``{\it The story of the pentaquark
shows how poorly we understand QCD}''~[14].~The absence of any
evidence for the $H$-dibaryon (among other things) led Jaffe to
observe that ``{\it The absence of exotics is one of the most
obvious features of QCD}''~[15].

\vspace*{8mm} {\color{ooorangec}\hrule}\vspace{2mm} \noindent
{\color{ooorangec}\large
\usefont{T1}{fradmcn}{m}{n}\xbt 3\quad What we do see }\vspace{3.5mm}

\noindent Although forty years of experimental searches has failed
to come up with compelling evidence for specifically  QCD-motivated
exotic hadrons, strong evidence for mesons that do not fit into the
simple $\qqbar$ scheme of the original quark model has been steadily
accumulating during the past decade.~These include a candidate for a
bound state of a proton and antiproton from the BESII
experiment~[16], so-called ``baryonium'', an idea that has been
around for a long, long time~[17], and the $XYZ$ mesons,
charmoniumlike and bottomoniumlike states that do not fit into any
of the remaining unfilled states in the $\ccbar$- and $\bbbar$-meson
level schemes~[18].

\end{multicols}

\vspace{3mm} \centerline{\psfig{figure=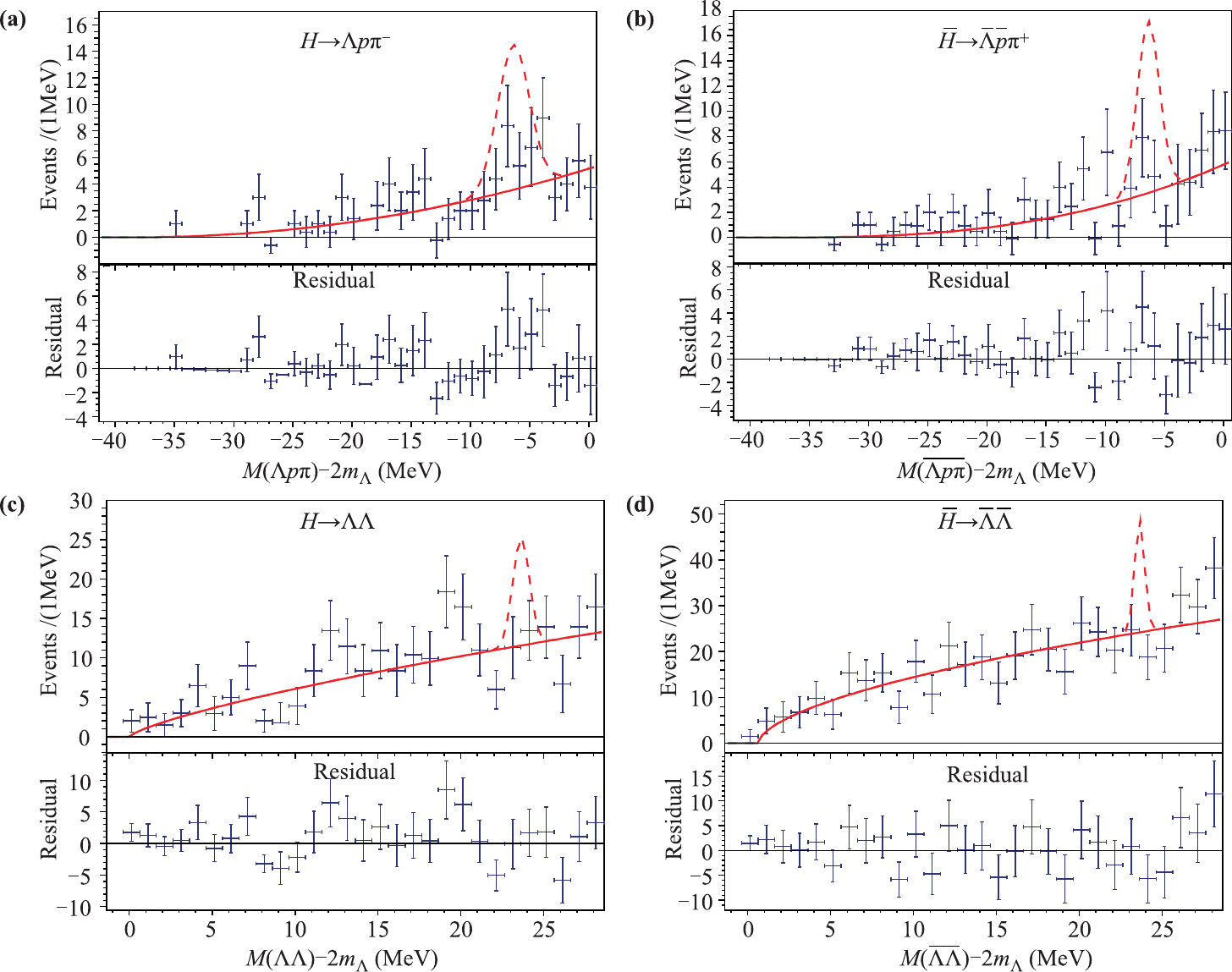}\vspace{1mm}}
 \parbox[c]{160mm}{\baselineskip
10.5pt\renewcommand{\baselinestretch}{1.05}\footnotesize \noindent
{\color{ooorangec}\bf Fig.~2}\quad {\bf(a)}--{\bf(d)} The $\lm
p\pim$, $\bar{\lm}\bar{p}\pip$, $\lm\lm$ and $\bar{\lm}\bar{\lm}$
invariant mass distributions.~The solid curves show background-only
fit results and the lower panels show the fit residuals.~The dashed
curves are expected signals for an $H$ production rate that is
1/20th that for $\bar{D}$'s.~Reproduced from Ref.~[12].

}\vspace{4mm}

\renewcommand\footnoterule{\vspace*{3pt}
\noindent
\def\temptablewidth{\textwidth}{\rule{0.3\temptablewidth}{0.5pt}}\vspace*{3pt}}

\begin{multicols}{2}
\setlength{\parindent}{1em}    

\noindent {\color{ooorangec}3.1\quad Baryonium in
radiative $\jp\rt\gamma\ppbar$ decays? }\vspace{3.5mm}

\noindent In 2003, the BESII experiment reported the observation of
a dramatic near-threshold mass enhancement in the $\ppbar$ invariant
mass spectrum in radiative $\jp\rt\gamma\ppbar$ decays, shown in the
top panel of Fig.~3(a) [16].~The lower panel in Fig.~3(a) shows the
$M(\ppbar)$ spectrum with the effects of phase-space divided out
(assuming an $S$-wave $\ppbar$ system).~It seems apparent from the
phase-space-corrected plot that the dynamical source for this
enhancement, whatever it may be, is at or below the $2m_p$ mass
threshold.~A fit with a Breit--Wigner (BW) line shape modified by a
kinematic threshold factor yielded a peak mass of
$1859^{+6}_{-27}$~MeV, about 18~MeV below $2m_{p}$, and an upper
limit of ${\it\Gamma}<30$~MeV  on the
width$^{2)}$\footnote{\hspace*{-5mm}$^{2)}$~This report only lists
quadrature sums of statistical and systematic errors.~Refer to the
cited papers for details.}.~It was subsequently pointed out that the
BW form used by BESII should be modified to include the effect of
final-state-interactions on the shape of the $\ppbar$ mass spectrum
[19, 20].~When this was done, it was found that the effects of FSI
are not sufficient to explain the observed structure and the peak
mass of the BW term shifted downward, from 1859~MeV to $1831\pm
7$~MeV while the range of allowed widths increased to
${\it\Gamma}<153$~MeV.

Soon after the BESII publication appeared,  Yan and Ding proposed a
Skyrme-like model for proton-antiproton interactions in which the
BESII $\ppbar$ mass-threshold enhancement is an $S$-wave $p\bar{p}$
bound state with binding energy around 20~MeV [21].~Since the $p$
and the $\bar{p}$ in such a system would annihilate whenever they
came within close proximity of each other, such a state would have a
finite width, creating a situation illustrated by the cartoon in
panel (b) of Fig.~3: for masses above the $2m_p$ threshold, the
state would decay essentially 100\% of the time by ``falling apart''
into a $p$ and $\bar{p}$; for masses below $2m_{p}$, the decay would
proceed via $p\bar{p}$ annihilation into mesons.~Since a preferred
channel for low-energy $S$-wave $\ppbar$ annihlation is
$\pipi\eta^{\prime}$, Yan and Ding advocated a search for
$\pipi\eta^{\prime}$ decays of this same state in radiative
$\jp\rt\gamma\pipi\eta^{\prime}$ decays.~A subsequent BESII study of
$\jp\rt\gamma\pipi\eta^{\prime}$ decays found a distinct peak at
$1834\pm 7$~MeV and width $68\pm 21$~MeV as shown in panel (c) of
Fig.~3 [22], in good agreement with the mass and width results from
the FSI-corrected fit to the $p\bar{p}$ mass spectrum.

The $\ppbar$ mass-threshold enhancement was confirmed at the same
mass with much higher statistics by BESIII [23].~The signal, shown
in panel (d) of Fig.~3, has a significance that is
$>30\sigma$$^{3)}$\footnote{\hspace*{-5mm}$^{3)}$~With apologies to
Tommaso~Dorigo:
http://www.science20.com/a\_quantum\_diaries\_survivor/a\_useful\_approximation\_for\_the\_tail\_of
$\frac{}{}~~~$\_a\_gaussian-141353}.~The large BESIII event sample
permitted the application of a partial wave analysis (PWA) that
established the $J^{PC}=0^{-+}$ quantum number assignment, in
agreement with baryonium expectations.~The $\pipi\eta^{\prime}$ peak
in $\jp\rt\pipi\eta^{\prime}$ decays was also confirmed and the
production-angle distribution was found to be consistent with a
$J^{PC}=0^{-+}$ assignment.~However, the situation still remains
unclear.~The BESIII measurements find a much larger width for the
$\pipi\eta^{\prime}$ peak than that found for the $\ppbar$ peak in
the BESIII partial wave analysis:
${\it\Gamma}_{\pipi\eta^{\prime}}=190\pm 38$~MeV versus
${\it\Gamma}_{\ppbar}<76$~MeV.~Another puzzling feature is the lack
of any evidence for the $\ppbar$ threshold enhancement in any other
channels, such as $\jp\rt\omega\ppbar$~[25],
$\yones\rt\gamma\ppbar$~[26] or in $B$ decays~[27].~BESIII is
actively looking at various other radiative $\jp$ decay channels for
evidence for or against other signs of resonance behaviour near
1835~MeV~[28].

\vspace*{5mm} \noindent {\color{ooorangec}3.2\quad The $XYZ$ mesons
}\vspace{3.5mm}

\noindent The $XYZ$ mesons are a class of (somewhat haphazardly
named) hadrons that are seen to decay to final states that contain a
heavy quark and a heavy antiquark, {i.e.}, $Q$ and $\bar{Q}$, where
$Q$ is either a $c$ or $b$ quark, but cannot be easily accommodated
in an unassigned $\QQbar$ level.~Since the $c$ and $b$ quarks are
heavy, their production from the vacuum in the fragmentation process
is heavily suppressed and, thus, any heavy quarks seen among the
decay products of a hadron must have been present among its original
constituents.~In addition, the heavy quarks in conventional $\QQbar$
``quarkonium'' mesons are slow and can be described reasonably well
by non-relativistic quantum mechanics.~Indeed it was the success of
non-relativistic charmonium potential-model descriptions of the
$\psi$ and $\chi_c$ states in the mid-1970's that led to the general
acceptance of the reality of quarks and the validity of the quark
model.~Quarkonium models specify the allowed states of a $\QQbar$
system; if a meson decays to a final state containing a $Q$ and a
$\bar{Q}$ but does not match the expected properties of any of the
unfilled levels in the associated $\QQbar$ spectrum, it is
necessarily exotic.

\vspace*{5mm} \noindent {\color{ooorangec}\it 3.2.1\quad
Charmoniumlike mesons}\vspace{3.5mm}

\noindent Charmoniumlike $XYZ$ mesons were first observed starting
in 2003 and continue to be found at a rate of about one or more new
ones every year.~There is a huge theoretical and experimental
literature on this subject that

\end{multicols}

\vspace{3mm} \centerline{\psfig{figure=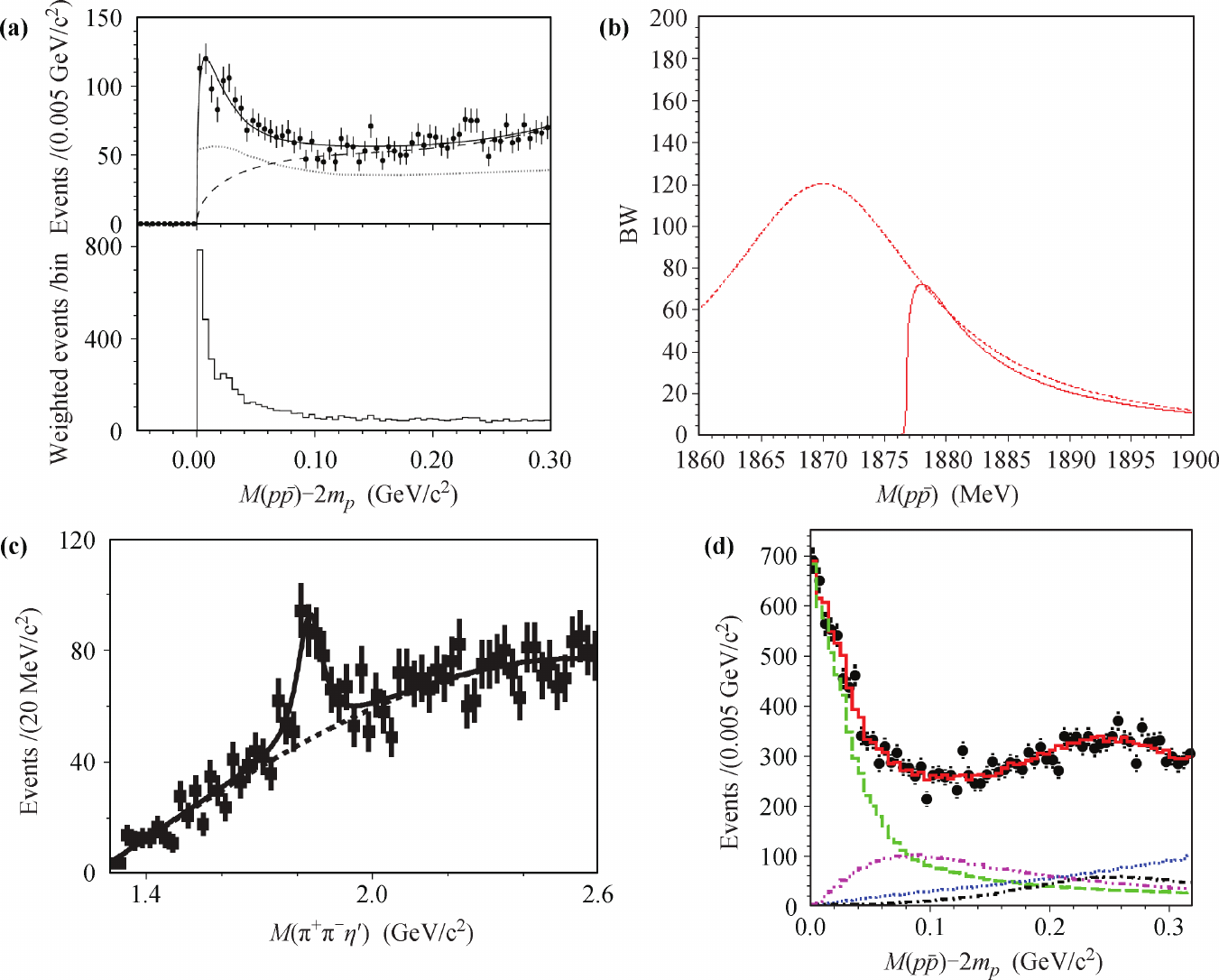}\vspace{1mm}}
 \parbox[c]{160mm}{\baselineskip
10.5pt\renewcommand{\baselinestretch}{1.05}\footnotesize \noindent
{\color{ooorangec}\bf Fig.~3}\quad {\bf(a)} The upper panel shows
the $M(\ppbar)$ distribution for $\jp\rt\gamma\ppbar$ decays from
BESII [16].~The lower panel shows the same distribution with the
kinematic threshold suppression factor removed.~{\bf(b)} A cartoon
showing a BW line shape and a threshold-attenuated BW line-shape for
a hypothesized baryonium state.~The solid curve shows the expected
$\ppbar$ line shape.~Below threshold, where the dashed curve
dominates, the state decays via $\ppbar$ annihilation into
mesons.~{\bf(c)} The $M(\pipi\eta^{\prime})$ mass distribution for
$\jp\rt\gamma\pipi\eta^{\prime}$ decays from Ref.~[22].~{\bf(d)} The
$M(\ppbar)$ distribution for $\jp\rt\gamma\ppbar$ from BESIII [23]
with PWA results shown as histograms.

}\vspace{1mm}

\begin{multicols}{2}
\setlength{\parindent}{1em}    

\noindent will not be repeated here [29,
30].~Instead I restrict myself to a few remark on subsets of the
measurements and outstanding issues.

Figure 4 shows the current state of the charmonium and
charmoniumlike meson spectrum below 4500~MeV.~Here the yellow boxes
indicate established charmonium states.~All of the (narrow) states
below the $2m_D$ open-charm threshold have been established and
found to have properties that are well described by the charmonium
model.~In addition, all of the $J^{PC}=1^{--}$ states above the
$2m_D$ open-charm threshold have also been identified.~The gray
boxes show the remaining predicted, but still unassigned, charmonium
states.~The red boxes show electrically neutral $X$ and $Y$ mesons
and the purple boxes show the charged $Z$ mesons, aligned according
to my best guess at their $J^{PC}$ quantum numbers.~In the
following, I briefly comment on each of the $XYZ$ entries in
(roughly) clockwise order, starting at the left with the $X(3940)$
and $X(4160)$.

\textbf{$\bm{X(3940)}$ {\bf and} $\bm{X(4160)}$}~~ The $X(3940)$ was
first seen by Belle~[31] as a peak in the distribution of masses
($M_X$) recoiling against a $\jp$ in inclusive $\ee\rt\jp X$
annihilations at $\sqrt{s}\simeq 10.6$~GeV shown in panel (a) of
Fig.~5.~In this figure, there are four distinct peaks: the lower
three are due to the exclusive  processes $\ee\rt\jp\eta_c$,
$\ee\rt\jp\chi_{c0}$ and $\ee\rt\jp\eta_c^{\prime}$.~The fourth peak
near 3940~MeV cannot be associated with any known or expected
charmonium state and has been named the $X(3940)$.~The curve shows
results of a fit that includes the three known charmonium states
plus a fourth state; the fit returned a mass $M=3943\pm 6$~MeV and
an upper limit on the total width of ${\it\Gamma} \le 52$~MeV.

Belle also did a study of exclusive $\ee\rt\jpsi
D^{(*)}\bar{D}^{(*)}$ decays in the same energy region.~Here, to
compensate for the low detetction efficiency for $D$ and $D^*$
mesons, a partial reconstruction technique was used that required
the reconstruction of the $\jpsi$ and only one $D$ or $D^*$
meson$^{4)}$\footnote{\hspace*{-5mm}$^{4)}$~In the remainder of this
report, the inclusion of charge conjugate states is always
implied.}, and determined the presence of the $\bar{D}$ or
$\bar{D}^*$ from energy momentum conservation [32].~With this
technique, the $X(3940)$ was seen in the
$D\bar{D}^*$\vspace{-0.5cm}\linebreak

\end{multicols}

\vspace{3mm} \centerline{\psfig{figure=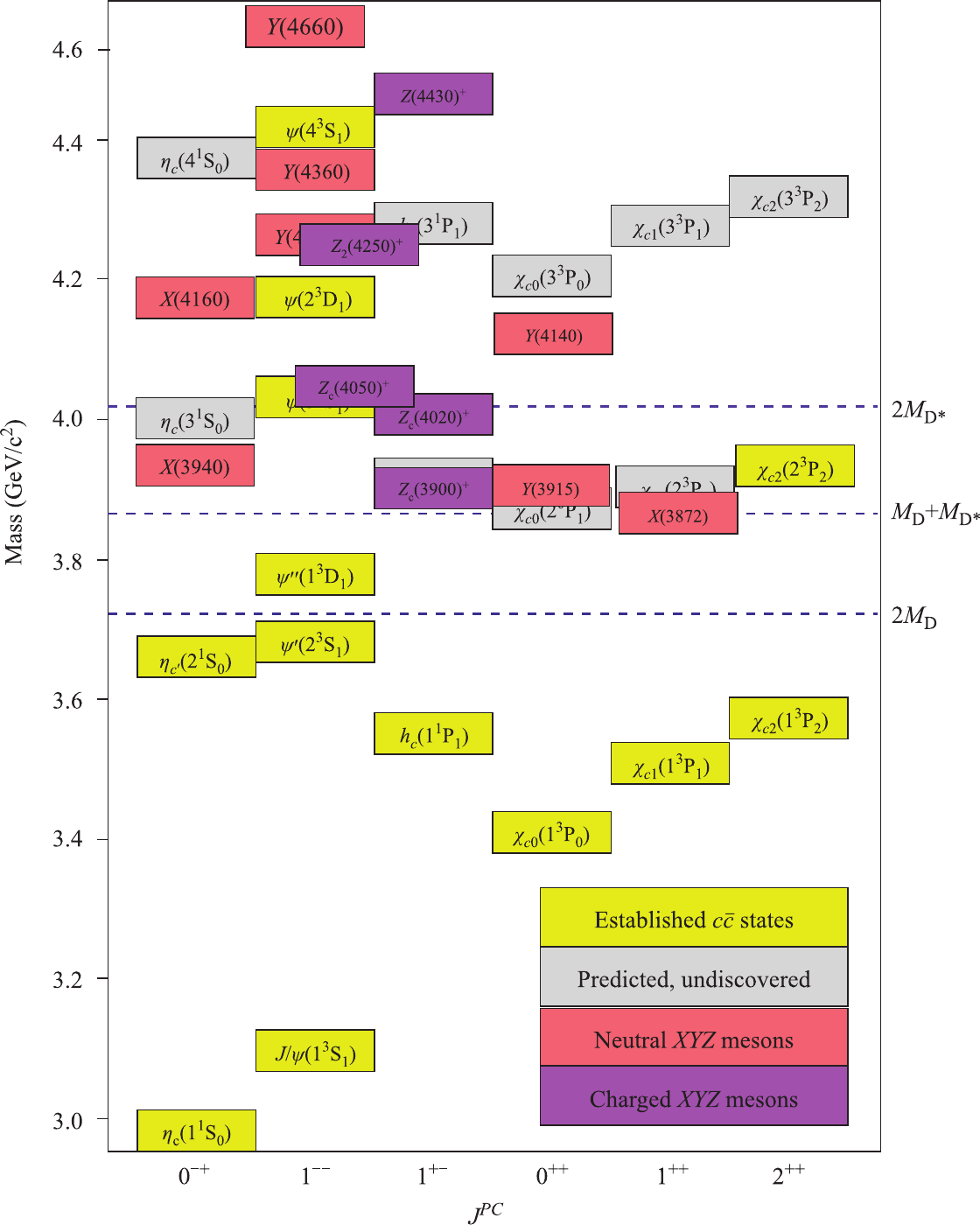}\vspace{1mm}}
 {\center\baselineskip
10.5pt\renewcommand{\baselinestretch}{1.05}\footnotesize \noindent
{\color{ooorangec}\bf Fig.~4}\quad The spectrum of charmonium and
charmoniumlike mesons

}\vspace{0mm}

\begin{multicols}{2}
\setlength{\parindent}{1em}    

\noindent invariant mass distribution in $\ee\rt\jpsi D\bar{D}^*$
annihilations [32], as can be seen in the lower panel of Fig.~5(b)
and the upper panel in Fig.~5(c).~Here the fits shown in the plots
return a mass and width of $M=3942\pm 9$~MeV and
${\it\Gamma}=37^{+27}_{-17}$~MeV.~The $\ee\rt \jpsi D^*\bar{D}^*$
study uncovered another, higher mass state decaying to
$D^*\bar{D}^*$ as can be seen in the lower panel of Fig.~5(c).~The
fitted mass and width of this state, which is called the $X(4160)$,
is $M=4156\pm 27$~MeV and ${\it\Gamma}= 139^{+113}_{-65}$~MeV~[32].

Neither the $X(3940)$ nor the $X(4160)$ show up in the $D\bar{D}$
invariant mass distribution for exclusive $\ee\rt\jpsi D\bar{D}$ at
the same energies.~Instead, the $M(D\bar{D})$ spectrum exhibits a
broad excess of events over an equally broad background as shown in
the upper panel of Fig.~5(b).~A fit to a resonant shape, shown as a
curve in the figure, returns a signal of marginal significance
($3.8\sigma$) with a peak mass of $M=3780\pm 48$~MeV and a width
${\it\Gamma}=347^{+316}_{-143}$~MeV.~Since the fitted values are
unstable under variations of the background shape parameterization
and the bin size, Belle makes no claims about this distribution
other than that it is inconsistent with phase space or pure
background.~On the other hand, Chao~[33] suggests that this may be
the $\chi^{\prime}_{c0}$, which is discussed below in conjunction
with the $X(3915)$.

The absence of signals for any of the known spin non-zero charmonium
states in the inclusive spectrum of Fig.~5(a) provides
circumstantial evidence for $J=0$ assignments for the $X(3940)$ and
$X(4160)$.~The $X(3940)\rt D^*\bar{D}$ decay mode then ensures that
its $J^{PC}$ values are $0^{-+}$.~The absence of any signal for
$X(4160)\rt D\bar{D}$ decay supports a $0^{-+}$ assignment for this
state as well.~In both cases, the measured masses are far below
expectations for the only available $0^{-+}$ charmonium levels: the
$\eta_c(3S)$ and $\eta_c(4S)$.~Since there are no strong reasons to
doubt the generally accepted identifications of the $\psi(4040)$
peak seen in the inclusive cross section for

\end{multicols}

\vspace{3mm} \centerline{\psfig{figure=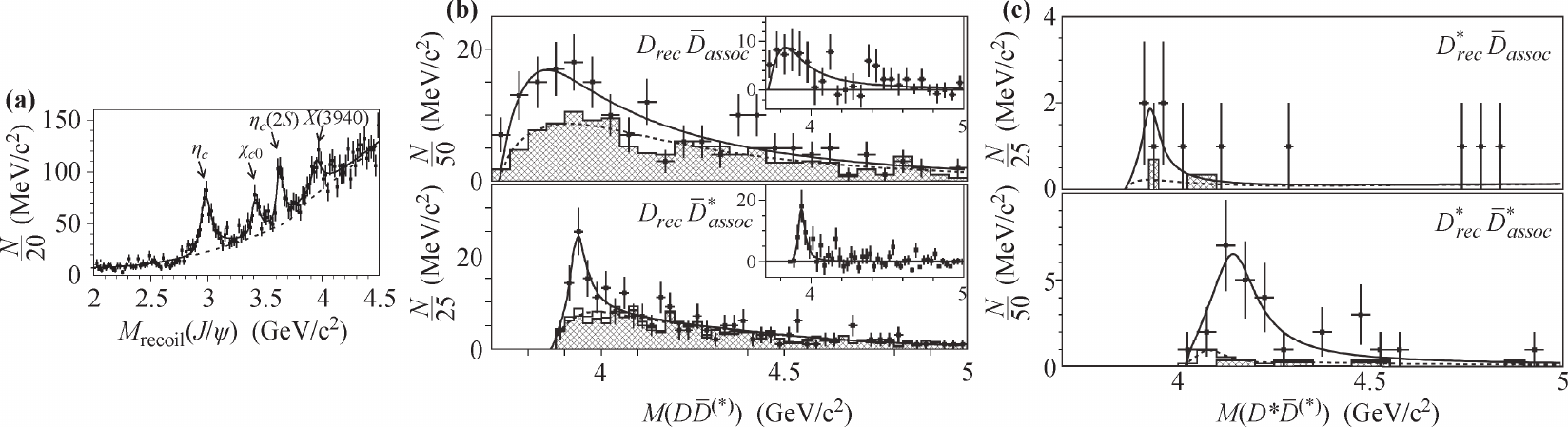}\vspace{1mm}}
 \parbox[c]{160mm}{\baselineskip
10.5pt\renewcommand{\baselinestretch}{1.05}\footnotesize \noindent
{\color{ooorangec}\bf Fig.~5}\quad {\bf(a)} The distribution of
masses recoiling from a $\jpsi$ in inclusive $\ee\rt \jpsi X$
annihilations near $\sqrt{s}\simeq 10.6$~GeV.~The curve is the
result of a fit described in the text.~{\bf(b)} The $D\bar{D}$ ({\it
upper}) and $D\bar{D}^*$ ({\it lower}) invariant mass distributions
from $\ee\rt \jpsi D\bar{D}^{(*)}$ annihilations near $\sqrt{s}
\simeq 10.6$~GeV.~Here the $\jpsi$ and a $D$ meson are
reconstructed, and the presence of the  $\bar{D}$ or $\bar{D}^*$ is
inferred from kinematics.~{\bf(c)} The $D^*\bar{D}$ ({\it upper})
and $D^*\bar{D}^*$ ({\it lower}) invariant mass distributions from
$\ee\rt \jpsi D^*\bar{D}^{(*)}$ annihilations near $\sqrt{s} \simeq
10.6$~GeV.~Here the $\jpsi$ and a $D^*$ are reconstructed and the
presence of the $\bar{D}$ or $\bar{D}^*$ is inferred from
kinematics.~The hatched histograms in the center and right panels
show backgrounds estimated from the $\jpsi$ and $D^{(*)}$ mass
sidebands; the insets in the center panels show
background-subtracted fits.

}\vspace{0mm}

\begin{multicols}{2}
\setlength{\parindent}{1em}    

\noindent $\ee\rt$ {\it hadrons} as the $\psi(3S)$ and the
$\psi(4415)$ peak as the $\psi(4S)$ [34, 35], these assignments
would imply hyperfine $n^3S-n^1S$ mass splittings that increase from
the measured value of $47.2\pm 1.2$~MeV for $n=2$ [13], to $\sim
100$~MeV for $n=3$ and $\sim 350$~MeV for $n=4$ [33].~This pattern
conflicts with expectations from potential models, where hyperfine
splittings are proportional to the square of the radial wavefunction
at $r=0$ and decrease with increasing $n$ [36].

\textbf{$\bm{Y(4260)}$, $\bm{Y(4360)}$ {\bf and} $\bm{Y(4660)}$:}
BaBar\vspace{-0.4cm}\linebreak

\end{multicols}

\vspace{0mm} \centerline{\psfig{figure=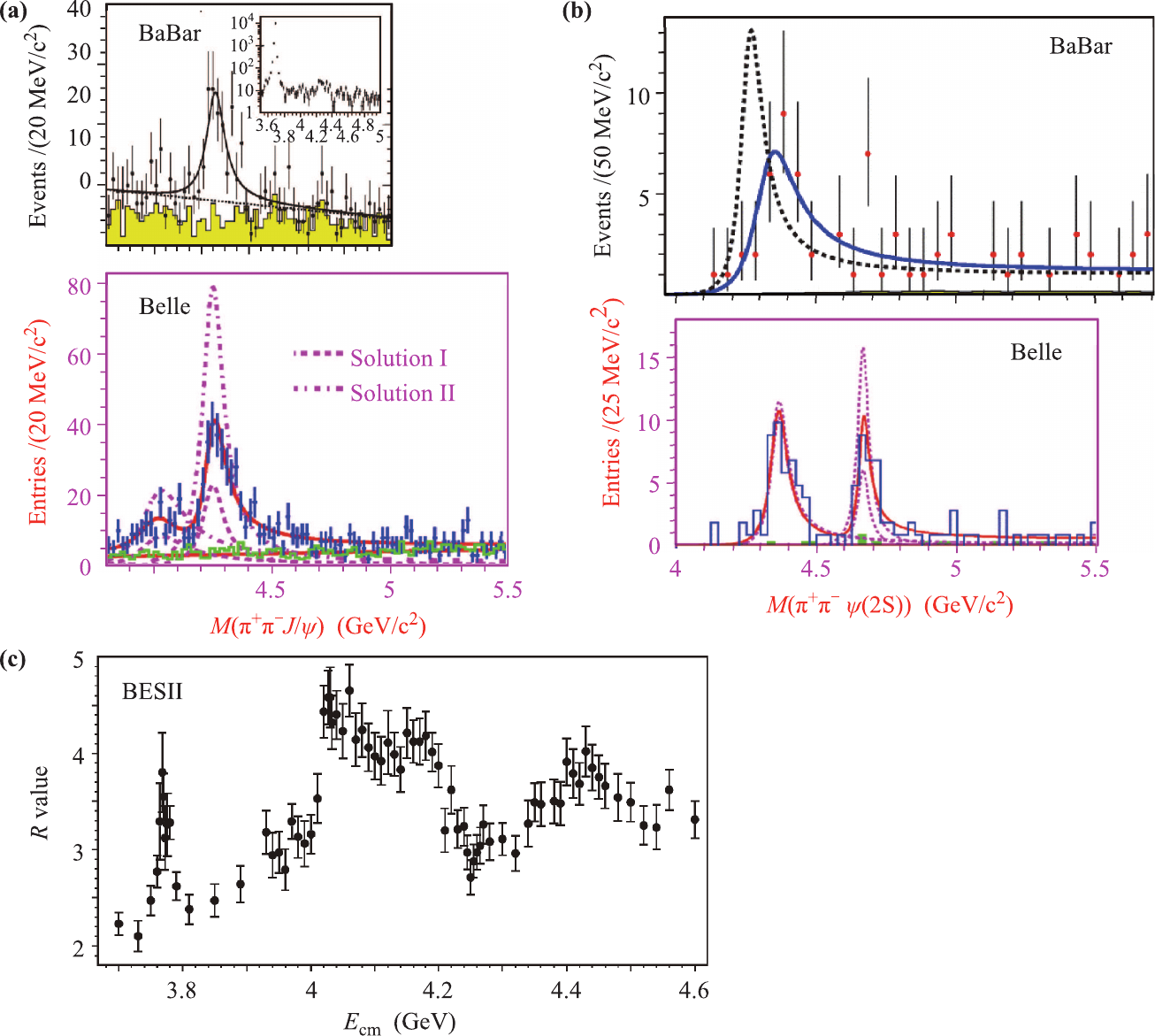}\vspace{1mm}}
 \parbox[c]{160mm}{\baselineskip
10.5pt\renewcommand{\baselinestretch}{1.05}\footnotesize \noindent
{\color{ooorangec}\bf Fig.~6}\quad {\bf(a)} The  $M(\pipi\jpsi)$
from $\ee\rt\gamma_{\rm isr} \pipi\jpsi$ events from BaBar ({\it
upper}) and Belle ({\it lower}).~{\bf(b)} The  $M(\pipi\psip)$ from
$\ee\rt\gamma_{\rm isr} \pipi\psip$ events from BaBar ({\it upper})
and Belle ({\it lower}).~{\bf(c)} The total inclusive Born cross
section for $\ee\rt~hadrons$ in units of $\sigma_{\rm
QED}(\ee\rt\mumu)$ from BESII.

}\vspace{4mm}

\begin{multicols}{2}
\setlength{\parindent}{1em}    

\noindent discovered the $Y(4260)$ as a peak near 4260~MeV in the
$M(\pipi\jp)$ distribution from initial-state-radiation
$\ee\rt\gamma_{\rm isr}\pipi\jp$ events [37] shown in the upper
panel of Fig.~6(a), and the $Y(4360)$ in the $\pipi\psi^{\prime}$
system produced via $\ee\rt\gamma_{\rm isr}\pipi\psip$ [38], shown
in the upper panel of Fig.~6(b).~These states were confirmed by
Belle, as shown in the lower panels of the same figures [39,
40].~Belle found another peak, the $Y(4660)$, at higher mass in the
$M(\pipi\psip)$ distribution.\ There is no sign of the $Y(4260)$ in
the $M(\pipi\psip)$ distributions: the expected shape for
$Y(4260)\rt \pipi\psip$ is shown as a dashed curve in the upper
panel of Fig.~6(b).~Neither is there any sign of the $Y(4360)$ or
$Y(4660)$ in the $M(\pipi\jpsi)$ distributions of Fig.~6(a).

These states have to be considered exotic because their production
mechanism ensures that $J^{PC}=1^{--}$ and all of the $1^{--}$
$\ccbar$ states near their masses have already been
assigned.~Moreover, there is no evidence for them in any exclusive
[41--44] or the inclusive [34, 35] charmed-meson production cross
section, where there is a pronounced dip at $\sqrt{s}\simeq
4.26$~GeV and no striking feature near 4.36~GeV, as can be seen in
BESII cross section data shown in Fig.~6(c) [34, 35].~This implies
large partial decay widths to $\pipi\jp(\psip)$; for example: a
specific analysis for the $Y(4260)$ finds ${\it\Gamma}(Y(4260)\rt
\pipi\jp)>1$~MeV~[45], which is huge by charmonium standards.~Some
authors have proposed that the $Y(4260)$ is a $\ccbar$-gluon hybrid
state [46, 47] while others have suggested that it is a
molecule-like $D\bar{D}_1(2420)$ bound-state [48].~The $Y(4260)$ is
discussed in more detail below.

\textbf{\bf The electrically charged  {$\bm{Z_c(3900)}$,
$\bm{Z_c(4020)}$, $\bm{Z_1(4050)}$, $\bm{Z_2(4250)}$} {\bf and}
{$\bm{Z(4430)}$} states:} Since the $Z_c(3900)$ [49, 50],
$Z_c(4020)$ [51, 52], $Z_1(4050)$ [53], $Z_2(4250)$ [53] and
$Z(4430)$ [54] are electrically charged and decay to hidden charm
final states, their minimal quark structure is a $\ccbar u\bar{d}$
four-quark combination and, therefore, if they are mesons, they must
be exotic.

\underline{The {\it Z(4430):}}~~ The first electrically charged,
charmoniumlike state to be reported was the $Z(4430)$, which was
found by Belle in 2008 in a study of $B\rt K\pip\psip$ decays
[54].~An $M^2(\pip\psip)$ {vs.} $M^2(K\pip)$ Dalitz plot for these
decays is shown in Fig.~7(a).~There, the most prominent feature is a
strong vertical band near $M^2(K\pip)\simeq 0.8$~GeV$^2$
corresonding to $B\rt K^*(890)\psip$ decays.~A second vertical band
near $M^2(K\pip)\simeq 2$~GeV$^2$ corresponds to $B\rt
K^*_2(1430)\psip$.~To reduce the influence of these structures,
Belle studied events with $M(K\pip)$ more than 100~MeV away from the
$K^*(890)$ and $K^*_2(1430)$ peak mass values (the $K^*$ veto).~The
$\pip\psip$ invariant mass distribution in $B\rt K\pip\psip$ decay
events that survive this $K^*$ veto,  shown an open histogram in
Fig.~7(b), exhibits a distinct peak near $M(\pip\psip)\simeq
4.43$~GeV.~A fit to a smooth background plus a BW signal function to
describe the peak returns a mass and width of $M=4433\pm 5$~MeV and
${\it\Gamma} = 45^{+35} _{-18}$~MeV; the statistical significance of
the peak is $6.5\sigma$.~However, a 2009 analysis by BaBar failed to
confirm the Belle signal [55].~Figure~7(c), taken from the BaBar
paper, presents a direct comparison between the Belle and BaBar
results with the same $K^*$ veto requirements.~A BW fit by BaBar
that used Belle's $Z(4430)$ mass and width values found a signal
with a statistical significance of only $\simeq 2\sigma$.

One concern about the $Z(4430)$ was that neither the BaBar nor the
2008 Belle analysis considered the possibility of interference
between $B\rt K Z(4430)$ and $B\rt K^*\psip$ amplitudes.~Because of
this, Belle did two subsequent analyses that explicitly accounted
for the possibility of such interference [56, 57].~The decay chain
$B\rt K\pip\psip$; $\psip\rt\leplep$ can be completely described by
four variables$^{5)}$\footnote{\hspace*{-5mm}$^{5)}$~This neglects
the finite width of the $\psip$ resonance.}, which can be taken as
$M(\pip\psip)$, $M(K\pip)$, the $\psip$ helicity angle $\theta$ and
the angle between the $K\pip$ and $\leplep$ planes in the $B$ rest
frame $\phi$.~Belle's first amplitude analysis integrated over
$\theta$ and $\phi$ and only considered the $K\pip$ and $\pip\psip$
masses [56]; the second analysis used all four variables [57].~Both
amplitude analyses found strong signals ($6.4\sigma$ in both cases)
for a resonance in the $\pip\psip$ channel, with a somewhat higher
peak mass and much broader width than that reported in the 2008
Belle paper; the four-dimensional analysis, based on $2010\pm 50$
$B\rt K\pip\psip$ signal events, reported $M=4485\pm
^{+36}_{-25}$~MeV and ${\it\Gamma} = 200^{+49}_{-58}$~MeV.~These
amplitude analyses show strong interference between the
$Z\rt\pip\psip$ and $K^*\rt K\pip$ channels, as is evident in
Fig.~8(a) where the $M^2(\pip\psip)$ distribution for the data
surviving the $K^*$ veto (points with errors) are shown with
projections of the four-dimensional fit results
superimposed$^{6)}$\footnote{\hspace*{-5mm}$^{6)}$~For the
four-dimensional analysis, only the $\psip\rt\leplep$ decay channel
was used.}.~The fit that includes the $Z^+\rt\pip\psip$ resonance is
shown as a solid histogram and the best fit with no $Z^+$ resonance
is shown as a dashed curve.~For masses below the $Z^+$ resonance
peak, the intereference is constructive and there is a clear
positive enhancement;  for masses above the peak the interference is
destructive and produces a depletion of events.~The original Belle
result, which was based on a one-dimensional fit to the
$M(\pip\psip)$ distribution, only fit the low-mass positive lobe and
this resulted in a lower mass and a narrower width.~The
four-dimensional analysis strongly favored the $J^P=1^+$ quantum
number assignment for\vspace{-0.4cm}\linebreak

\end{multicols}

\vspace{3mm} \centerline{\psfig{figure=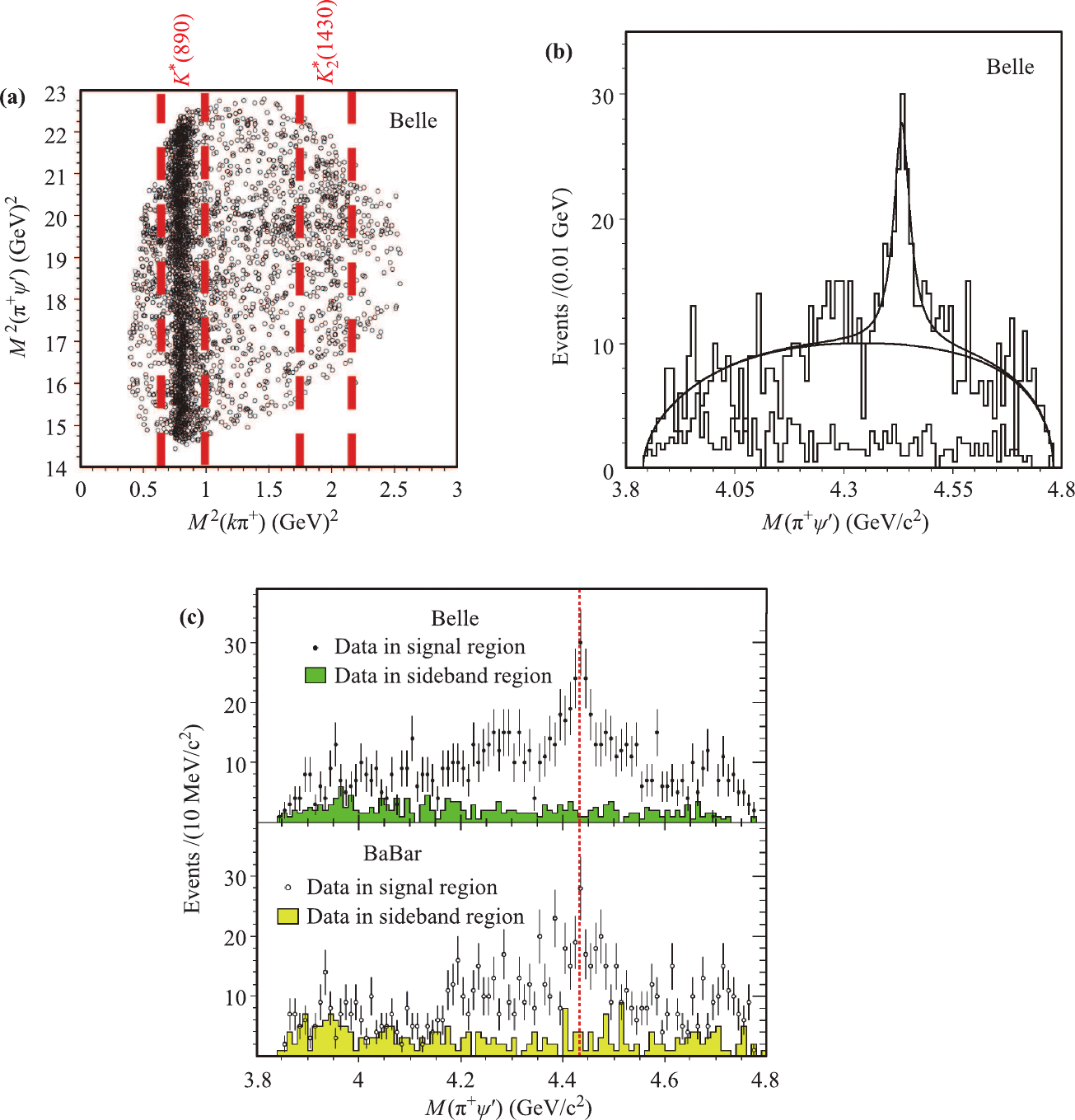}\vspace{1mm}}
 \parbox[c]{160mm}{\baselineskip
10.5pt\renewcommand{\baselinestretch}{1.05}\footnotesize \noindent
{\color{ooorangec}\bf Fig.~7}\quad {\bf(a)} An $M^2(\pip\psip)$
(\emph{vertical}) {vs.} $M^2(K\pip)$ (\emph{horizontal}) Dalitz plot
for $B\rt K\pip\psip$ events from Belle.~The vertical dashed red
lines indicate the boundaries of $K^*$ veto requirement described in
the text.~{\bf(b)} The $\pip\psip$ invariant mass distribution from
$B\rt K\pip\psip$ decays from Belle [54] for events with the $K^*$
veto requirement applied is shown as the open histogram.~The shaded
histogram is non-$\psip$ background, estimated from the $\psip$ mass
sidebands.~The curves represent results of a fit described in the
text.~{\bf(c)}  A comparison of the Belle data ({\it upper}) and
BaBar data ({\it lower}) [55] with the the $K^*$ and $K^*_2$ vetoed.

}\vspace{2mm}

\begin{multicols}{2}
\setlength{\parindent}{1em}    

\noindent  the $Z(4430)$.

In 2014, the LHCb experiment repeated the Belle four-dimensional
analysis with a sample of more than 25 K $\bar{B}^0\rt K^-\pip\psip$
decays [58] and confirmed the Belle signal with a significance
greater than $14\sigma$.~A comparison of their $M(\pip\psip)$ data
distribution with projections of their fit results superimposed is
shown in Fig.~8(b), where strong interference effects, similar in
character to those reported by Belle, are apparent.~The LHCb results
for the mass, width and $J^P$ values, $M=4475^{+17}_{-26}$~MeV,
${\it\Gamma} = 172^{+39}_{-36}$~MeV and $J^P=1^+$, agree well with
the Belle four-dimensional fit results but with smaller errors.~With
their ten-fold larger event sample, the LHCb group was able to relax
the assumption of a BW form for the $Z^+$ amplitude and directly
measure the real and imaginary parts of the amplitude as a function
of mass.~The results are shown as data points in Fig.~8(c).~There
the phase motion near the resonance peak agrees well with
expectations for a BW amplitude as indicated by the nearly circular
red curve superimposed on the plot.~This rapid phase motion is a
clear signature for a BW-like resonance behavior but, by itself,
does not necessarily rule out explanations of the $Z(4430)$
structure as being due to a rescattering process [59, 60].

The Belle and LHCb product branching fractions for $Z(4430)^+$
production in $\bar{B}^0$ decays are

 \vspace{0.25cm}$\displaystyle {\mathcal B}(\bar{B}^0\rt
K^- Z(4430)^+)\times {\mathcal B}(Z(4430)^+\rt \pip\psip) $

\vspace{0.15cm} $~~~~~~~ =  6.0 ^{+3.0}_{-2.4}\times 10^{-5}~{\rm
Belle,\ and}$


\end{multicols}

\vspace{3mm} \centerline{\psfig{figure=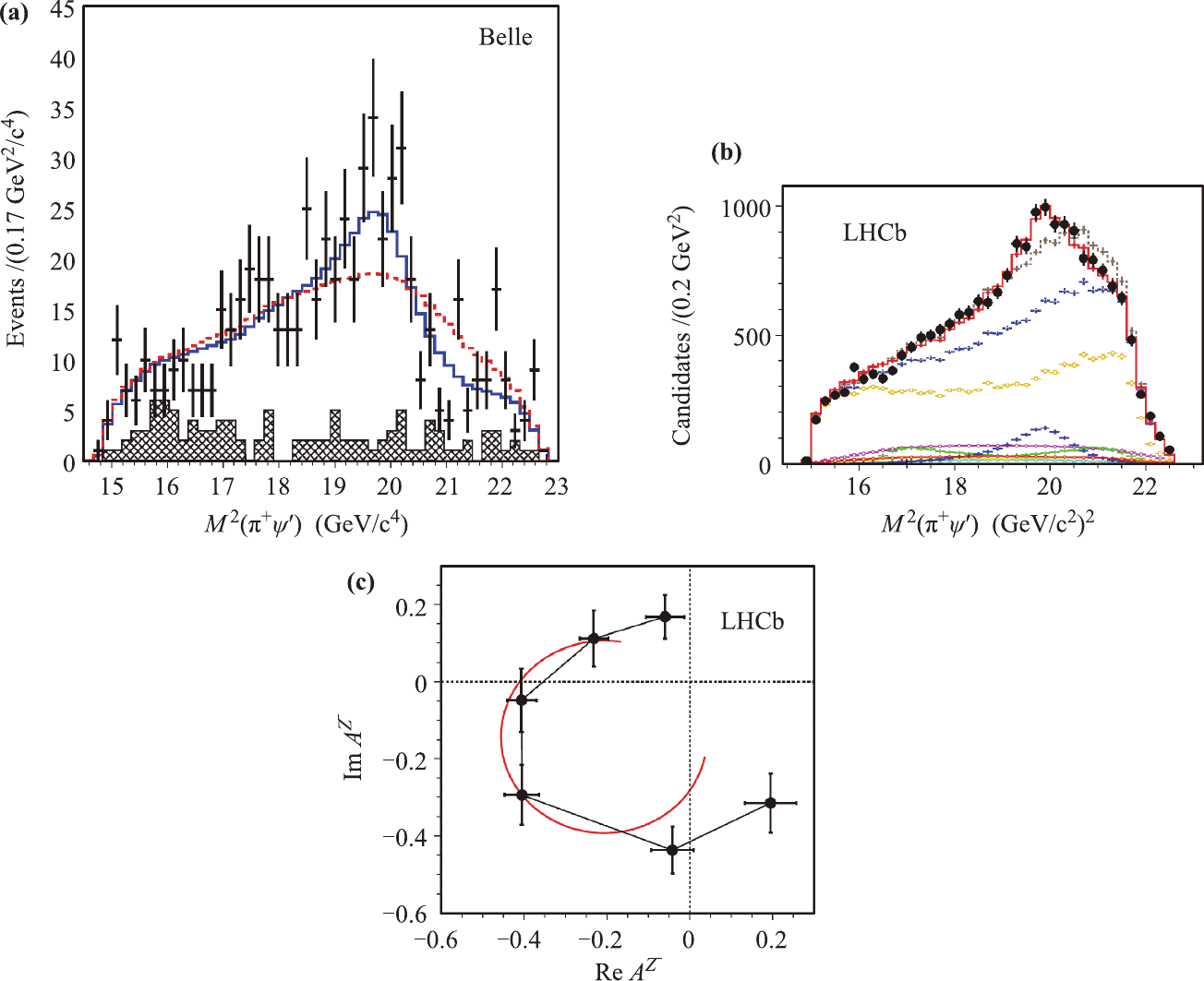}\vspace{1mm}}
 \parbox[c]{160mm}{\baselineskip
10.5pt\renewcommand{\baselinestretch}{1.05}\footnotesize \noindent
{\color{ooorangec}\bf Fig.~8}\quad {\bf(a)} The data points show the
Belle $M^2(\pip\psip)$ distribution with the $K^*$ veto applied.~The
solid (blue) histogram shows four-dimensional fit results with a $
Z^+\rt\pip\psip $ resonance included.~The dashed  (red) curve shows
best fit results with no resonance in the $\pip\psip$ channel.~{\bf
(b)} The LHCb group's $M^2(\pip\psip)$ distribution for all events
(no $K^*$ veto), together with projections from the four-dimensional
fits.~The solid red histogram shows the fit that includes a
$Z^+\rt\pip\psip$ resonance term; the dashed brown histogram shows
the fit with no resonance in the $\pip\psip$ channel.~{\bf (c)} The
Real (\emph{horizontal}) and Imaginary (\emph{vertical}) parts of
the ($1^+$) $Z^+\rt\pip\psip$ amplitude for different mass bins
spanning the $4430$~MeV mass region from LHCb [58].~The red curve
shows expectations for a BW resonance amplitude.

}\vspace{4mm}

\begin{multicols}{2}
\setlength{\parindent}{1em}    

\vspace{0.15cm} ${\mathcal B}(\bar{B}^0\rt K^- Z(4430)^+)\times
{\mathcal B}(Z(4430)^+\rt \pip\psip) $

\vspace{0.15cm} $~~~~~~~ =  3.4 ^{+1.1}_{-2.4}\times 10^{-5}~{\rm
LHCb}; $

\vspace{0.25cm} \noindent the weighted average is $4.4\pm 1.7 \times
10^{-5}.$   There is no experimental value for ${\mathcal
B}(\bar{B}^0\rt K^- Z(4430)^+)$, although we expect that it cannot
be larger than ${\mathcal B}(B^-\rt K^- X(3872))$, for which BaBar
has set a 90\% CL upper limit of $3.2\times 10^{-4}$ [61].~This
seems reasonable because the leading quark-line diagram for $B\rt K
X(3872)$ is a ``factorizable'' weak interaction process that is
favored in $B$ meson decays, while the leading-order diagram for
producing a charged charmoniumlike state is ``non-factorizable'',
and expected to be suppressed in $B$ meson decays [63--65].~Thus, it
is probably safe to expect that ${\mathcal B}(\bar{B}^0\rt K
Z(4430))<{\mathcal B}(B^-\rt K^- X(3872))<3.2\times 10^{-4}$.~In
that case, the Belle-LHCb average given above give a {\it lower}
limit ${\mathcal B}(Z(4430)^+\rt \pip\psip) > 5.3\%$ that, when
coupled with our average of the Belle and LHCb measurements of the
$Z$ total width (${\it\Gamma}_{Z(4430)} = 181\pm 31$~MeV), gives a
lower limit on the partial width
${\it\Gamma}(Z(4430)^+\rt\pip\psip)>7.5$~MeV, which is very large by
charmonium standards: for example, the largest measured hadronic
transition width between well established charmonium states is
${\it\Gamma}(\psip\rt\pipi\jpsi) = 102\pm 3$~keV
[13]$^{7)}$\footnote{\hspace*{-5mm}$^{7)}$~Hadronic transitions
between charmonium states are OZI suppressed.~This is not the case
for four-quark charmoniumlike states.}.

Another striking feature of the $Z(4430)$ is that its $\pip\psip$
partial decay width is much stronger than that for $\pip\jpsi$.~In
2009, BaBar reported ${\mathcal B}(\bar{B}^0\rt K^- Z(4430))\times
{\mathcal B}(Z(4430)\rt \pip\jpsi) < 0.4\times 10^{-5}$, an order of
magnitude lower than the Belle-LHCb average value for $Z(4430)\rt
\pip\psip$.~Belle recently reported the first observation of the
$Z(4430)\rt\pip\jpsi$ decay mode in $B\rt K\pip\jpsi$ decays with a
product branching fraction of $5.4 ^{+4.1}_{-1.3}\times 10^{-6}$
[62], which is near the BaBar upper limit and about an order of
magnitude smaller than the corresponding rate for $Z(4430)^+$
production in the $B\rt K\pip\psip$ channel.

\underline{\it The $Z_1(4050)$ and $Z_2(4250):$} The $Z_1(4050)$ and
$Z_2(4250)$ are resonances in the $\pip\chicone$ channel that were
found by Belle in a two-dimensional ($M(K\pip)$ {\it vs.}
$M(\pip\chicone)$) amplitude analysis of $B\rt K\pip\chicone$ decays
[53].~Figure~9(a) shows the $M^2(\pip\chicone)$ distribution for
events with $1.0\ {\rm GeV} < M(K\pip) < 1.32\ {\rm GeV}$ ({\it
upper}) and $M(K\pip)> 1.54$~GeV ({\it lower}) with the best fit for
a model with all known $K\pi$ resonances but no resonance in the
$\pip\chicone$ channel.~Figure~9(b) shows the same data with a fit
that includes one resonance in the $\pip\chicone$ channel.~Here the
fit quality improves substantially; the statistical significance of
the $Z\rt\pip\chicone$ term is more than $10\sigma$.~The use of
different parameterizations of the $K\pi$ scalar amplitude or the
inclusion of an additional $J^P=1^-$ or $2^+$ $K^*$ resonance with a
mass and width left as free parameters do not reduce the
significance of the $Z^+\rt \pip\chicone$ term to below $6\sigma$.

It is evident in Fig.~9(b) that the inclusion of a single $Z\rt
\pi\chicone$ amplitude does not reproduce the details of the
data.~Because of this, Belle repeated the fit with two BW amplitudes
in the $Z\rt\pip\chicone$ channel.~This model fits the data well, as
can be seen in Fig.~9(c); the significance of the two $Z$ model
relative to the single $Z$ model is $5.7\sigma$.~The masses and
widths of the two $Z$s are

\vspace{0.25cm}$\displaystyle Z_1(4050){:}~~ M_1       = 4051
^{+24}_{-43} \ {\rm MeV}, ~~ {\it\Gamma}_1 = 82^{+51}_{-28} \ {\rm
MeV}$

\vspace{0.15cm}$\displaystyle Z_2(4250){:}~~ M_2      = 4248
^{+185}_{-45} \ {\rm MeV},~~ {\it\Gamma}_2 = 177^{+321}_{-72} \ {\rm
MeV.} $

\vspace{0.25cm} \noindent

A BaBar study of $B\rt K\pip\chicone$ decays did not confirm the
Belle claim of resonances in the $\pip\chicone$ channel
[66].~Figure~10(a) shows BaBar's background-subtracted
$M(\pip\chicone)$ distribution for events with $1.0\ {\rm GeV} <
M(K\pip) < 1.32\ {\rm GeV}$.~The dashed curve shows the projection
of the BaBar fit with no $Z^+$ resonances in the $\pip\chicone$
channel.~This fit obviously overshoots the data in the
$M(\pip\chicone)\sim 3.9$~GeV region and undershoots the data near
$M(\pip\chicone)\sim 4.1$~GeV.~The red curve shows the result when
two $Z^+\rt\pip\chicone$ resonance terms with the Belle mass and
width values are added;  this improves the agreement with the data,
but not markedly.~The BaBar report claims that, overall, the
$M(\pip\chicone)$ distributions in the data [including those in the
regions of $M(K\pip)$ that are not shown in Fig.~10(a)] are
adequately described by resonances in the $K\pi$ sector only, and
the inclusion of $Z^+$ resonances does not signficantly improve the
fit quality.~Based on this they set 90\% CL product branching
fraction upper limits of [66]

\vspace{0.25cm}$\displaystyle {\mathcal B}(\bar{B}^0\rt K^-
Z_1(4050)^+)\times {\mathcal B}(Z_1(4050)^+$

\vspace{0.15cm}$\displaystyle ~~~~~~~~\rt \pip\chicone)  < 1.8\times
10^{-5}~{\rm and}$

\vspace{0.15cm}$\displaystyle  {\mathcal B}(\bar{B}^0\rt K^-
Z_2(4250)^+)\times {\mathcal B}(Z_2(4250)^+$

\vspace{0.15cm}$\displaystyle ~~~~~~~~\rt \pip\chicone)  < 4.0\times
10^{-5}, $

\vspace{0.25cm} \noindent which are below, but not in strong
contradiction with,

\end{multicols}

\vspace{1mm} \centerline{\psfig{figure=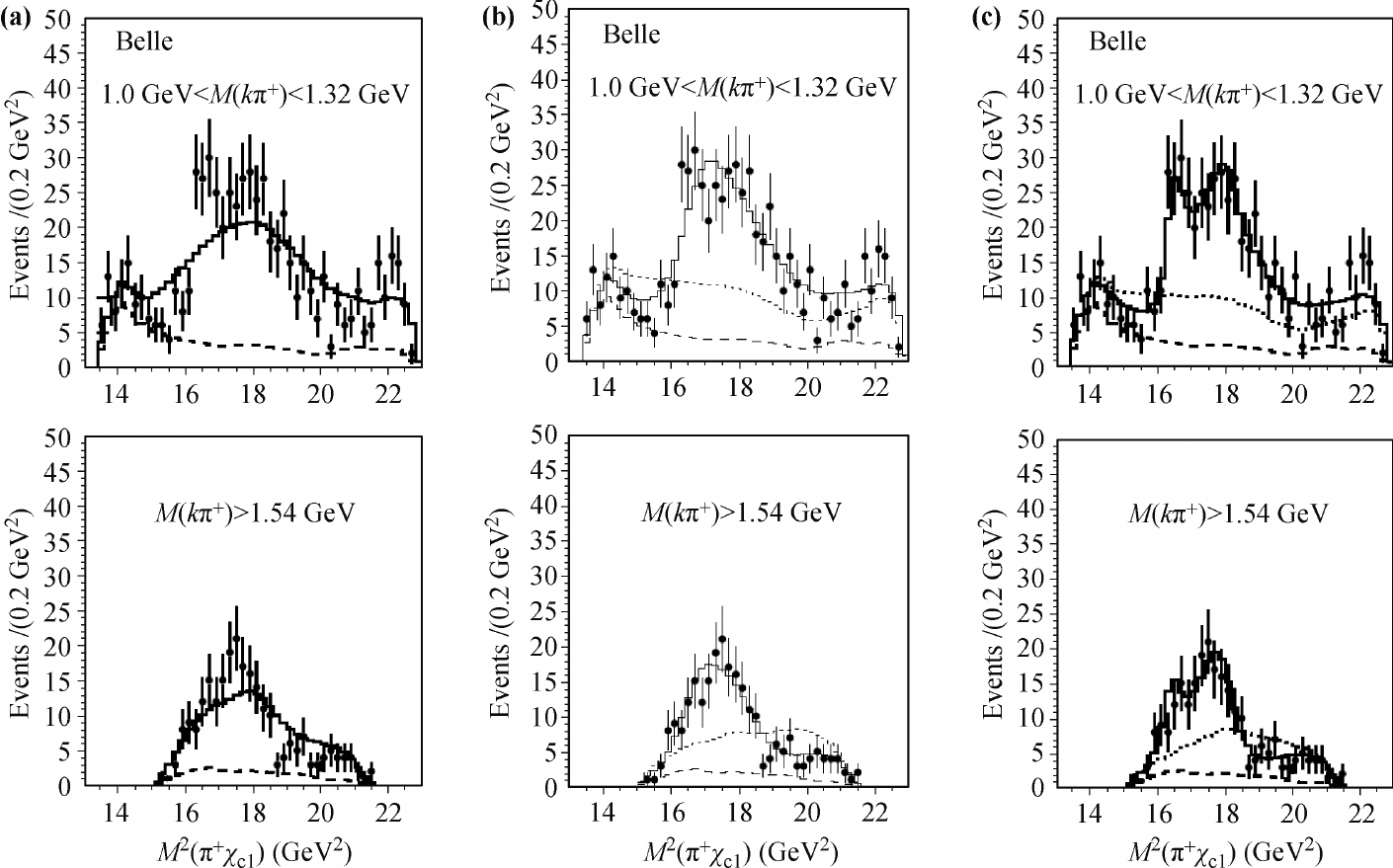}\vspace{1mm}}
 \parbox[c]{160mm}{\baselineskip
10.5pt\renewcommand{\baselinestretch}{1.05}\footnotesize \noindent
{\color{ooorangec}\bf Fig.~9}\quad {\bf(a)} The $M^2(\pip\chicone)$
distribution for $B\rt K\pip\chicone$ events with $1.0\ {\rm
GeV}<M(K\pip)< 1.32\ {\rm GeV}$ ({\it upper}) and
$M(K\pip)>1.54$~GeV ({\it lower}) from Belle.~The solid histograms
are projections of the two-dimensional fit with no resonance in the
$\pip\chicone$ channel.~The dashed histogram indicates the
background level determined from the $\chicone$ mass
sidebands.~{\bf(b)}  The same data with results of the fit with a
single $Z^+ \rt \pip\chicone$ resonance amplitude.~Here the dotted
histogram indicates the sum of all fit components other than the
$Z^+$.~{\bf(c)}  The same data with results of the fit that includes
two $Z^+\rt \pip\chicone$ resonances with mass and width values
given in the text.

}\vspace{0mm}

\begin{multicols}{2}
\setlength{\parindent}{1em}    

\noindent
 the Belle measurements [53]

\vspace{0.25cm}$\displaystyle {\mathcal B}(\bar{B}^0\rt K^-
Z_1(4050)^+)\times {\mathcal B}(Z_1(4050)^+$

\vspace{0.1cm}$\displaystyle ~~~~~~~~\rt \pip\chicone)  = 3.0
^{+4.0}_{-1.8}\times 10^{-5}~{\rm and} $

\vspace{0.1cm}$\displaystyle {\mathcal B}(\bar{B}^0\rt K^-
Z_2(4250)^+)\times {\mathcal B}(Z_2(4250)^+$

\vspace{0.15cm}$\displaystyle ~~~~~~~~\rt \pip\chicone)  = 4.0
^{+19.8}_{-1.0}  \times 10^{-5}.$

\vspace{0.25cm} \noindent

For comparison, Fig.~10(b), from Ref.~[53], shows the Belle data
with the same $M(K\pip)$ requirement and the results of a fit with
no $Z^+$ resonances (dashed blue curve) and the fit with the two
$Z^+$ resonances described above.~The Belle and BaBar data, and the
results of their no $Z^+$ fits are qualitatively very similar, while
the results of the fits that include $Z^+$ terms are quite
different.~This is because in Belle fits, the $Z^+$ amplitudes are
coherent with those for the $K\pip$ channel and, as in the case of
the $Z(4430)^+$ analysis, interference  between the $Z^+$ and
$K\pip$ amplitudes is very strong, in this case resulting in strong
destructive interference below and above the peak region.~A
comparison of Figs.~10(a) and (b) clearly shows that by neglecting
the possibility of interference between the $Z^+$ and $K\pip$
amplitudes, the BaBar analysis underestimates the strength of
possible $Z^+_1$ and $Z^+_2$ signals in their data.

\underline{\it The $Z_c(3900):$}~~ The discovery and early
measurements of the $Y(4260)$ were based on measurements of the
initial state radiation process, $\ee\rt \gamma_{isr} Y(4260)$ at
$\sqrt{s}\simeq 10.6$~GeV.~This reaction requires that either the
incident $e^-$ or $e^+$ radiates a $\sim 4.5$~GeV photon prior to
annihilating, which results in a strong reduction in event
rate.~However, since the PEPII and KEKB $B$-factories ran with such
high luminosity (${\mathcal L} > 10^{34}\ {\rm cm}^{-2}\cdot{\rm
s}^{-1}$), the measurements were feasible.~A more efficient way to
produce $Y(4260)$ mesons would be to run a high luminosity $\ee$
collider as a ``$Y(4260)$ factory,'' {i.e.}, at a cm energy of
4260~MeV, corresponding to the peak mass of the $Y(4260)$.~This was
done at the upgraded, two-ring Beijing electon-positron collider
(BEPCII) [67] in 2013, and large numbers of $Y(4260)$ decays were
detected in the new BESIII spectrometer [68].

The first channel to be studied with these data was
$\ee\rt\pipi\jp$, where a distinct peak, called the $Z_c(3900)$, was
seen near $3900$~MeV in the distribution of the larger of the two
$\pi^{\pm}\jp$ invariant mass combinations in each event ($M_{\rm
max}(\pi\jp)$), as can be seen in Fig.~11(a) [49].~A fit using a
mass-independent-width BW function to represent the $\pi^{\pm}\jp$
mass peak yielded a mass and width of $M_{Z_c(3900)}=3899.0\pm
6.1$~MeV and ${\it\Gamma}_{Z_c(3900)}= 46 \pm 22$ ~MeV, which is
$\sim 24$~MeV above the $m_{D^{*+}} + m_{\bar{D}^{0}}$ (or $m_{D^+}
+ m_{\bar{D}^{*0}}$) threshold.~The $Z_c(3900)$ was observed by
Belle in isr data at about the same time [50].

A subsequent BESIII study of the $(\DDbar^*)^{\pm}$ systems produced
in $(\DDbar^*)^{\pm}\pi^{\mp}$ final states in the same data sample
found very strong near-threshold peaks in  both the $D^0D^{*-}$ and
$D^+\bar{D}^{*0}$ invariant mass distributions [69], as shown in
Figs.~11(b) and (c), respectively.~The curves in the two figures
show the results of fits to the data with threshold-modified BW line
shapes to represent the peaks.~The average values of the mass and
widths from these fits are used to determine the resonance pole
position ($M_{\rm pole} + \zi{\it\Gamma}_{\rm pole}$) with real and
imaginary values of

\end{multicols}

\vspace{3mm}
\centerline{\psfig{figure=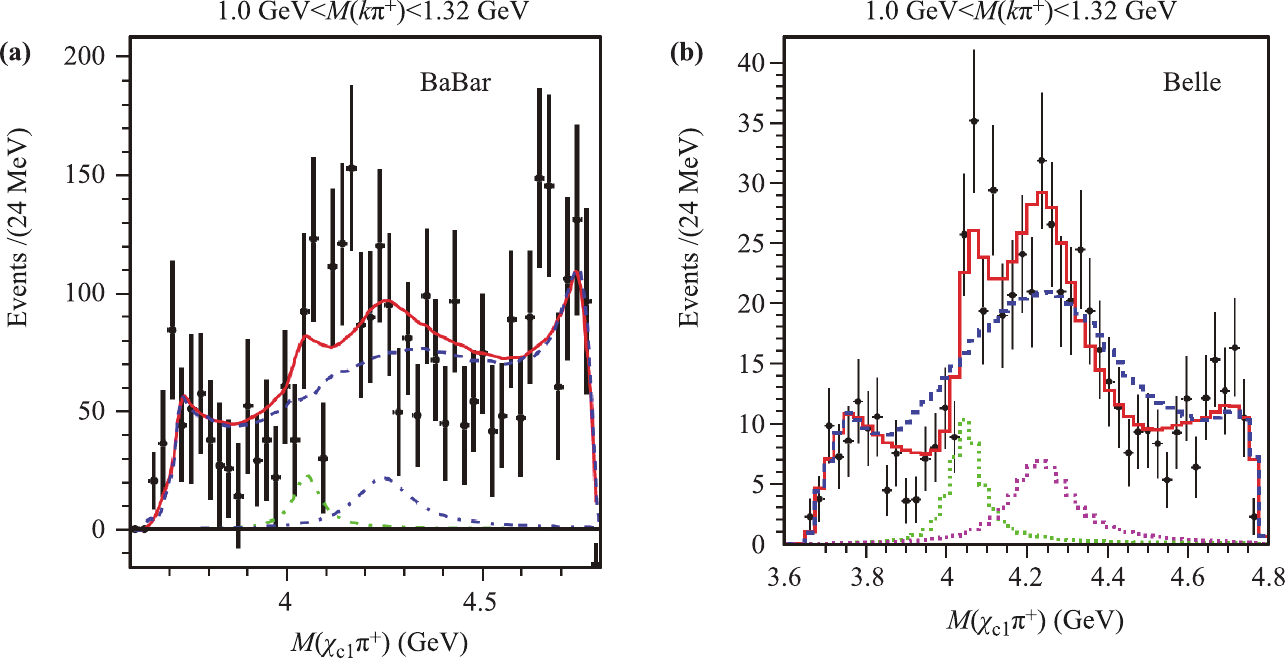}\vspace{1mm}}
 \parbox[c]{160mm}{\baselineskip
10.5pt\renewcommand{\baselinestretch}{1.05}\footnotesize \noindent
{\color{ooorangec}\bf Fig.~10}\quad {\bf(a)} The data points show
the background-subtracted $M^2(\pip\chicone)$ distribution for $B\rt
K\pip\chicone$ events with $1.0\ {\rm GeV}<M(K\pip)< 1.32\ {\rm
GeV}$ from BaBar.~The dashed (blue) curve shows results from BaBar's
fit with no $Z^+$ resonances; the red curve shows the results from a
fit that includes two incoherent $Z^+\pip\chicone$ resonances with
Belle's fitted mass and width values.~{\bf(b)} The corresponding
$M(\pip\chicone)$ distribution from Belle.~The solid red histogram
is a projection Belle's fit with two coherent resonances in the
$Z^+\rt\pip\chicone$ channel.~The dashed (blue) curve indicates the
best fit with no $Z^+$.

}\vspace{0mm}

\vspace{1mm}
\centerline{\psfig{figure=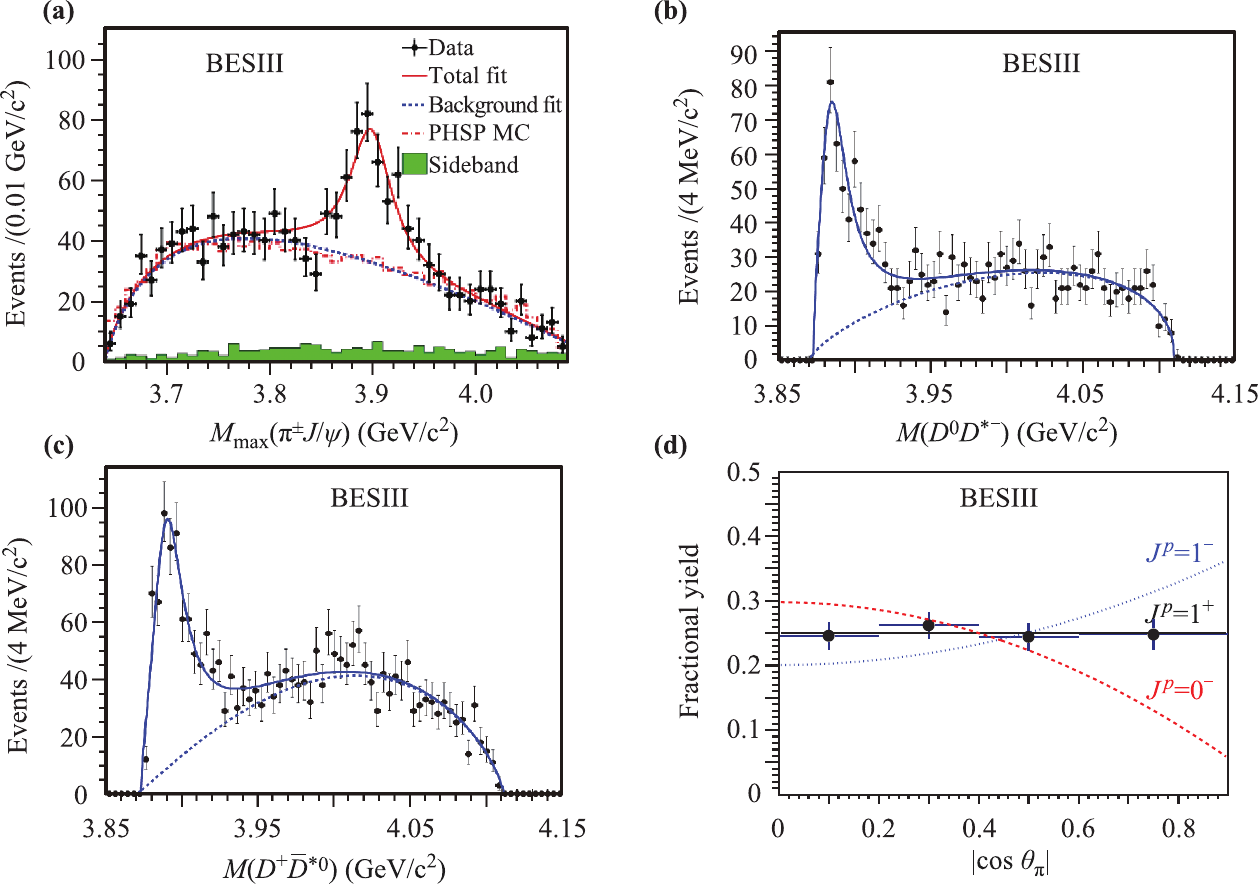}\vspace{1mm}}
 \parbox[c]{165mm}{\baselineskip
10.5pt\renewcommand{\baselinestretch}{1.05}\footnotesize \noindent
{\color{ooorangec}\bf Fig.~11}\quad {\bf(a)} Invariant mass
distributions for $\pip\jp$ from $\ee\rt\pipi\jp$ events from
Ref.~[49]; {\bf(b)} $M(D^0 D^{*-})$ and {\bf(c)}
$M(D^+\bar{D}^{*0})$ for $\ee\rt (D\bar{D}^*)^+\pim$ events from
Ref.~[69]; {\bf(d)} The efficiency corrected production angle
distribution compared with predictions for $J^P = 0^-$
(\emph{dashed-red}), $J^P=1^-$ (\emph{dotted blue}) and $J^P=1^+$
(\emph{solid black}) quantum number assignments.

}\vspace{4mm}

\begin{multicols}{2}
\setlength{\parindent}{1em}    

\noindent $M_{\rm pole}= 3883.9\pm 4.5$~MeV and ${\it\Gamma}_{\rm
pole}=24.8\pm 12$~MeV.

Since the pole mass position is $\simeq 2\sigma$ lower than the
$Z_c(3900)$ mass reported in Ref.~[49], BESIII cautiously named this
$D\bar{D}^*$ state the $Z_c(3885)$.~In the mass determinations of
both the $Z_c(3885)$ and $Z_c(3900)$, effects of possible
interference with a coherent component of the background are
ignored, which can bias the measurements by amounts comparable to
the resonance widths, and this might account for the different mass
values.~In any case, we consider it highly likely that the
$Z_c(3885)$ is the $Z_c(3900)$ in a different decay channel.~If this
is the case, the partial width for $Z_c(3900)\rt D\bar{D}^*$ decays
is $6.2\pm 2.9$ times larger than that for $\jp\pip$, which is small
compared to open-charm {vs.} hidden-charm decay-width ratios for
established charmonium states above the open-charm threshold, such
as the $\psi(3770)$ and $\psi(4040)$, where corresponding ratios are
measured to be more than an order-of-magnitude larger [13].

Since the $Z(3885)\rt D\bar{D}^*$ signals are so strong, the $J^P$
quantum numbers can be determined from the dependence of its
production on $\theta_{\pi}$, the bachelor pion production angle
relative to the beam direction in the $\ee$ cm system.~For
$J^P=0^-$, $\zd N/\zd|\cos\theta_{\pi} |$ should go as
$\sin^2\theta_{\pi}$; for $1^-$ it should follow
$1+\cos^2\theta_{\pi}$ and for $1^+$ it should be flat ($0^+$ is
forbidden by Parity).~Figure~11d shows the efficiency-corrected
$Z_c(3885)$ signal yield as a function of $|\cos\theta_{\pi}|$,
together with expectations for $J^P=0^+$ (dashed red), $1^-$ (dotted
blue) and $J^P=1^+$.~The $J^P=1^+$ assignment is clearly preferred
and the $0^-$ and $1^-$ assignments are ruled out with high
confidence.

\underline{\it The $Z_c(4020):$}~~ With data accumulated at the
peaks of the $Y(4260)$, $Y(4360)$ and nearby energies, BESIII made a
study of $\pipi h_c(1P)$ final states.~Exclusive $h_c(1P)$ decays
were detected via the $h_c\rt\gamma\eta_c$ transition, where the
$\eta_c$ was reconstructed in 16 exclusive hadronic decay
modes.~With these data, BESIII observed a distinct peak near
4020~MeV in the $M_{\rm max}(\pi^{\pm}h_c)$ distribution that is
shown in Fig.~12(a).~A fit to this peak, which the BESIII group
called the $Z_c(4020)^+$, with a signal BW function (assuming
$J^P=1^+$) plus a smooth background,
 returns a $\sim 9\sigma$ significance signal with a fitted
mass of $M_{Z_c(4020)}=4022.9\pm 2.8$~MeV, about $ 5$~MeV above
$m_{D^{*+}}+m_{\bar{D}^{*0}}$, and a width of
${\it\Gamma}_{Z_c(4020)}= 7.9\pm 3.7$ ~MeV [51].~The product
$\sigma(\ee\rt\pim Z_c(4020)^+)\times {\mathcal
B}(Z_c(4020)^+\rt\pip h_c) $ is measured to be $7.4\pm 2.7 \pm
1.2$~pb at $\sqrt{s}=4260$~MeV, where the second error reflects the
uncertainty of ${\mathcal B}(h_c\rt \gamma \eta_c$).

The inset in Fig.~12(a) shows the result of including a
$Z_c(3900)^+\rt \pip h_c$ term in the fit.~In this case, a marginal
$\sim 2\sigma$ signal for $Z_c(3900)^+\rt \pip h_c$ is seen to the
left of the $Z_c(4020)$ peak.~This translates into  an upper limit
on the product $\sigma(\ee\rt\pim Z_c(3900)^+)\times {\mathcal
B}(Z_c(3900)^+\rt\pip h_c) $ of 11~pb.~Since the product
$\sigma(\ee\rt\pim Z_c(3900)^+)\times {\mathcal
B}(Z_c(3900)^+\rt\pip\jpsi)$ is measured to be $62.9\pm 4.2 $ pb
[49], this limit implies that the $Z_c(3900)^+\rt\pip h_c$ decay
channel is suppressed\vspace{-0.4cm}\linebreak

\end{multicols}

\vspace{0mm}
\centerline{\psfig{figure=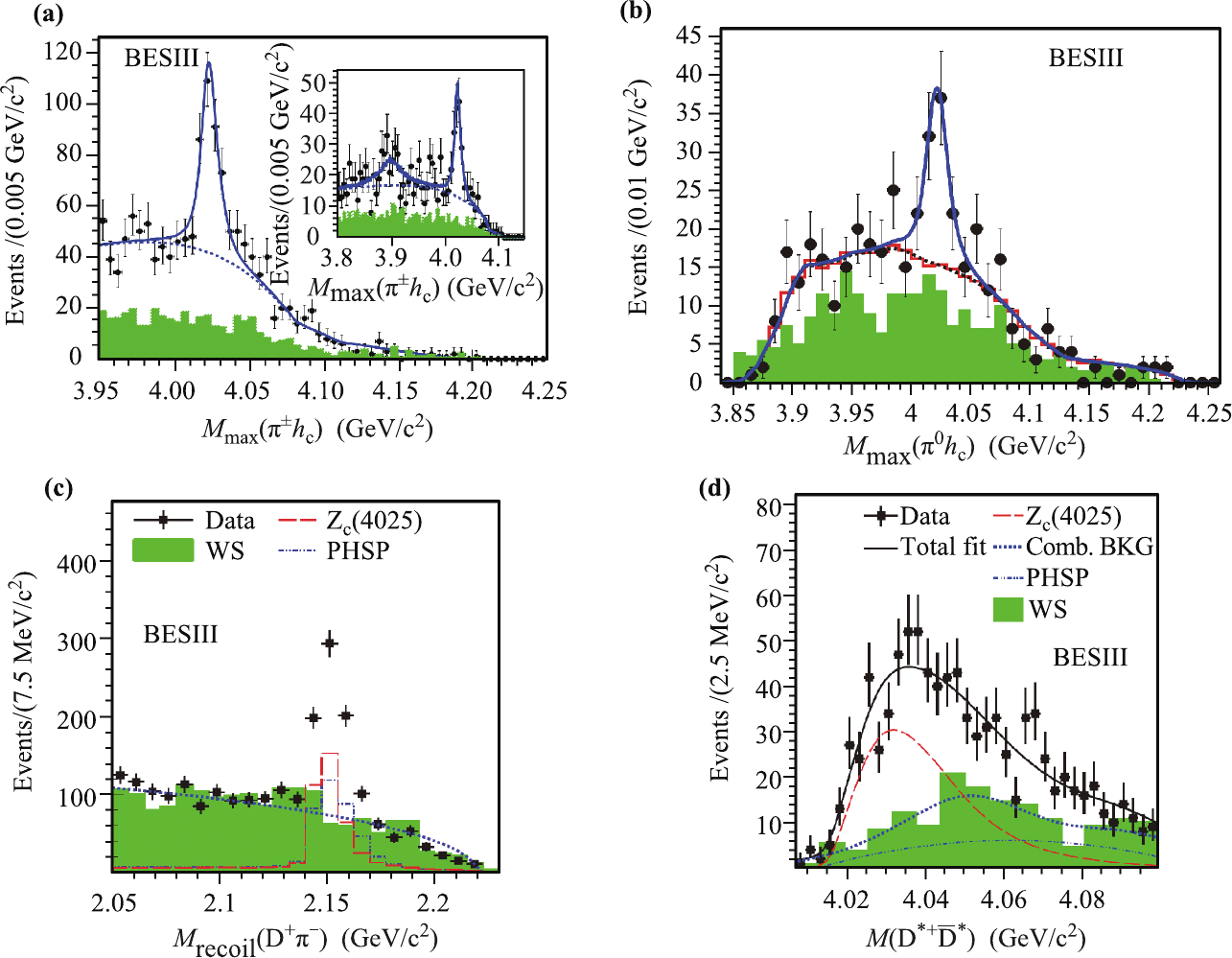}\vspace{1mm}}
 \parbox[c]{160mm}{\baselineskip
10.5pt\renewcommand{\baselinestretch}{1.05}\footnotesize \noindent
{\color{ooorangec}\bf Fig.~12}\quad {\bf(a)} The $M_{\rm max}(\pip
h_c)$ distribution for $\ee\rt\pipi h_c$ events from BESIII.~The
shaded histogram is background estimated from the $h_c$ mass
sidebands.~The curves are results of fits described in the text.
{\bf(b)} The corresponding $M_{\rm max}(\piz h_c)$ distribution for
$\ee\rt \pi^0\pi^0 h_c$ events from BESIII.~{\bf(c)} The
distribution of masses recoiling from a detected $D^+$ and $\pim$
for $\ee\rt D^+\pim \piz X$ events at $\sqrt{s}=4260$~MeV.~The peak
near 2.15~MeV corresponds to $\ee\rt \pim D^{*+}\bar{D}^{*0}$
events.~The red dashed histogram shows the expected recoil mass
distribution for $\ee\rt \pim Z_c$, with $M_{Z_c}=4025$~GeV;  the
open, dash-dot histogram shows results for MC $\pim
D^{*+}\bar{D}^{*0}$ three-body phase-space events.~The shaded
histogram is combinatoric background from wrong-sign combinations in
the data.~{\bf(d)} $M(D^*\bar{D}^*)$ for $\ee\rt
(D^*\bar{D}^*)^+\pim$ events, {i.e.}, events in the 2.25~MeV peak in
panel (c).~The curves are described in the text.

}\vspace{2mm}

\begin{multicols}{2}
\setlength{\parindent}{1em}    

\noindent
 relative to that for $\pip\jpsi$ by at least a
factor of five.

BESIII recently reported observation of the neutral member of the
$Z_c(4020)$ isospin triplet [70].~The $M_{\rm max}(\piz h_c )$
distribution for $\ee\rt\pi^0\pi^0 h_c$ events in the same data set
used for the $Z_c(4020)^{\pm}$, shown in Fig.~12(b), looks
qualitatively like the $M_{\rm max}(\pip h_c )$ distribution with a
distinct peak near $4020$~MeV.~A fit to the data that includes a BW
term with a width fixed at the value measured for the $Z_c(4020)^+$
and floating mass returns a mass of $4023.9\pm 4.4$~MeV; this and
the signal yield are in good agreement with expectations based on
isospin symmetry.

A study of $\ee\rt D^{*+}\bar{D}^{*0}\pim$ events in the
$\sqrt{s}=4.26$~GeV data sample using a partial reconstruction
technique that only required the detection of the bachelor $\pim$,
the $D^+$ from the $D^{*+}\rt \piz D^+$ decay and one $\piz$, either
from the $D^{*+}$ or the $\bar{D}^{*0}$ decay, to isolate the
process and measure the $D^{*+}\bar{D}^{*0}$ invariant mass
[52].~The signal for real $D^{*+}\bar{D}^{*0}\pim$ final states is
the distinct peak near 2.15~MeV in the $D^+\pim$ recoil mass
spectrum shown on Fig.~12(c).~The measured $D^*\bar{D}^*$ invariant
mass distribution for events in the 2.15~MeV peak, shown as data
points in Fig.~12(d), shows a strong near-threshold peaking behavior
with a shape that cannot be described by a phase-space-like
distribution, shown as a dash-dot blue curve, or by combinatoric
background, which is determined from wrong-sign (WS) events in the
data ({i.e.}, events where the bachelor pion and charged $D$ meson
have the same sign) that are shown as the shaded histogram.~The
solid black curve shows the results of a fit to the data points that
includes an efficiency weighted $S$-wave BW function, the WS
background shape scaled to measured non-$D^{*+}\bar{D}^{*0}\pim$
background level under the signal peak in Fig.~12(d), and a
phase-space term.~The fit returns a $13\sigma$ signal with mass and
width $M=4026.3\pm 4.5 $~MeV and ${\it\Gamma}=24.8\pm 9.5$~MeV,
values that are close to those measured for the $Z_c(4020)^+\rt\pip
h_c $ channel.~Although BESIII cautiously calls this
$(D^*\bar{D}^*)^+$ signal the $Z_c(4025)$, in the following we
assume that this is another decay mode of the $Z_c(4020)$.

From the numbers provided in Ref.~[52], we determine
$\sigma(\ee\rt\pim Z_c(4020))\times {\mathcal B}(Z_c(4020)\rt
D^*\bar{D}^*)=89\pm 19$~pb.~This implies that the partial width for
$Z_C(4020)\rt D^*\bar{D}^*$ is larger than that for $Z_c(4020)\rt\pi
h_c$, but only by a factor of $12\pm 5$, not by the large factors
that are characteristic of open charm decays of conventional
charmonium.

There is no sign of $Z_c(4020)\rt D\bar{D}^*$ in either Figs.~11(b)
or (c), where $\sim 500$ and $\sim 700$ event $Z_c(3885)$ signals in
the $D^0 D^{*-}$ and $D^+\bar{D}^{*0}$ distributions, respectively,
correspond to the product $\sigma(\ee\rt\pim Z_c(3900))\times
{\mathcal B}(Z_c(3900)\rt D\bar{D}^*) = 84\pm 23$~pb, about the same
as the corresponding product for the $Z_c(4020) D^*\bar{D}^*$ signal
mentioned in the previous paragraph.~This absence of any evident
signals for $Z_c(4020)\rt D\bar{D}^*$ in Figs.~11(b) and (c)
suggests that the $Z_c(4020)\rt D\bar{D}^*$ partial width is
considerably smaller than that for $Z_c(4020)\rt D^*\bar{D}^*$,
which may have some relation to the $Z_c(4020)$'s proximity to the
$2m_{D^*}$ threshold.~However, until BESIII provides a limit on a
$Z_c(4020)\rt D\bar{D}^*$ signal at $\sqrt{s}=4260$~MeV, this effect
cannot be quantified.

\textbf{$\bm{X(3915)}$}~~The $X(3915)$ is an $\omega\jp$ mass peak
with $M=3918\pm 2$~MeV with ${\it\Gamma}= 20\pm 5$~MeV [13]  seen in
$B\rt K\omega\jp$ decays by both Belle [71] and BaBar [72, 73] [see
Figs.~13(a) and (b)] and in the two-photon process
$\twog\rt\omega\jp$ also by Belle [74] and BaBar [75]
[Fig.~13(c)]$^{8)}$\footnote{\hspace*{-5mm}$^{8)}$~Some of the
literature refers to this state as the $Y(3940)$.}.~BaBar measured
its $J^{PC}$ quantum numbers to be $0^{++}$.~The PDG currently
assigns this as the $\chi_{c0}(2P)$ charmonium level ({i.e.}, the
$\chip_{c0}$), an assignment that has a number of problems~[77]:

{i}) The $J=2$ member of the $\chip_{c0,1,2}$ multiplet, the
$\chi_{c2}(2P)$ ($\chip_{c2}$), was seen by Belle [78] and confirmed
by

\end{multicols}

\vspace{0mm}
\centerline{\psfig{figure=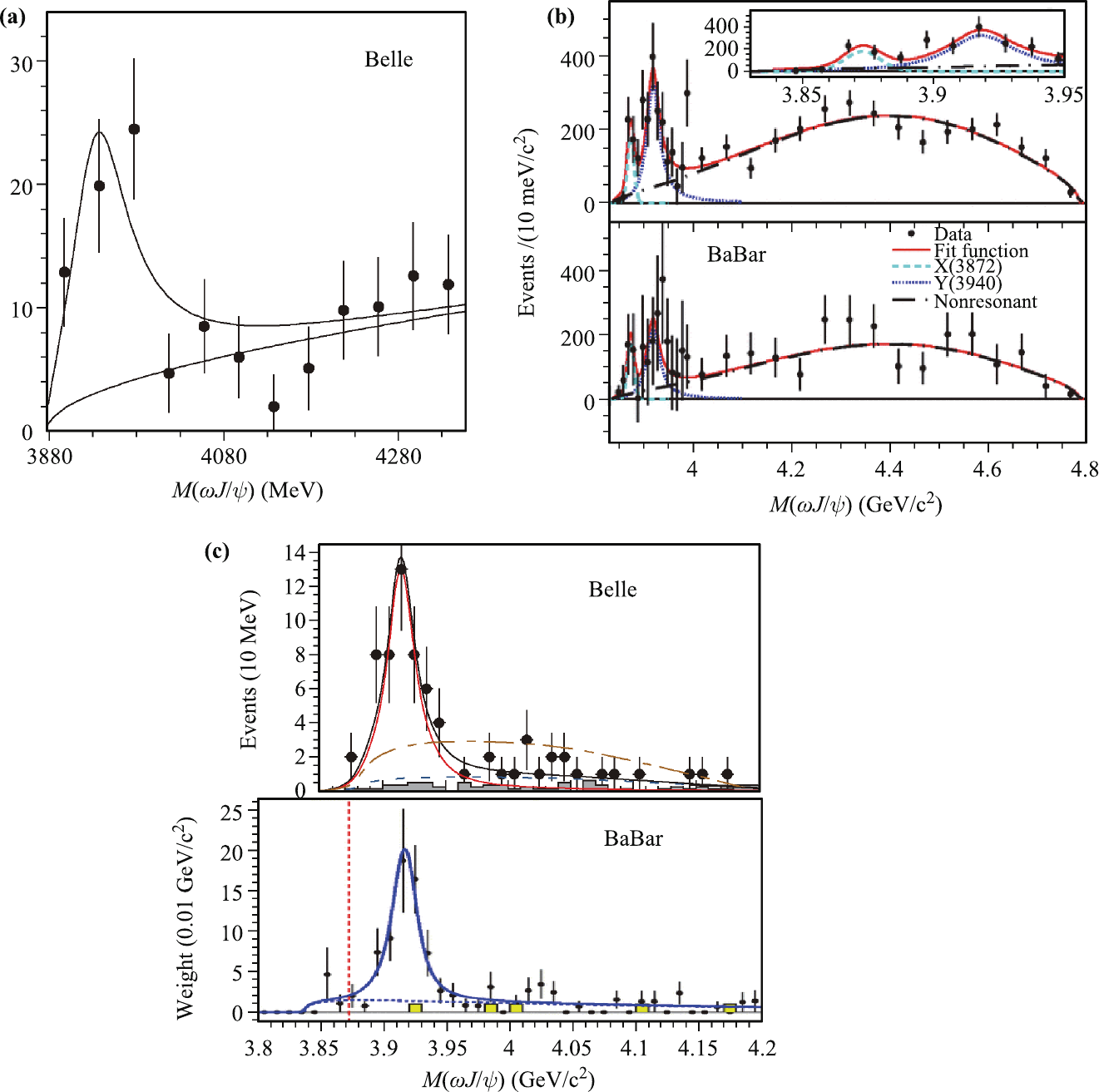}\vspace{1mm}}
 \parbox[c]{160mm}{\baselineskip
10.5pt\renewcommand{\baselinestretch}{1.05}\footnotesize \noindent
{\color{ooorangec}\bf Fig.~13}\quad {\bf(a)} The
$X(3915)\rt\omega\jpsi$ signals in $B\rt K\omega\jpsi$ decays from
Belle.~{\bf(b)} The corresponding distributions from BaBar, where
the upper panel shows results for $B^+\rt K^+\omega\jpsi$  and the
lower panel shows those for $B^0\rt K_S \omega\jpsi$.~The inset in
the upper panel shows an expanded view of the low end of the
$\omega\jpsi$ mass scale, where the smaller, low-mass peak is due
the $X(3872)\rt\omega\jpsi$ and the larger, higher mass peak is the
$X(3915)\rt\omega\jpsi$ signal.~{\bf(c)} The $X(3915)\rt\omega\jpsi$
signals in $\gamma\gamma\rt\omega\jpsi$ fusion reactions from Belle
({\it upper}) and BaBar ({\it lower}).

}\vspace{2mm}

\begin{multicols}{2}
\setlength{\parindent}{1em}    

\noindent BaBar [79] as a distinct peak in the $D\bar{D}$ invariant
mass distribution near 3930~MeV in $\gamma\gamma\rt D\bar{D}$
events.~The mass, production rate and production angle dependence
agree quite well with expectations for the $\chip_{c2}$ [see
Figs.~14(a) and (b)], and there are no known reasons to doubt this
assignment.~The weighted averages of the Belle and BaBar mass and
width measurements are $M=3927\pm 3$~MeV and ${\it\Gamma}=24\pm
6$~MeV [13].~With the $\chip_{c2}$ well established with a reliably
known mass, the $X(3915)=\chip_{c0}$ assignment implies a $2^3P_2 -
2^3P_0$ fine splitting of only $9\pm 4$~MeV,  which is very small in
comparison with the corresponding $n=1$ splitting of $141.4\pm
0.3$~MeV [13] and theoretical predictions for $n=2$ that range from
69~MeV [36] to 80~MeV [76].

{ii}) There is no sign of $X(3915)\rt \DDbar$ decay, which is
expected to be a favored decay mode.~Belle [80] and BaBar [81] have
studied the process $B\rt K \DDbar$ and both groups see  prominent
signals for $\psi(3770)\rt \DDbar$, but no hint of a $\DDbar$ mass
peak near 3915~MeV [see Fig.~14(c)].~Since neither group reported a
$X(3915)\rt\DDbar$ limit, I formed my own conservative estimate of
an upper bound by scaling the total number of Belle events in the
two mass bins surrounding 3915~MeV in the upper panel of Fig.~14(c)
to the $\psi(3770)$ signal while assuming constant acceptance.~This
gives ${\mathcal B}(X(3915)\rt \omega\jp)>{\mathcal B}(X(3915)\rt
D^0\bar D^0)$, which strongly contradicts the theoretical
expectation that the $\chip_{c0} \rt \DDbar$ decay channel should
dominant [76].

{iii)}~Since branching fractions cannot exceed unity, the measured
quantity ${\it\Gamma}^{\chip_{c2}}_{\twog}\times {\mathcal
B}(\chip_{c2}\rt \DDbar) =0.21\pm0.04$~keV [78, 79], translates to a
$\sim$90\%~CL lower limit
${\it\Gamma}^{\chip_{c2}}_{\twog}>0.16$~keV.~Using the assumption
${\it\Gamma}^{\chip_{c0}}_{\twog}/{\it\Gamma}^{\chip_{c2}}_{\twog}=
{\it\Gamma}^{\chi_{c0}}_{\twog}/{\it\Gamma}^{\chi_{c2}}_{\twog} =
4.5\pm 0.6$ [13], which is valid for pure charmonium states, I infer
a lower limit ${\it\Gamma}^{\chip_{c0}}_{\twog}>0.80$~keV that, when
taken with the measured value
${\it\Gamma}^{X(3915)}_{\twog}{\mathcal
B}(X(3915)\rt\omega\jp)=0.054\pm 0.009$~keV [82], gives an upper
limit ${\mathcal B}(\chi_{c0}^{\prime}\rt\omega\jp)< 5.6\%$
if\vspace{-0.4cm}\linebreak

\end{multicols}

\vspace{1mm}
\centerline{\psfig{figure=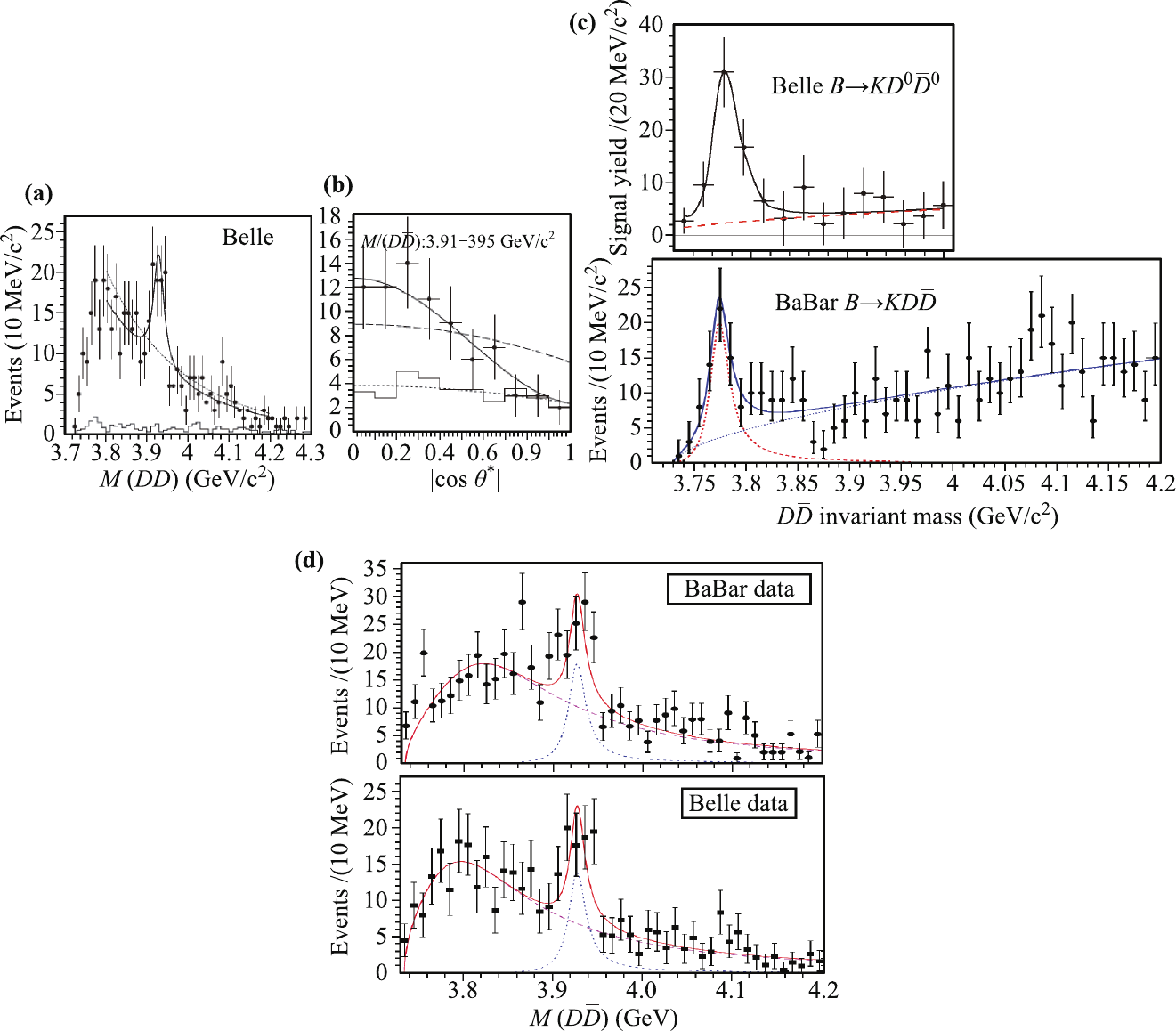}\vspace{1mm}}
 \parbox[c]{160mm}{\baselineskip
10.5pt\renewcommand{\baselinestretch}{1.05}\footnotesize \noindent
{\color{ooorangec}\bf Fig.~14}\quad {\bf(a)} The $D\bar{D}$
invariant mass distribution for $\gamma\gamma\rt D\bar{D}$ events in
Belle.~The narrow peak is attributed to $\chip_{c2}\rt D\bar{D}$.
{\bf(b)} The $|\cos\theta^*|$ distribution for events in the region
of the $\chip_{c2}$ peak.~The solid line show expectations for
spin=2, helicity=2; the dashed line shows the spin zero expectation.
{\bf(c)} The $M(D^0\bar{D}^0)$ distribution from $B\rt
KD^0\bar{D}^0$ decays from Belle [80] ({\it upper}); the
$M(D\bar{D})$ distribution from $B\rt KD\bar{D}$ decays from BaBar
[81] ({\it lower}).~The near-threshold peak in both plots is due to
the $\psi(3770)$.~{\bf(d)} The $M(D\bar{D})$ distribution from
$\twog\rt \DDbar$ two-photon annihilation from BaBar [79] ({\it
upper}); the same distribution from Belle [78] ({\it lower}).~The
narrow peak is due to the $\twog\rt\chip_{c2}\rt\DDbar$ and the
curves show results of fits described in Ref.~[77].

}\vspace{4mm}

\begin{multicols}{2}
\setlength{\parindent}{1em}    

\noindent $X(3915)=\chip_{c0}$.~The PDG value for the $X(3915)$
total width of $20\pm 5$~MeV and the above-mentioned limit
${\mathcal B}(X(3915)\rt \DDbar) < {\mathcal
B}(X(3915)\rt\omega\jp)$ imply an upper limit
${\it\Gamma}(\chip_{c0}\rt \DDbar)< 1.5$~MeV for the
$X(3915)=\chip_{c0}$ assignment.~This upper limit is below the range
of theoretical predictions that range from $7$~MeV {[83]} to 187~MeV
[84].

{iv)}~The average of the Belle and BaBar measurements for $X(3915)$
production in $B$ meson decay is ${\mathcal B}(B\rt KX_{3915})\times
{\mathcal B}(X_{3915}\rt\omega\jpsi)=(3.0^{+0.9}_{-0.7})\times
10^{-5}$.~If I make the (reasonable?) assumption that ${\mathcal
B}(B\rt K\chip_{c0})< {\mathcal B}(B\rt K\chi_{c0})=(1.3\pm
0.2)\times 10^{-4}$ [13], I find the {\it lower} limit ${\mathcal
B}(\chi_{c0}^{\prime}\rt \omega\jpsi)>14\%$, in strong contradiction
to the 5.6\% {\it upper} limit established in the previous
paragraph.

Some authors have identified another candidate for the
$\chip_{c0}$.~Guo and Meissner [77] propose that the broad
background under the $\twog\rt \chip_{c2}\rt\DDbar$ peak in the
$M(D\bar{D})$ distribution in $\gamma\gamma\rt D\bar{D}$ events as
seen in Fig.~14(a) is due to $\twog\rt\chip_{c0}\rt\DDbar$.~Their
fits to BaBar and Belle data, shown as dashed curves Fig.~14(d),
yield a ``$\chip_{c0}$'' mass and width of $M=3838 \pm 12$~MeV and
${\it\Gamma}=221\pm 21$~MeV, which is in general agreement with
charmonium model expectations.~This fit is almost certainly biased
because the authors do not consider contributions to the $D\bar{D}$
signal from $\chi_{c2}^{\prime}\rt D\bar{D}^*$ decays where the
$\pi$ or $\gamma$ from $\bar{D}^*\rt\bar{D}\pi (\gamma)$ decay is
missed.~Nevertheless, their suggestion that a broad and light
$\chi_{c0}^{\prime}$ might be responsible for at least part of the
measured $D\bar{D}$ ``background'' should be carefully
considered.~With enough data, contributions from
$\chi_{c0}^{\prime}\rt D\bar{D}$ and $\chi_{c2}\rt D\bar{D}^*$ can
be sorted out from the relative rates for $D^0\bar{D}^0$,
$D^{*+}D^{*-}$ and $D^0D^{*-}$ final states and the shapes of their
associated $p_T$ distributions.~This is a good candidate for a ``day
one'' measurement for the BelleII experiment [85].

Chao [33] suggested that $\chip_{c0} \rt D\bar{D}$ may be the source
of the broad peak in the $M(\DDbar)$ spectrum seen by Belle in
$\ee\rt \jp \DDbar$ events [32], as discussed above and shown in the
upper panel of Fig.~5(b).~The Belle group's fit to this distribution
yielded a broad signal with a peak mass at $3780\pm 48$~MeV and a
significance of $\simeq 3.8\sigma$.~This mass is closer to
expectations for the $\chip_{c0}$ than that of the $X(3915)$.~Belle
is now investigating this reaction with about twice as much data
than was used in the Ref.~[32] analysis.

Although these suggestions are not definitive, they, plus the
arguments from the previous paragraph, show that the PDG's
assignment of the $X(3915)$ to the $\chip_{c0}$ charmonium level is
premature and possibly misleading.

If the $X(3915)$ is not an ordinary $c\bar{c}$ charmonium state,
what is it?  Molina and Oset proposed a molecule-like picture with
states that are dynamically generated from vector-vector
interactions.~They find a $J^{PC}=0^{++}$ state near 3940~MeV with a
17~MeV width [86] that is dominantly $D^*\bar{D}^*$ and $D_s^*
D_s^*$.~The $X(3915)$ mass is below the threshold for decays to
these final states and the authors find a  strong decay to
$\omega\jpsi$ plus a strong suppression of decays to $D\bar{D}$,
consistent with what is seen experimentally.~This paper also
predicts other states, including some with isospin=1, but none of
these can be as easily associated with any of the $XYZ$ peaks that
are seen in experiment.

\textbf{$\bm{Y(4140)}$}~~In 2009, the CDF group reported $3.8\sigma$
evidence for a near-threshold peak in the $\phi\jpsi$ invariant mass
distribution from $B^+\rt K^+\phi\jpsi$ decays based on an analysis
of a 2.7~fb$^{-1}$ data sample [87].~CDF subsequently reproduced
this signal with a larger, 6.0~fb$^{-1}$ data sample and with a
significance that increased to more  than $5\sigma$ [88] [see
Fig.~15(a)].~The mass and width were reported as $M=4143.4 \pm
3.0$~MeV and ${\it\Gamma}= 15.3^{+10.7}_{-6.6}$~MeV.~The CDF group's
fit, shown as a red curve in Fig.~15(a), includes a second structure
with mass and width $M= 4274.4^{+8.6}_{-7.0}$~MeV and
${\it\Gamma}=32.3^{+23.2}_{-17.1}$~MeV, with a significance
estimated to be $3.1\sigma$.

The existence of the $Y(4140)$ has been somewhat controversial
because it was not seen in $B\rt K \phi\jpsi$ decays by LHCb [89],
which set an upper limit on the $Y(4140)$ production rate in $B$
decays that is about a factor of two below the CDF group's central
value for the same quantity.~Figure~15(b) shows the LHCb group's
$M(\phi\jpsi)$ distribution as a histogram with expectations for the
two peaks reported by CDF superimposed as blue dashed curves.~While
the agreement between the LHCb data and the CDF-inspired curves is
pretty poor, there does seem to be excesses in the LHCb distribution
in the vicinity of both of the CDF peaks.~The excess near threshold
is pretty broad and suggests that if there is a state decaying to
$\phi\jpsi$ in that mass region, it may have a broader width than
the CDF group's measured ${\it\Gamma}$ value.

This year, the CMS group published the $M(\phi\jpsi)$ distribution
in $B^+\rt K^+\phi\jpsi$ shown in Fig.~15(c), where there are
distinct signals near the positions of both of the CDF peaks
[90].~Their fit to the data, shown as red curves in the figure, give
a $Y(4140)$ mass an width of $M=4148.0 \pm 6.7$~MeV and ${\it\Gamma}
= 28^{+24}_{-22} $~MeV, which agree within errors with the CDF
measurements.~The estimated significance of the $Y(4140)$ peak is
greater than $5\sigma$.~Note that the central value of the CMS width
measurement, while consistent with the CDF value, is higher by
nearly a factor of two, and has a large positive-side error.

The fitted mass of the second peak, $M=4314 \pm 9$~MeV, is about
$3\sigma$ higher than the CDF value, while the fitted width,
${\it\Gamma} = 38^{+34}_{-22} $~MeV, is in good
agreement,\vspace{-0.4cm}\linebreak

\end{multicols}

\vspace{3mm}
\centerline{\psfig{figure=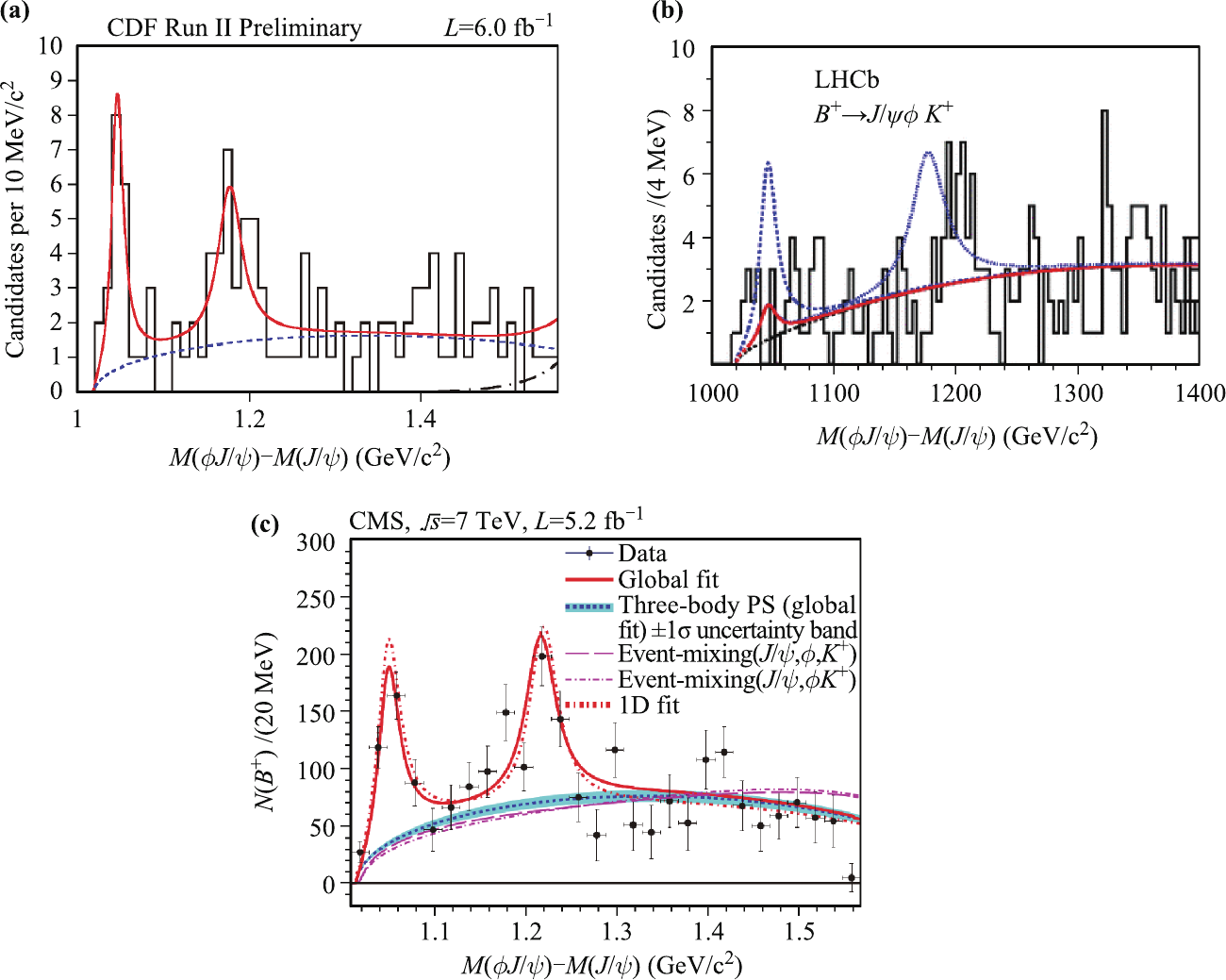}\vspace{1mm}}
 \parbox[c]{160mm}{\baselineskip
10.5pt\renewcommand{\baselinestretch}{1.05}\footnotesize \noindent
{\color{ooorangec}\bf Fig.~15}\quad {\bf(a)} The CDF 6.0~fb$^{-1}$
$M(\phi\jpsi)-M(\jpsi)$ distribution for $B^+\rt K^+\phi\jpsi$
decays is shown as a histogram.~The red curve shows the results of a
fit that includes the $Y(4140)$, a possible second resonance near
4275~MeV and a phase-space non-resonant term (shown as a dashed
line).~{\bf(b)} The same distribution from LHCb, with expectations
based on the CDF results shown as a dashed blue curve.~The LHCb
upper limit is based on the results of the fit to their data shown
as a solid red curve.~{\bf(c)} The same distribution from the CMS
group with the results of their fits superimposed.

}\vspace{2mm}

\begin{multicols}{2}
\setlength{\parindent}{1em}    

\noindent albeit with large errors.~The CMS paper also provides a
plot that compares the $K^+K^-K^+$ invariant mass distribution for
$B\rt K^+K^-K^+\jpsi$ decays with phase-space expectations [see
Fig.~16(a)], where there is some indication of structure around
$1.7\sim 1.8$~GeV.~The CMS group warns that if such a structure
actually exists, it could create a reflection peak in the
$M(K^+K^-\jpsi)$ distribution near $4.3$~GeV that could distort the
measured parameters of the second $\phi\jpsi$ mass peak.~As was the
case for the $Z(4430)$ parameter extraction from $B\rt K\pip\psip$
decays, a better understanding of this second peak probably needs an
experiment with a data sample that is large enough to support an
amplitude analysis.

The BaBar group recently reported on a search for the $Y(4140)$ in
$B\rt K\phi\jpsi$ decays [91].~Figure~16(b) shows their
efficiency-corrected $M(\phi\jpsi)$ distribution together with
results of a fit that includes two resonances with the CDF group's
mass and width values.~Although there are signs of signals both for
the $Y(4140)$ and the $Y(4275)$, the statistical significance in
each case is less that $2\sigma$ after systematic errors are
considered.

The $Y(4140)$ shows up just above the $m_{\phi} + m_{\jpsi}$ mass
threshold and e.g., the $X(3915)$ mass is just above the
$m_{\omega}+m_{\jpsi}$ mass threshold.~This led to some speculation
that the two states may be related: e.g., the $X(3915)$ may be a
$D^*\bar{D}^*$ molecule and the $Y(4140)$ is its hidden charm
counterpart $D_s^{*+}D_s^{*-}$ molecule [92, 93].~In this case the
$Y(4140)$ would have the same $J^{PC}$ quantum numbers as the
$X(3915)$, namely $0^{++}$, and this implies that the $Y(4140)$
could be accessible in $\gamma\gamma$ fusion reactions.~The
$M(\phi\jpsi)$ distribution from a Belle study of
$\gamma\gamma\rt\phi\jpsi$ near the $m_{\phi} + m_{\jpsi}$ threshold
is shown in Fig.~16(c), where there is no sign of the $Y(4140)$ but
$\sim 3\sigma$ evidence for a peak at $M=4350.6^{+4.6}_{-5.1}$~MeV
with width ${\it\Gamma}=$ $13^{+18}_{-10}$~MeV [94].~The mass value
is $\sim 4\sigma$ above that of the second peak seen by CDF and CMS
in $B\rt K\phi\jpsi$ decays and, so, if both observations are from
real meson states, they are not likely to be from the same
source.~The absence of any events in the $Y(4140)$ region translates
into the upper limit ${\it\Gamma}_{\gamma\gamma}^{Y(4140)}\times
{\mathcal B}(Y(4140)\rt\phi\jpsi)<41$~eV for $J^{PC}=0^{++}$, which
is smaller, but not much smaller, than the measured value ($54\pm
9$~eV) of the corresponding quantity for the $\gamma\gamma \rt
X(3915) \rt \omega\jpsi$ signal discussed above.~However, it is much
lower than theoretical expectations for a $D_s^{*+}D_s^{*-}$
molecule: $176^{+137}_{-93}$~eV for $J^{PC}=0^{++}$ or
$189^{+147}_{-100}$~eV for $J^{PC}=2^{++}$ [94].


\end{multicols}

\vspace{1mm}
\centerline{\psfig{figure=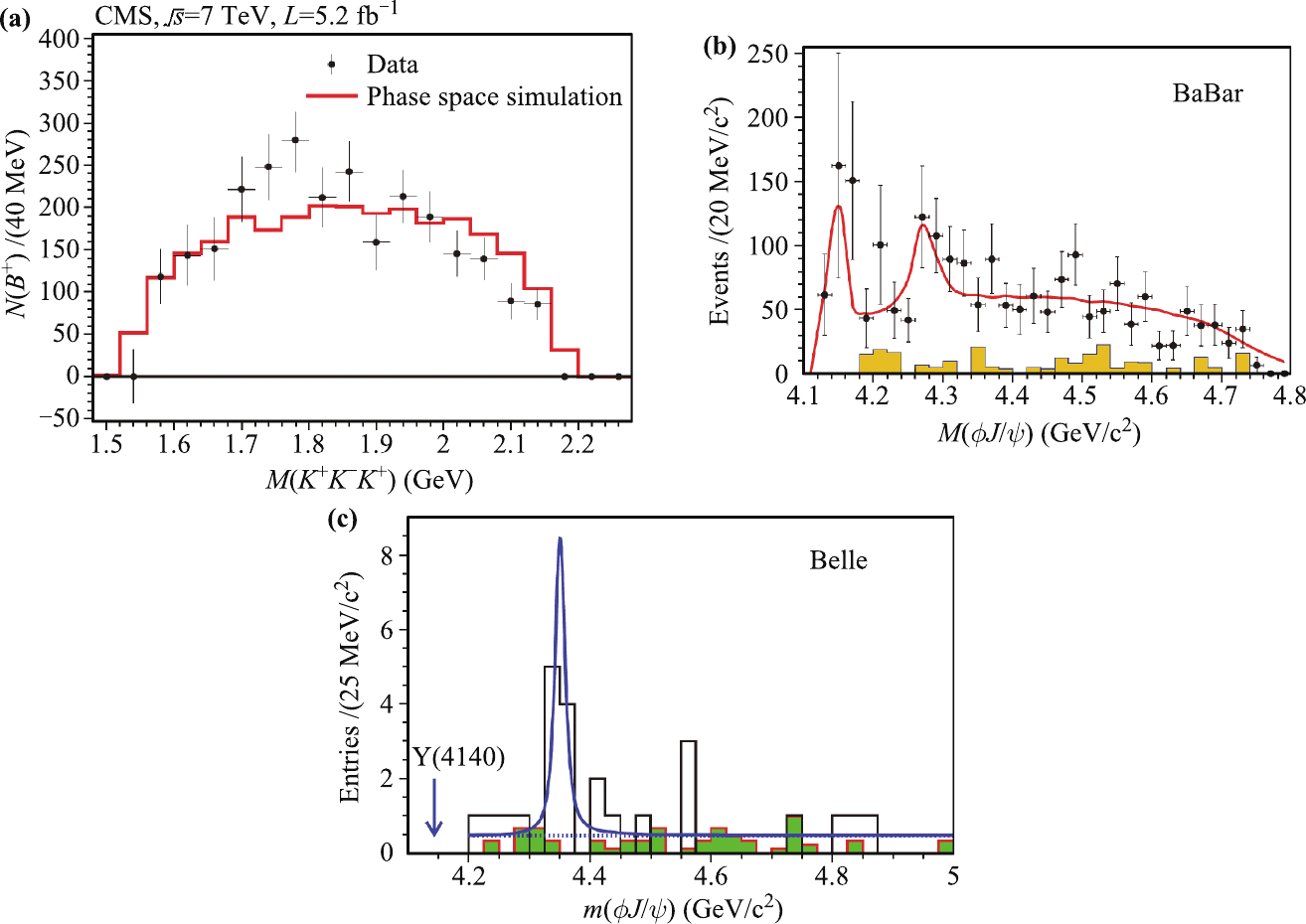}\vspace{1mm}}
 \parbox[c]{160mm}{\baselineskip
10.5pt\renewcommand{\baselinestretch}{1.05}\footnotesize \noindent
{\color{ooorangec}\bf Fig.~16}\quad {\bf(a)} The data points show
the CMS experiment's $M(K^+K^-K^+)$ distribution from $B^+\rt
K^+K^-K^+\jpsi$ decays; the histogram shows the expectations for a
four-body phase-space distribution.~{\bf(b)} The efficiency
corrected $M(\phi\jpsi)$ distribution for $B\rt K\phi\jpsi$ decays
from BaBar.~The red curve shows the results of a fit that includes
two resonances with mass and width fixed at the CDF values.~{\bf(c)}
The open histogram shows the $M(\phi\jpsi)$ distribution for
$\gamma\gamma\rt \phi\jpsi$ events from Belle; the shaded histogram
shows the background estimated from the $\phi$ and $\jpsi$ mass
sidebands.~The solid curve shows the results of the fit described in
the text.

}\vspace{4mm}

\begin{multicols}{2}
\setlength{\parindent}{1em}    


 The LHCb results discussed here were based on an analysis
of a $0.17$~fb$^{-1}$ data sample, which is only about one tenth of
their current total.~Hopefully new results based on all of the
available data will soon be available.

\textbf{$\bm{X(3872)}$}~~The $X(3872)$ was first seen in 2003 as a
peak in the $\pipi\jp$ invariant mass distribution in $B\rt
K\pipi\jp$ decays [95] as shown in Fig.~17(a).~It is well
established and has been seen and studied by a number of experiments
[96--100].~The PDG average mass value, $3871.68\pm 0.17$~MeV, is
indistinguishable from that for the $m_{D^0} + m_{D^{*0}}=3871.84\pm
0.27$~MeV threshold [13]; a recent result from LHCb is
$M_{X(3872)}-(m_{D^0}+m_{D^{*0}}) = -0.09\pm 0.28$~MeV [101].~It is
also narrow, Belle has reported a 95\% C.L.~upper limit on its total
width of ${\it\Gamma} <1.2$~MeV [102].

While LHCb and CDF have conclusively established the $J^{PC}$
quantum numbers as $1^{++}$ [103, 104], its isospin is less well
understood.~Both CDF [105] and Belle [102]

\end{multicols}

\vspace{3mm}
\centerline{\psfig{figure=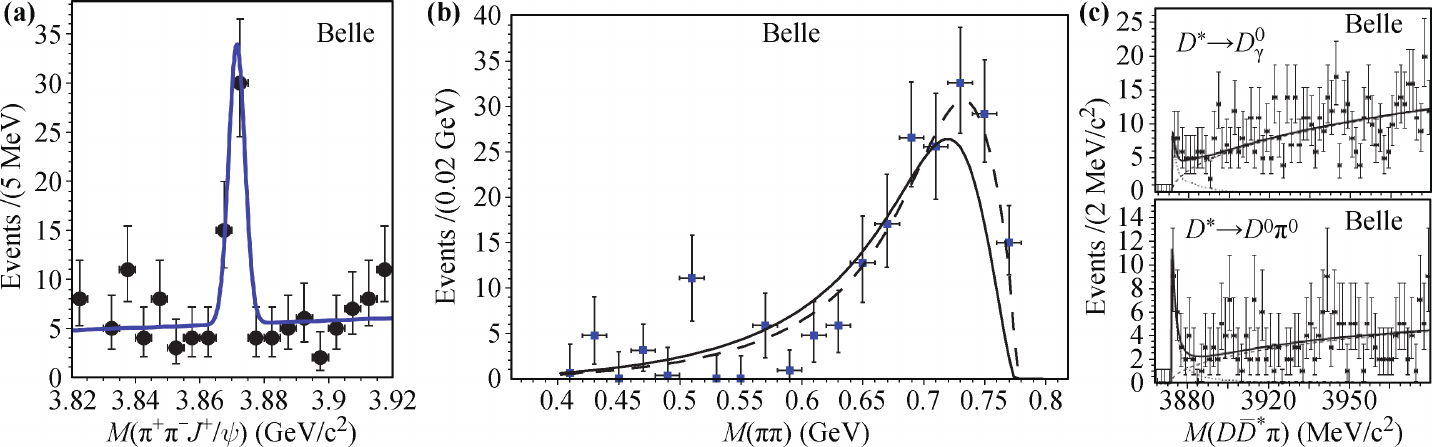}\vspace{1mm}}
 \parbox[c]{160mm}{\baselineskip
10.5pt\renewcommand{\baselinestretch}{1.05}\footnotesize \noindent
{\color{ooorangec}\bf Fig.~17}\quad {\bf(a)} The $\pipi\jpsi$
invariant mass distribution for $B\rt K\pipi\jpsi$ events from
Belle's original $X(3872)$ paper.~{\bf(b)} The $\pipi$ invariant
mass distribution for $X(3872)\rt\pipi\jp$ events in Belle.~The
curves shows results of fits to a $\rho\to\pipi$ line shape
including $\rho$-$\omega$ interference.~The dashed (solid) curve is
for even (odd) $X(3872)$ parity.~{\bf(c)} $M(D^0\bar{D}^{*0})$
distributions from $B\rt K D^0\bar{D}^{*0}$ decays.~The upper plot
is for $\bar{D}^{*0}\rt \bar{D}^0\gamma$ decays, the lower plot is
for $\bar{D}^{*0}\rt \bar{D}^0\piz$ decays.~The peaks near threshold
are attributed to $X(3872)\rt D^0\bar{D}^{*0}$ decays.

}\vspace{4mm}

\begin{multicols}{2}
\setlength{\parindent}{1em}    

\noindent have shown that the $\pipi$ system in $X\rt\pipi\jpsi$
decays originates from $\rho^0\rt\pipi$ decay, which is consistent
with the $J^{PC}=1^{++}$ assignment reported by LHCb.~The decay
$X\rt\rho\jpsi$ implies that either the $X(3872)$ has isospin $I=1$
or $0$: if it has isospin zero, its decay to $\rho\jpsi$ is isospin
violating; if it has $I=1$, it should have charged
partners.~Searches for narrow, charged partner states decaying to
$\rho^{\pm}\jpsi$ by BaBar [106] and Belle [102] set branching ratio
limits well below expectations for isospin conservation.~However,
these limits should probably be viewed with some caution because if
charged partners of the neutral $X(3872)$ exist and have masses that
are even just a few MeV higher, their decays to $D\bar{D}^*$ final
states would probably be significantly stronger than those for their
neutral partner; these higher rates could suppress the
$\rho^{\pm}\jpsi$ decay branching fraction below isospin
expectations and produce a significantly broader width.~However,
other evidence against $I=1$ comes from observations by Belle [107]
and BaBar [72, 73] of $X(3872)\rt\omega\jpsi$ decays with a
branching fraction that is nearly equal to that for $\rho\jpsi$
decays; the PDG average is ${\mathcal
B}(X(3872)\rt\omega\jpsi)/{\mathcal B}(X(3872)\rt\pipi\jpsi)=0.8\pm
0.3$ [13].~Since $M_{X(3872)}-m_{\jpsi} \simeq 775$~MeV, and $\sim
7$~MeV below $m_{\omega}$, the decay $X(3872)\rt \omega\jpsi$ can
only proceed via the low energy tail of the $\omega$ peak and is,
therefore, kinematically suppressed.~I estimate a suppression factor
of $\sim 1/120$$^{9)}$\footnote{\hspace*{-5mm}$^{9)}$~This estimate
neglects the possible effects of $\rho$-$\omega$ interference.}.~The
fact that this strongly suppressed $I=0$ final state is accessed at
nearly the same rate as the $I=1$, $\rho\jpsi$ final state -- which
is also kinematically suppressed, but only by a factor that I
estimate to be $\sim 1/5$ -- suggests that the $X(3872)$ is (mostly)
an isospin singlet and that the $\rho\jpsi$ decay mode violates
isospin symmetry.

The $X(3872)$ has also been seen to decay to $D^0\bar{D}^{*0}$
[108--110].~Since its $J^{PC}=1^{++}$, the $X(3872)$ couples to a
$D\bar{D}^*$ pair in an $S$- and/or  $D$-wave.~The $D^0\bar{D}^{*0}$
system is right at threshold and, so, the $S$-wave should be
dominant, in which case some very general universal theorems apply
[111--113].~One consequence of these theorems is that, independently
of its dynamical origin, the $X(3872)$ should exist for some
fraction of the time as a $D\bar{D}^*$ molecule-like state (either
bound or virtual) with size determined by its scattering length
$a_{00}$, which in turn depends upon the mass difference $\delta
m_{00} = m_{X(3872)} - m_{D^0}-m_{D^{*0}}$ ($=-0.09\pm0.28$~MeV) as
$a_{00}=\hbar/\sqrt{\mu\delta m_{00}}$, where $\mu$ is the
$D\bar{D}^*$ reduced mass.~This, taken with the experimentally
allowed range  of $\delta m_{00}$ values, suggests that for the
$D^0\bar{D}^{0*}$ configuration, the mean $D^0$-$\bar{D}^{*0}$
separation is huge, of order $a_{00}\sim 10$~fm, and much larger
than the extent of the $D^{+}D^{*-}$ configuration, for which
$\delta m_{+-}=m_{X(3872)}-m_{D^+}-m_{D^{*-}}\simeq 8$~MeV and
$a_{+-}\simeq 2$~fm.~The very different properties of the
$D^{0}\bar{D}^{*0}$ and $D^{+}D^{*-}$ configurations imply that the
$X(3872)$ isospin is not very well defined [114].

The BESIII experiment recently reported the observation of the
process $\ee\rt\gamma X(3872)$ at cms energies in the region of the
$Y(4260)$ peak [115].~The $X(3872)$ was detected via its
$\pipi\jpsi$ decay channel; a $\pipi\jpsi$ invariant mass plot
summed over the data at all of the energy points is shown in
Fig.~18(a), where a $6.3\sigma$ peak at the mass of the $X(3872)$ is
evident.~Figure~18(b) shows the energy dependence of the $X(3872)$
production rate where there is some indication that the observed
signal is associated with the $Y(4260)$.~Assuming that
$Y(4260)\rt\gamma X(3872)$ decays are the source of this signal, and
using the PDG lower limit ${\mathcal B}(X(3872)\rt\pipi\jpsi)>0.026$


\end{multicols}

\vspace{2mm}
\centerline{\psfig{figure=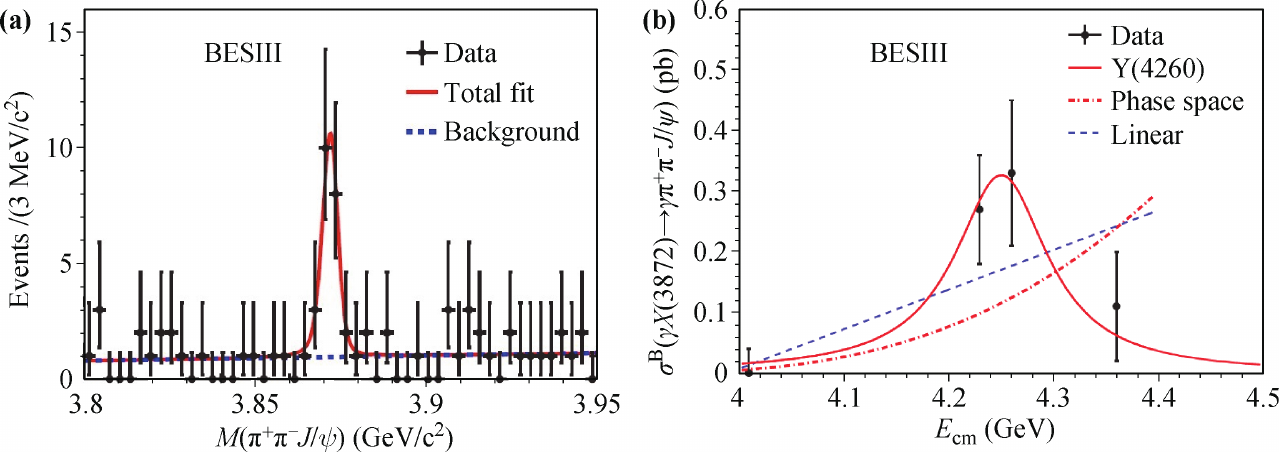}\vspace{1mm}}
 \parbox[c]{160mm}{\baselineskip
10.5pt\renewcommand{\baselinestretch}{1.05}\footnotesize \noindent
{\color{ooorangec}\bf Fig.~18}\quad {\bf(a)} The data points show
the BESIII experiment's $M(\pipi\jpsi)$ distribution for
$\ee\rt\gamma\pipi\jpsi$ events at energies near the $Y(4260)$
resonance.~The fitted peak has a mass and width of $M=3871.9\pm
0.7$~MeV and ${\it\Gamma} = 0.0^{+1.7}_{-0.0}$~MeV ($<2.4$~MeV),
which are in good agreement with the PDG world average values for
the $X(3872)$.~{\bf(b)} The energy dependence of the BESIII
$\sigma^B(\ee\rt\gamma X(3872))\times {\mathcal
B}(X(3872)\rt\pipi\jpsi)$, measurement where $\sigma^B$ denotes the
Born cross section.~The solid curve is the $Y(6260)$ line shape
fitted to the data; the dashed curves show phase-space and linear
production model expectations.

}\vspace{2mm}

\begin{multicols}{2}
\setlength{\parindent}{1em}    

\noindent {}[13], we determine

\vspace{0.3cm}$\displaystyle \frac{{\mathcal B}(Y(4260)\rt\gamma
X(3872))}{{\mathcal B}(Y(4260)\rt \pipi\jpsi)}>0.05, $\hfill (1)

\vspace{0.3cm} \noindent which is substantial and suggestive of some
commonality in the nature of the $Y(4260)$, $X(3872)$ and
$Z_c(3900)$.

Various interpretations for the $X(3872)$ have been proposed.~The
only viable candidate for a charmonium assignment is the
$2^{3}P_{1}$ state, {i.e.}, the $\chip_{c1}$, which is the only
charmonium level with the right quantum numbers and an expected mass
that is anywhere near 3872~MeV.~The measured mass is below the range
of the theoretically expected values discussed in Ref.~[76], where a
computation using a non-relativistic potential model (NR) gives
3925~MeV while that based on a relativized potential model (GI) [36]
gives 3953~MeV.~Potential model calculations of masses of charmonium
states that are above the open-charm thresholds at $2m_D$ and $m_D +
m_{D^*}$ do not explicitly take into account effects of virtual, but
on-mass-shell $D\bar{D}$ and $D\bar{D}^*$ states and, so, it is not
too surprising that both models miss the mass of the well measured
multiplet partner $\chip_{c2}$ by a substantial amount: NR gives
3972~MeV and GI gives 3979~MeV, while the PDG average of measured
values is $\sim 50$~MeV lower at $3927.2\pm 2.7$~MeV.~In the absence
of any computation that incorporates the measured $\chip_{c2}$ as
input, I estimate a range of expected $\chip_{c1}$ mass values
pegged to the $m_{\chip_{c2}}$ measurement to be $3880~{\rm MeV}\le
M(\chip_{c1}) \le 3901$~MeV by subtracting the NR and GI calculated
$\delta m_{2-1}= m(\chip_{c2})- m(\chip_{c1})$ splittings, namely
$\delta m_{2-1}{\rm (NR)}=47$~MeV and $\delta m_{2-1}{\rm (GI)} =
26$~MeV, from the $\chip_{c2}$ measurement.~The $X(3872)$ mass is
below this range, but not by much.~The biggest problem with the
$\chip_{c1}$ charmonium assignment for the $X(3872)$ is the strength
of the $\rho\jpsi$ discovery decay channel, which, for charmonium,
is isospin- and OZI-violating and expected to be strongly
suppressed.

The close match between the $X(3872)$ mass and $m_{D^0}+m_{D^{*0}}$
-- they are equal to within 1 part in~$10^4$ -- suggests that there
is some close relation between the $X(3872)$ and the
$D^0\bar{D}^{*0}$ system.~In 1994, T{\" o}rnqvist proposed the
existence of deuteron-like $D\bar{D}^*$ isoscalar states bound by
pion exchange with $J^{PC}=0^{-+}$ and $1^{++}$ and mass near
3870~MeV that he called ``deusons'' [116].~In August 2003, shortly
after the $X(3872)$ discovery was announced by Belle at the Fermilab
Lepton-Photon 2003 meeting, he identified it with the $1^{++}$
$D\bar{D}^*$ deuson predicted in his 1994 paper.~Because of its
proximity to the $D^{0}\bar{D}^{*0}$ threshold, the deuson wave
function is predominantly $D^0\bar{D}^{*0}$, and not a very pure
isospin eigenstate as discussed above [114], and this can account
for the strong isospin violation.~There has been a considerable
number of subsequent reports that describe the $X(3872)$ as a
molecule [118--129], an idea that predates even T{\" o}rnqvist's
1994 paper [130--133].~However, the CDF group reported that only
($16.1 \pm 5.3$)\% of their $X(3872)$ signal are produced via $B$
meson decays; most of their observed signal is produced promptly in
high energy $p\bar{p}$ interactions [134].~In addition, the D0
experiment showed that the many of the characteristics of $X(3872)$
production in these high energy $p\bar{p}$ collisions, such as the
rapidity- and $p_T$-dependence of the inclusive production cross
section, are very similar to those for the $\psip$.~The LHCb [99]
and CMS [135] experiments report similar characteristics for
inclusive $X(3872)$ production in $\sqrt{s}=7$~TeV proton-proton
collisions.~If the $X(3872)$ is a large and fragile molecule with a
miniscule binding energy, why would its production characteristics
in high energy $pp$ and $p\bar{p}$ collisions match those of the
nearly point-like and tightly bound $\psip$ $\ccbar$ state?  This
issue is discussed in detail in Ref.~[136], where the authors use
the powerful software tools that successfully model particle
production in high energy hadron colliders to show that the prompt
production cross section for a loosely bound $D^0\bar{D}^{*0}$
molecule should be about two-orders-of-magnitude smaller than the
production rates measured by CDF.

Both Belle [138] and BaBar [137] reported similar strengths for
$X(3872)$ decay to $\gamma\jpsi$ final states; the average of their
measured branching fractions correspond to a $X(3872)\rt\gamma\jpsi$
partial width that is less than that for $X(3872)\rt\pipi\jpsi$ by a
factor of $0.24 \pm 0.05$.~This is in good agreement with a
calculation based on a molecular picture by Aceti, Molina and Oset
[139], where they find this ratio to be 0.18.~In addition to
$D^0\bar{D}^{*0}$, this calculation includes $D^{+}D^{*-}$ and
$D_s^+ D_s^{*-}$ components in the $X(3872)$ wave
function.~Interestingly, the authors find that the
${\it\Gamma}(\gamma\jpsi)$/${\it\Gamma}(\pipi\psi)$ partial width
ratio is very sensitive to these extra terms: if only the
$D^0\bar{D}^{*0}$ term is considered, the ratio becomes very much
smaller,  of order $\sim 10^{-4}$.~Since many calculations based on
a molecule picture ignore the $D^{+}D^{*-}$ and (especially) the
$D_s^+ D_s^{*-}$ terms, their reliability may be questionable.

Swanson pointed out that the relative partial widths for the
$X(3872)\rt\gamma\psip$ and $X(3872)\rt\gamma\jpsi$ decay channels
would be a powerful diagnostic for the $X(3872)$ [140].~For a pure
$\chip_{c1}$ charmonium state, he estimates that this ratio would be
in the range 0.7--6.8, while for  pure $D\bar{D}^*$ molecular state
the expectations are distinctly smaller, at $\sim 3\times
10^{-3}$.~Until recently, the situation with $X(3872)\rt\gamma\psip$
was not very clear.~BaBar reported $3.5\sigma$ evidence for this
mode at a rate that is $3.4 \pm 1.4$ times that for $\gamma\jpsi$
[137]; Belle reported no signal and a 90\%~CL upper limit on this
ratio of 2.1 [138].~This year LHCb study of this mode measured a
$4.4\sigma$ signal for $X(3872)\rt\gamma\psip$, and a ratio relative
to an $X(38972)\rt\gamma\jpsi$ signal in the same data of $2.5\pm
0.7$ [141], much higher than expectations for a pure $D\bar{D}^*$
molecular state.

Maiani and collaborators advocate a QCD-type tetraquark structure
({i.e.}, a $cq$ diquark and a $\bar{c}\bar{q}$ diantiquark, where
$q=d$~or~$u$) for the $X(3872)$ [142].~In their first paper on the
subject, they predicted that the $X(3872)$ would be one of two
neutral states that differ in mass by $8\pm 3$~MeV, a value that is
related to the $u$- and $d$-quark mass difference; one of the states
was predicted to be produced in $B^+\rt K^+X_1(3872)$ and the other
in $B^0\rt K^0X_2(3872)$ decays.~They also predicted the existence
of charged partner states, $X^{\pm}$, nearby in mass.\ BaBar [106]
and Belle [102] studied these possibilities and found no evidence
for different $X(3872)$ mass differences in $B^+$ and $B^0$ decays
and put limits on ${\mathcal B}(B\rt K X^{\pm})\times {\mathcal
B}(X^{\pm}\rt \pi^{\pm}\piz\jpsi$ that are well below isospin-based
predictions.~However, the recent discovery of the
$Z_c(3900)^{\pm,0}$ triplet has supplied the ``missing'' particles,
albeit with different $G$-parity and at somewhat higher masses.~This
revived interest in QCD-tetraquark ideas [143].~Among other things,
this model has a natural explanation for the $Z(4430)$ [144] and,
since QCD-tetraquarks are tightly bound, compact objects, their
prompt production in high hadron collisons with properties can be
expected to be similar to that for the $\psip$.~On the other hand,
Ref.~[143] proposed a second $Z_c$ state at about 100~MeV below the
$Z_c(3900)$ mass and contrasted this prediction with a molecule
expectation for a second $Z_c$ existence of another $\pip \jpsi$
near the $D^*\bar{D}^*$ threshold.~The subsequent observation by
BESIII [51] of the higher mass $Z_c(4020)$ near the $2m_{D^*}$
threshold clearly favors the molecule scheme.

QCD-hybrid mesons are color-octet quark-antiquark combinations with
an excited gluon degree of freedom [145].~While there is
considerable literature that identifies the $Y(4260)$ as a
charmonium hybrid, {i.e.}, a $\ccbar$-gluon structure [46, 47],
there have been no suggestions that this idea may apply to the
$X(3872)$ as well.~This may be partly due to the fact that LQCD
calculations find the lowest $1^{++}$ charmonium hybrid mass to be
near 4400 MeV and far above that of the $X(3872)$ [146].

Hadrocharmonium is a model in which a $Q\bar{Q}$ pair forms a
tightly bound system embedded in a light mesonic cloud that it
interacts with via QCD analogs of Van der Waals forces [130,
147].~The $Q\bar{Q}$ core states have wave functions that are
closely related to conventional Quarkonium states.~This would
explain, for example, the dominance of $Z(4430)\rt\pi\psip$ over
$\pi\jpsi$ decays.~The proponents of this model have not made any
$X(3872)$-specific predictions.

The defects in all of the above-mentioned pictures for the $X(3872)$
have inspired models that incorporate combinations of the different
ideas.~For example, a charmonium-molecular
hybrid$^{10)}$\footnote{\hspace*{-5mm}$^{10)}$~This ``hybrid'' is
not the same as the previously discussed QCD hybrid.} model by
Takizawa and Takeuchi finds a specific $X(3872)$ wave function
[148]:

\vspace{0.25cm}$\displaystyle |X(3872)\rangle=0.237|\ccbar\rangle
-0.944|D^0\bar{D}^{* 0}\rangle $

\vspace{0.15cm} $~~~~~~~~~~~~~~~~-0.228|D^+D^{*-}\rangle , $\hfill
(2)

\vspace{0.25cm} \noindent which translates into about 6\% $\ccbar$,
69\% of isoscalar $D\bar{D}^*$ and 26\% isovector
$D\bar{D}^*$.~Hadronic production is hypothesized to proceed via the
$\ccbar$ core component and this could explain why $X(3872)$
production properties are similar to those of the $\psip$.~This
calculation uses a high $\chip_{c1}$ input mass, namely 3950~MeV,
which may result in an underestimate of the $\ccbar$ component, but
does not negate the basic idea.~Note that similar numerical results
are found in Ref.~[113].

There are two related states that are orthogonal to the one given in
Eq.~(2).~Reference [148] discusses one that is mostly $\ccbar$ and
probably should  be considered to be the $\chip_{c1}$ charmonium
state.~This is pushed up in mass to well above both $D\bar{D}^*$
thresholds and becomes wide and not easily identifiable.~Presumably
the other state would be dominantly isovector $D\bar{D}^*$, and this
might be the $Z_c(3900)^0$, the neutral $I_z=0$ member of the
$Z_c(3900)^{\pm,0}$ isospin triplet.~If so, the residual $\ccbar$
core and isoscalar $D\bar{D}^*$ components may result in a
measureable isospin violating differences between $Z_c(3900)^0\rt
D^0\bar{D}^{*0}$  and $Z_c(3900)^0\rt D^+ D^{*-}$ decays.

The $X(3872)$ has a long interesting story that we cannot do justice
to in this brief report and, instead, refer the readers to a recent
review [149].

\vspace*{5mm} \noindent {\color{ooorangec}\it 3.2.2\quad
Bottomoniumlike mesons}\vspace{3.5mm}

\noindent Figure~19 indicates the recent status of the $\bbbar$
bottomonium and bottomoniumlike mesons.~Here the beige boxes
indicate the well established bottomonium mesons, the green boxes
show those that were recently established, and the red and purple
boxes indicate anomalous states that are discussed below.

The large $Y(4260)\rt\pipi\jp$ signal discovered in the charmonium
mass region by BaBar motivated a Belle search for similar behavior
in the bottomonium system [150].~This uncovered anomalously large
$\pipi\Upsilon(nS)$ $(n=1,2,3)$ production rates that peak around
$\sqrt{s}=10.89$~GeV as shown in the upper panel of Fig.~20(a)
[151].~This peak energy is about $2\sigma$ higher than that of the
peak in the $\ee\rt~hadron$ cross section at
$\sqrt{s}\simeq10.87$~GeV that is usually associated with the
conventional $\Upsilon(5S)$ bottomonium meson (and shown in the
lower panel of Fig.~20(b).~If the peak in the $\pipi\Upsilon(nS)$
cross section is attributed to $\Upsilon(5S)$ decays, it implies
$\Upsilon(5S)\rt \pipi\Upsilon(nS)$ ($n=1,2,3$) partial widths that
are hundreds of times larger than theoretical predictions [152], and
the corresponding measured rates for the $\Upsilon(4S)$ [13].~This
suggests the presence of a new, non-$\bbbar$ $b$-quark-sector
equivalent of the $Y(4260)$ with mass around $10.89$~GeV [153] and
referred to in the following as the $\Upsilon(10890)$.

The large cross sections for $\ee\rt\pipi\Upsilon(nS)$ ($n=1,2,3$)
near the $\Upsilon(5S)$ motivated Belle to take a large sample of
data at this energy in order to investigate the source of this
anomaly.~By the end of Belle operations, a 121.4~fb$^{-1}$ data
sample had been accumulated near the peak of the $\Upsilon(5S)$
($\sqrt{s}=10.87$~GeV) and data taken in energy scans around the
peak accounted for an additional sample of 12~fb$^{-1}$.~One of the
first studies done with this $\Upsilon(5S)$ data sample was an
investigation of the inclusive missing-mass spectrum recoiling from
$\pipi$ pairs.

The upper right panel of Fig.~20 shows the distribution of masses
recoiling against all of the $\pipi$ pairs in events collected near
the peak of the $\Upsilon(5S)$ resonance [154].~The combinatoric
background is huge -- there are typically $10^6$ entries in each
1~MeV bin -- and the statistical errors are tiny ($\sim 0.1\%$).~The
data were fit piece-wise with sixth-order polynomials, and the
residuals from the fits are shown in the lower panel of Fig.~20(b),
where, in\vspace{-0.4cm}\linebreak

\end{multicols}

\vspace{-1mm}
\centerline{\psfig{figure=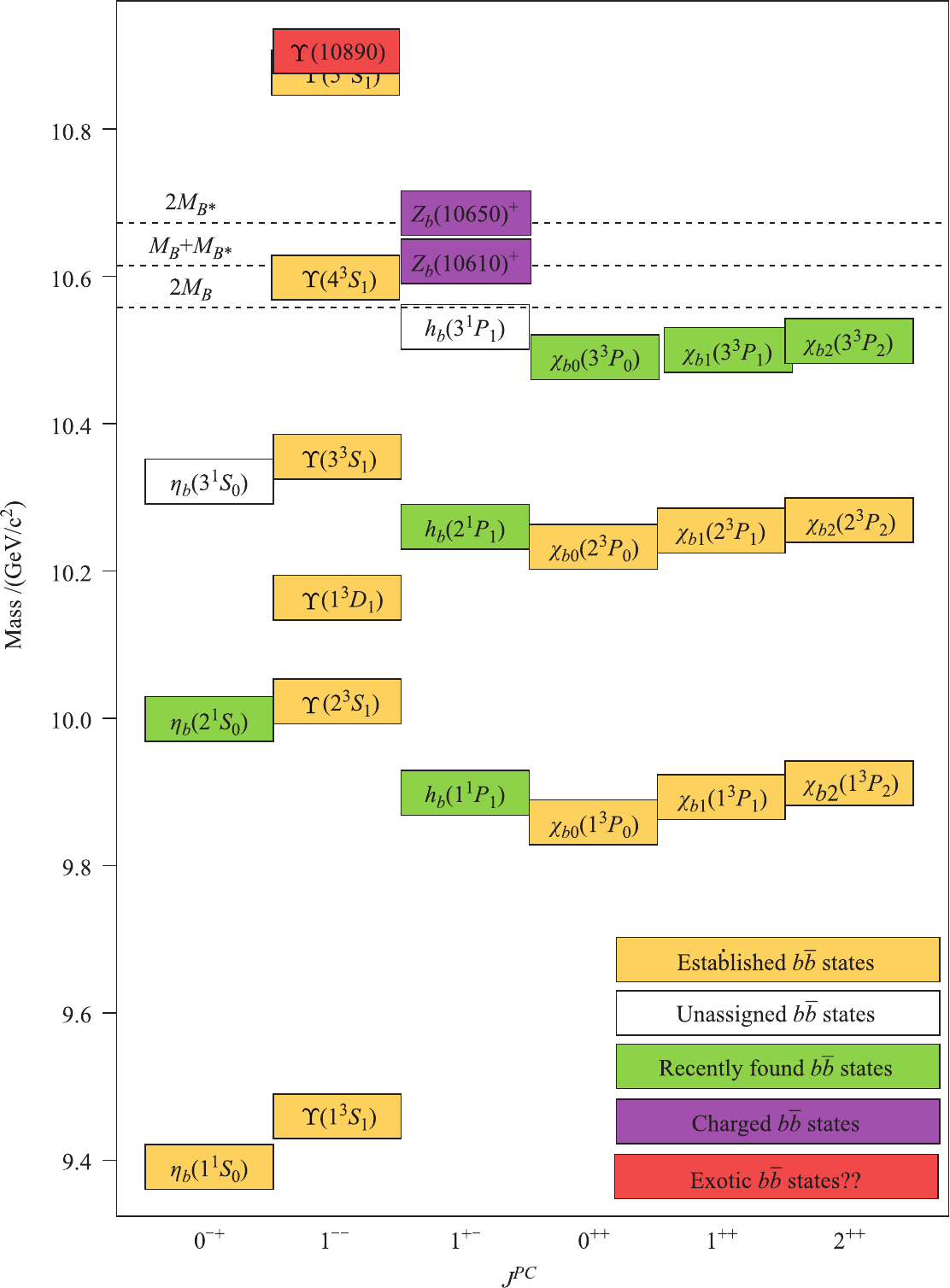}\vspace{1mm}}
 {\center\baselineskip
10.5pt\renewcommand{\baselinestretch}{1.05}\footnotesize \noindent
{\color{ooorangec}\bf Fig.~19}\quad The spectrum of bottomonium and
bottomoniumlike mesons.

}\vspace{2mm}

\vspace{3mm}
\centerline{\psfig{figure=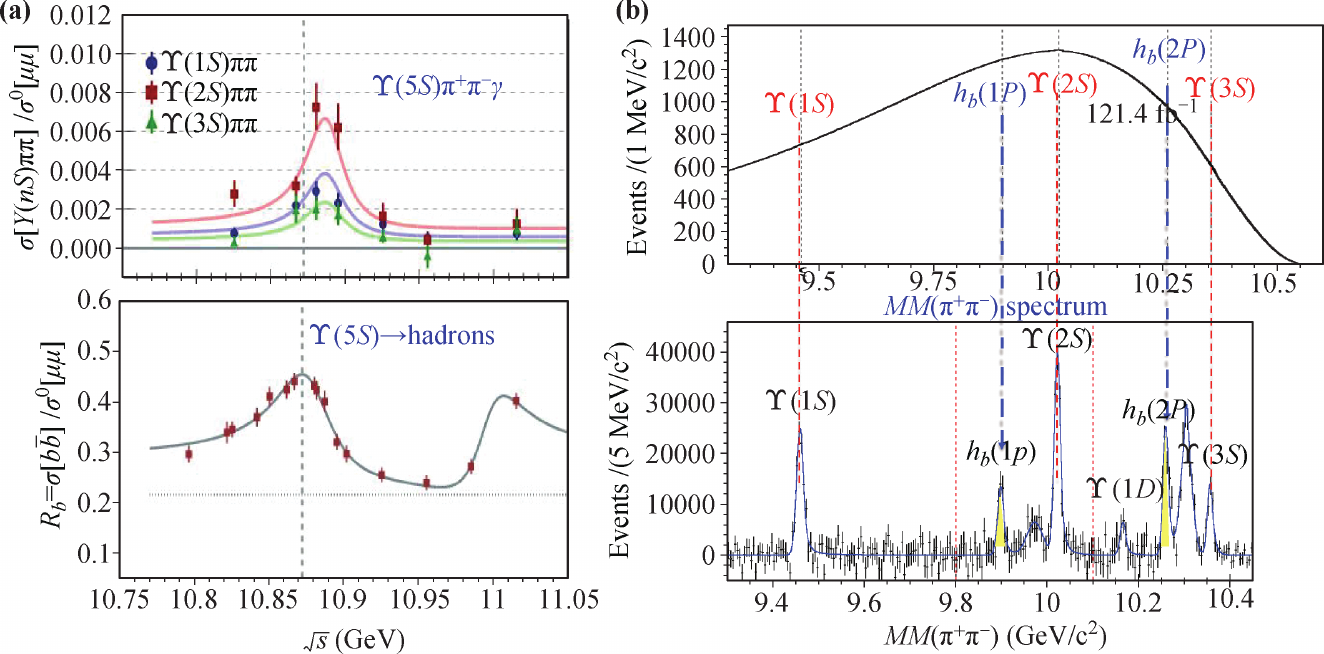}\vspace{1mm}}
 \parbox[c]{160mm}{\baselineskip
10.5pt\renewcommand{\baselinestretch}{1.05}\footnotesize \noindent
{\color{ooorangec}\bf Fig.~20}\quad {\bf(a)} Cross sections for
$\ee\rt \pipi\Upsilon (nS)$ ($n=1,2,3$) ({\it upper}) and
$\ee\rt~hadrons$ ({\it lower}) in the vicinity of the $\Upsilon(5S)$
resonance (from Ref.~[151]).~{\bf(b)} Distribution of masses
recoiling against $\pipi$ pairs at cms energies near $10.87$~GeV
({\it upper}) and residuals from piecewise fits to the data with
smooth polynomials ({\it lower}) (from Ref.~[154]).~The $h_b(1P)$
and $h_b(2P)$ peaks, shaded in yellow, are first observations.

}\vspace{1mm}

\begin{multicols}{2}
\setlength{\parindent}{1em}    

\noindent addition to peaks at the $\yones$, $\ytwos$,
$\Upsilon(3S)$ masses and some expected reflections, there are
unambiguous signals for the $h_b(1P)$ and $h_b(2P)$, the $1^1P_1$
and $2^1P_1$ bottomonium states.~This was the first observation of
these two elusive levels [154].~One puzzle is that the $\pipi
h_b(mS)$, ($m=1,2$) final states are produced at rates that are
nearly the same as those for $\pipi\Upsilon(nS)$ ($n=1,2,3$), even
though the $\pipi h_b$ transition requires a heavy-quark spin flip
that should result in a strong suppression.~This motivated a more
detailed investigation of these channels.

Figure~21(a) shows the $\pipi h_b$ yields versus the maximum $h_b
\pi^{\pm}$ invariant mass for $h_b=h_b(1P)$ (upper) and
$h_b=h_b(2P)$ (lower), where it can be seen that all of the $\pipi
h_b$ events are concentrated in two $M_{\rm max}(h_b\pi)$ peaks near
$10~610$~MeV and and $10~650$~MeV [155].~Studies of fully
reconstructed $\pipi \Upsilon(nS)$, ($n=1,2,3)$
$\Upsilon(nS)\rt\ell^+\ell^-$ events in the same data sample found
peaks at the same masses in the $M_{\rm max}(\Upsilon(nS)\pi)$
distributions for all three $\Upsilon(nS)$ states; these are shown
in the three center panels of Fig.~21.~Here the fractions of
$\pipi\Upsilon(nS)$ events in the two peaks are substantial -- $\sim
6\%$ for the $\yones $, $\sim 22\%$ for the $\ytwos$ and $\sim 43\%$
for the $\Upsilon(3S)$ -- but, unlike the case for the $\pipi
h_b(mP)$ channels, they do not account for all of the
$\pipi\Upsilon(nS)$ event yield [156].~Fitted values of the peak
masses and widths in all five channels are consistent with each
other; the weighted average mass and width values are

\vspace{0.2cm}$\displaystyle Z_1(10610){:}~~ M_1       = 10607.2 \pm
2.0 \ {\rm MeV} ,~~ $

\vspace{0.1cm}$\displaystyle\hspace{19mm}{\it\Gamma}_1 = 18.4 \pm
2.4 \ {\rm MeV}$;

\vspace{0.1cm}$\displaystyle Z_2(10650){:}~~ M_2      = 10552.2 \pm
1.5 \ {\rm MeV},~~~ $

\vspace{0.1cm}$\displaystyle\hspace{19mm}{\it\Gamma}_2 = 11.5 \pm
2.2 \ {\rm MeV.} $

\vspace{0.25cm} \noindent A study of $\piz\piz\Upsilon(nS)$,
$(n=1,2,3)$ found a $6.5\sigma$ signal for the neutral
$Z_b(10610)^0$ isospin partner state with a mass $M_{Z_b(10610)^0}=
10609\pm 6$~MeV and a production rate that is consistent with
isospin-based expectations [157].

The lower mass state is just $2.6\pm 2.2$~MeV above the
$m_B+m_{B^*}$ mass threshold and the higher mass state is only
$2.0\pm 1.6$~MeV above $2m_{B^*}$.~Dalitz-plot analyses of the
$\pipi\Upsilon(nS)$ final states establish $J^P=1^+$ quantum number
assignments for both states [158, 159].~The close proximity of the
$B\bar{B}^*$ and $B^*\bar{B}^*$ thresholds and the $J^P=1^+$ quantum
number assignment is suggestive of virtual $S$-wave molecule-like
states.

The $B^{(*)}\bar{B}^{*}$ molecule picture is supported by a Belle
study of $\ee\rt B^{(*)}\bar{B}^*\pi$ final states in the same data
sample [160], where $B\bar{B}^*$ and $B^*\bar{B}^*$ invariant mass
peaks are seen at the $Z_b(10610)$ and $Z_b(10650)$ mass values,
respectively, as shown in the right panels of Fig.~21.~From these
data, preliminary values of the branching fractions ${\mathcal
B}(Z_b(10610)\rt B^+\bar{B}^{*0} + \bar{B}^0B^{*+} + c.c.) =
(86.0\pm 3.6)\%  $ and ${\mathcal B}(Z_b(10610)\rt
B^{*+}\bar{B}^{*0} + c.c.) = (73.4\pm 7.0)\%  $ are inferred.~The
$B^{(*)}\bar{B}^{*}$ ``fall apart'' modes are stronger than the sum
total of the $\pip\Upsilon(nS)$ and $\pip h(mP)$ modes, but only by
factors of $\sim 6$ for the $Z_b(10610)$ and $\sim 3$ for the
$Z_b(10650)$.~The measured branching fraction for $Z_b(10610)\rt
B^{*}\bar{B}^{*}$ is consistent with zero.~This pattern, where
$B\bar{B}^*$ decays dominate for the $Z_b(10610)$ and $B^*\bar{B}^*$
decays are dominant for the $Z_b(10650)$ are consistent with
expectations for molecule-like structures.

\vspace*{5mm} \noindent {\color{ooorangec}3.3\quad Comments
}\vspace{3.5mm}

\noindent Table~1 provides a tabulation of recently
discovered\vspace{-0.4cm}\linebreak

\end{multicols}

\vspace{3mm}
\centerline{\psfig{figure=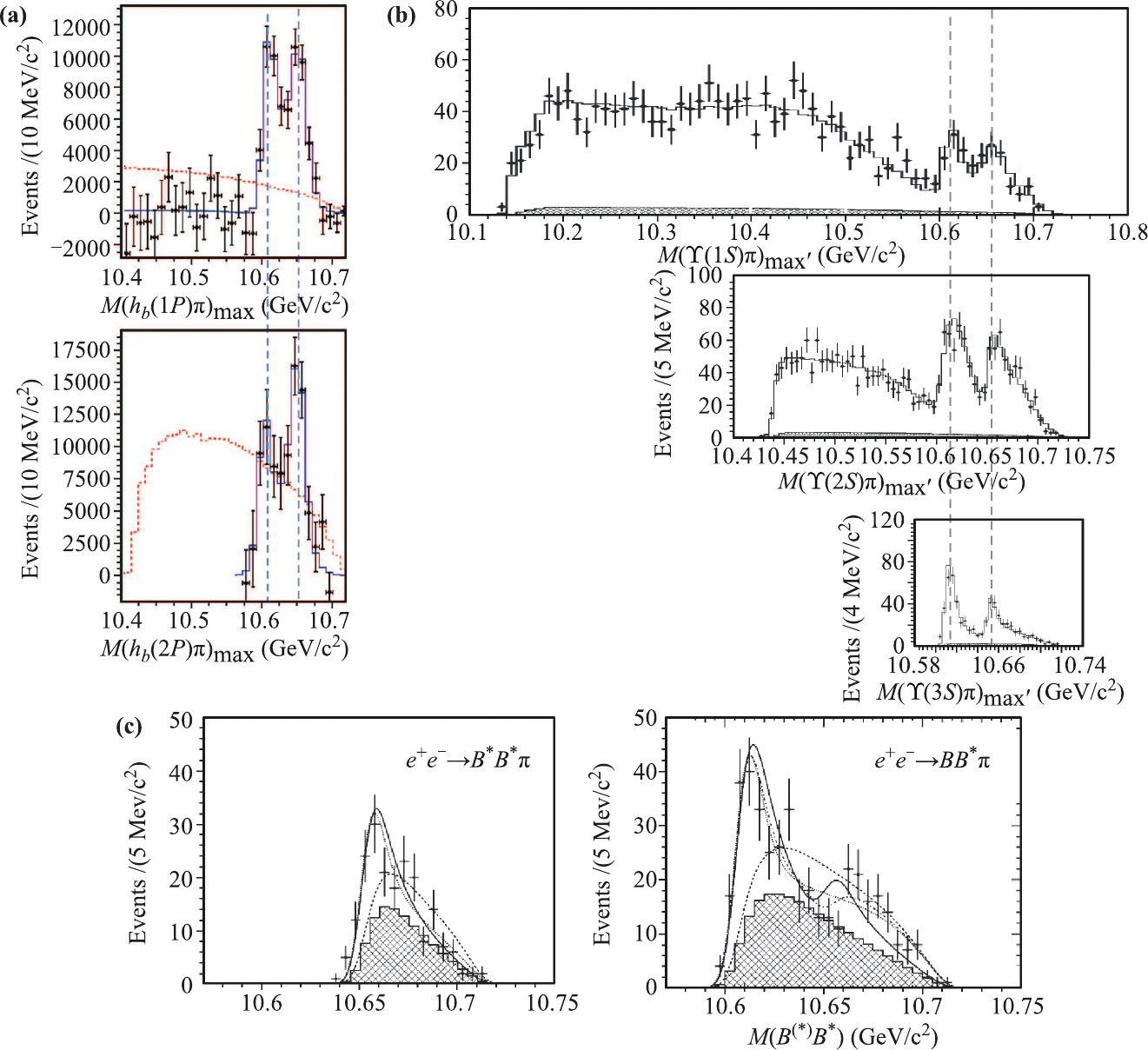}\vspace{1mm}}
 \parbox[c]{160mm}{\baselineskip
10.5pt\renewcommand{\baselinestretch}{1.05}\footnotesize \noindent
{\color{ooorangec}\bf Fig.~21}\quad {\bf(a)} Invariant mass
distributions for $h_b(1P)\pip$ ({\it upper}) and $h_b(2P)\pip$
({\it lower}) from $\ee\rt\pipi h_b(nP)$ events.~{\bf(b)} Invariant
mass distributions for $\yones\pip$ ({\it upper}), $\ytwos\pip$
({\it center}) and $\Upsilon(3S)\pip$ ({\it lower}) in
$\ee\rt\pipi\Upsilon(nS)$ events.~The figures are from Ref.~[155],
and scaled to make the make the horizontal scales (almost) match.
{\bf(c)} The $B^*\bar{B}^*$ ({\it upper}) and $B\bar{B}^*$ ({\it
lower}) invariant mass distributions for $\ee\rt
B^{(*)}\bar{B}^*\pi$ events near $\sqrt{s}=10.86$~GeV (from
Ref.~[160]).

}\vspace{1mm}

\begin{multicols}{2}
\setlength{\parindent}{1em}    

\noindent mesons (and candidate mesons), together with observed
production \& decay channels, averages of mass and width
measurements and the $J^{PC}$ values when they are known.~For
simplicity, I assume the $Z_c(3900)\rt \pip\jpsi$ and $Z_c(3885)\rt
(D\bar{D}^*)^+$ are the same state and average the measured mass and
width values, the $Z_c(4020)\rt\pip h_c$ and $Z_c(4025)\rt
(D^*\bar{D}^*)^+$ measurements are treated in the same way.

\vspace*{5.5mm} \noindent {\color{ooorangec}\it 3.3.1\quad
Molecules?}\vspace{3.5mm}

\noindent The recent BESIII findings, taken together with previous
experimental results, establishes a concentration of charmoniumlike
states crowding the $D\bar{D}^*$  and $D^*\bar{D}^*$  mass threshold
regions and bottomoniumlike isospin triplets near the $B\bar{B}^*$
and $B^*\bar{B}^*$ thresholds, which are suggestive of molecule-like
structures [118--129].

If we assume that the $Z_c(3900)$ and $Z_c(3885)$ are the same
state, it corresponds to an isospin triplet with a pole mass that is
about $\sim$15~MeV above the $D^0D^{*-}$ threshold with $J^P=1^+$
and a significant coupling to $D\bar{D}^*$.~This is consistent with
a virtual $S$-wave $D\bar{D}^*$ molecule of the form
$(D\bar{D}^*-D^*\bar{D})/\sqrt{2}$.~The $X(3872)$ also has $J^P=1^+$
and couples to $D\bar{D}^*$ with mass right at the ${D^0}
{\bar{D}^{*0}}$ threshold and seems to be an isospin singlet
[102].~This suggests an $S$-wave $D\bar{D}^*$ molecule of the form
$(D\bar{D}^*+D^*\bar{D})/\sqrt{2}$.~(Note that although they both
have $J^P=1^+$, the $X(3872)$ has even $C$ parity, while the $I_z=0$
member of the $Z_c(3900)$ triplet must have odd $C$ parity.)

The $Z_c(4020)/Z_c(4025)$ observations (assuming they are the same
state) establish the existence of another isospin triplet just above
the $D^*\bar{D}^*$ threshold.~If this is the charmed-sector
equivalent of the $Z_b(10650)$, it also has $J^P=1^+$ and could be
considered as a virtual $D^*\bar{D}^*$ molecule for which the
$I_z=0$ component must have odd $C$ parity.~This suggests that there
might be an $X(3872)$-like, even $C$ parity isospin singlet
$D^*\bar{D}^*$ nearby, and labeled as $X_{c2}$ in the lower part of
the level diagram shown in the left panel of Fig.~22.


\end{multicols}

\vspace{3mm}\noindent {\baselineskip
10.5pt\renewcommand{\baselinestretch}{1.05}\footnotesize \noindent
{\color{ooorangec}\bf Table 1}\quad New $c\bar{c}$ and $b\bar{b}$
mesons above open-flavor threshold.~The masses $M$ and widths
${\it\Gamma}$ are weighted averages of measurements with
uncertainties added in quadrature.~For the $X(3872)$, only $
\pipi\jpsi $ decays are used in the mass average.\ Ellipses (...)
indicate an inclusive reaction.~In the $J^{PC}$ column, question
marks indicate my educated guess or no information.\ For charged
states, $C$ is that of the neutral isospin
partner.~$``\Upsilon(5S)''$ is in quotes to reflect the suspicion
that the anomalous $\pipi\Upsilon(nS)$ events originate not from the
$\Upsilon(5S)$, but from another $1^{--}$ meson with a nearby mass,
{i.e.}, the $\Upsilon(10890)$.~This Table is a modification of one
from Ref.~[161] via Ref.~[29, 30].

} \vspace{1mm}
{\renewcommand{\arraystretch}{1.38}\tabcolsep=2pt\footnotesize\noindent
\begin{tabular}{@{\hspace{2.8mm}}lcccll}\hline
 State & $M$~/MeV & ${\it\Gamma}$~/MeV & $J^{PC}$ & Process~(decay mode) &     Experiment  \\[-3mm]
 \multicolumn{6}{@{\hspace{0mm}}c@{\hspace{0mm}}}{\def\temptablewidth{\textwidth}{\rule{\temptablewidth}{1pt}}}\\[-1mm]
    ${X(3872)}$& 3871.68$\pm$0.17 & $<1.2$ &    $1^{++}$    & $B \to K + (J/\psi\, \pi^+\pi^-)$ &    Belle~[95, 102], \babar~[98], LHCb~[103]  \\
    & & & & $p\bar p \to (J/\psi\, \pi^+\pi^-)+ ...$ &    CDF~[96, 104, 105, 160],    \DZero~[97]  \\
    & & &   & $B \to K + (J/\psi\, \pi^+\pi^-\pi^0)$ &    Belle~[107],    \babar~[72, 73]  \\
    & & & & $B \to K + (D^0 \bar D^0 \pi^0)$ &    Belle~[108, 109],    \babar~[110]\\
    & & & & $B \to K + (J/\psi\, \gamma)$ &    \babar~[137],    Belle~[138],    LHCb~[141] \\
    & & & & $B \to K + (\psip \, \gamma)$ &    \babar~[137],    Belle~[138],    LHCb~[141] \\
    & & & &  $pp \to (J/\psi\, \pi^+\pi^-)+ ...$   &    LHCb~[99],    CMS~[100]  \\
    ${X(3915)}$ & $3917.4\pm2.7$ & 28$^{+10}_{-~9}$ & $0^{++}$ &    $B \to K + (J/\psi\, \omega)$ &    Belle~[71],    \babar~[72, 73] \\
    & & & & $e^+e^- \to e^+e^- + (J/\psi\, \omega)$ &    Belle~[74],    \babar~[75] \\
    ${\chi_{c2}(2P)}$ & $3927.2\pm2.6$ & 24$\pm$6 & $2^{++}$ &     $e^+e^-\to e^+e^- + (D\bar{D})$ &     Belle~[78],     \babar~[79] \\
    $X(3940)$ & $3942^{+9}_{-8}$ & $37^{+27}_{-17}$ & $0(?)^{-(?)+}$ &     $e^+e^- \to J/\psi + (D^* \bar D)$ &     Belle [32] \\
    &&&& $e^+e^- \to J/\psi + (...)$ &     Belle~[31] \\
    ${G(3900)}$ & $3943\pm21$ & 52$\pm$11 & $1^{--}$ &     $e^+e^- \to \gamma + (D \bar D)$ &     \babar~[163],      Belle~[164]      \\
    $Y(4008)$ & $4008^{+121}_{-\ 49}$ & 226$\pm$97 & $1^{--}$ &     $e^+e^- \to \gamma + (J/\psi\, \pi^+\pi^-)$ &    Belle~[39]      \\
    $Y(4140)$ & $4144\pm3$  & $17 \pm 9$ & $?^{?+}$ &     $B \to K + (J/\psi\, \phi)$ &     CDF~[87, 88],     CMS~[90]\\
    $X(4160)$ & $4156^{+29}_{-25} $ & $139^{+113}_{-65}$ &$0(?)^{-(?)+}$ &     $e^+e^- \to J/\psi + (D^* \bar D)$ &     Belle [32] \\
    ${Y(4260)}$ & $4263^{+8}_{-9}$ & 95$\pm$14 & $1^{--}$ & $e^+e^- \to \gamma + (J/\psi\, \pi^+\pi^-)$ & \babar~[37, 165],  CLEO~[166], Belle~[39]\\
    & & & & $e^+e^-\to (J/\psi\, \pi^+\pi^-)$ &CLEO~[167]\\
    & & & & $e^+e^-\to (J/\psi\, \pi^0\pi^0)$ & CLEO~[167]\\
    $Y(4274)$ & $4292 \pm 6$ & $34\pm 16$ & $?^{?+}$ &     $B\to K + (J/\psi\, \phi )$ &     CDF~[88],     CMS~[90] \\
    $X(4350)$ & $4350.6^{+4.6}_{-5.1}$ & $13.3^{+18.4}_{-10.0}$ &0/2$^{++}$ &     $e^+e^-\to e^+e^- \,(J/\psi\, \phi)$ &     Belle~[94]  \\
    ${Y(4360)}$ & $4361\pm13$ & 74$\pm$18 & $1^{--}$ &     $e^+e^-\to\gamma + (\psip \, \pi^+\pi^-)$ &     \babar~[38],     Belle~[40] \\
    $X(4630)$ & $4634^{+\ 9}_{-11}$ & $92^{+41}_{-32}$ & $1^{--}$ &     $e^+e^-\to\gamma\, (\Lambda_c^+ \Lambda_c^-)$ &     Belle~[168]\\
    $Y(4660)$ & 4664$\pm$12 & 48$\pm$15 & $1^{--}$ &     $e^+e^-\to\gamma + (\psip \, \pi^+\pi^-)$ &     Belle~[40] \\ \hline
    ${Z_c^+(3900)}$ & $3890\pm 3$ & $33\pm 10$ & $1^{+-}$ &     $Y(4260) \to \pi^- + (J/\psi\, \pi^+)$ &     BESIII~[49],~Belle~[50]  \\
     &  &  &  &     $Y(4260)\to \pi^- + (D\bar{D}^{*})^+$ &     BESIII~[69] \\
    ${Z_c^+(4020)}$ & $4024\pm 2$ & $10\pm 3$ & $1(?)^{+(?)-}$ &     $Y(4260) \to \pi^- + (h_c\, \pi^+)$ &     BESIII~[51] \\
     &  &  &  &     $Y(4260) \to \pi^- + (D^*\bar{D}^{*})^+$ &     BESIII~[52]  \\
    $Z_1^+(4050)$ & $4051^{+24}_{-43}$ & $82^{+51}_{-55}$ & $?^{?+}$&     $ B \to K + (\chi_{c1}\, \pi^+)$ &     Belle~[53],     \babar~[66] \\
    $Z^+(4200)$ & $4196^{+35}_{-32}$ & $370^{+99}_{-149}$ & $1^{+-}$&     $ B \to K + (\jpsi \, \pi^+)$ &     Belle~[62] \\
    $Z_2^+(4250)$ & $4248^{+185}_{-\ 45}$ &     177$^{+321}_{-\ 72}$ &  $?^{?+}$ &     $ B \to K + (\chi_{c1}\, \pi^+)$ & Belle~[53], \babar~[66] \\
    $Z^+(4430)$ & $4477\pm 20$ & $181 \pm 31$ & $1^{+-}$&     $B \to K + (\psip \, \pi^+)$ &     Belle~[54, 56, 57],     LHCb~[58]\\
     &  &  & &     $B \to K + (J\psi\, \pi^+)$ &     Belle~[62] \\ \hline
    $Y_b(10890)$ & 10888.4$\pm$3.0 & 30.7$^{+8.9}_{-7.7}$ & $1^{--}$ &      $e^+e^- \to (\Upsilon(nS)\, \pi^+\pi^-)$ &      Belle~[152] \\ \hline
    $Z_{b}^+(10610)$ & 10607.2$\pm$2.0 & 18.4$\pm$2.4 & $1^{+-}$ &
       $``\Upsilon(5S)'' \to \pi^- + (\Upsilon(nS)\,\pi^+)$,~$n=1,2,3$ &      Belle~[155, 158, 159] \\
     &  &  &  &      $`` \Upsilon(5S)'' \to \pi^- + (h_b(nP)\,\pi^+)$,~$n=1,2$ &       Belle~[155] \\
     &  &  &  &      $`` \Upsilon(5S)'' \to \pi^- + (B\bar{B}^*)^+$,~$n=1,2$ &       Belle~[160] \\
    $Z_{b}^0(10610)$ & 10609$\pm$ 6 &  & $1^{+-}$ &       $``\Upsilon(5S)'' \to \pi^0 + (\Upsilon(nS)\,\pi^0)$,~$n=1,2,3$ &      Belle~[157] \\
    $Z_{b}^+(10650)$ & 10652.2$\pm$1.5 & 11.5$\pm$2.2 & $1^{+-}$ &       $`` \Upsilon(5S)'' \to\pi^- + (\Upsilon(nS)\,\pi^+)$,~$n=1,2,3$ & Belle~[155] \\
     &  &  &  &       $`` \Upsilon(5S)'' \to \pi^- + (h_b(nP)\,\pi^+)$,~$n=1,2$ &       Belle~[155] \\
     &  &  &  &      $`` \Upsilon(5S)'' \to \pi^- + (B^*\bar{B}^*)^+$,~$n=1,2$ &       Belle~[160] \\
 \hline
\end{tabular}}
\vspace{3mm}

\vspace{3mm}
\centerline{\psfig{figure=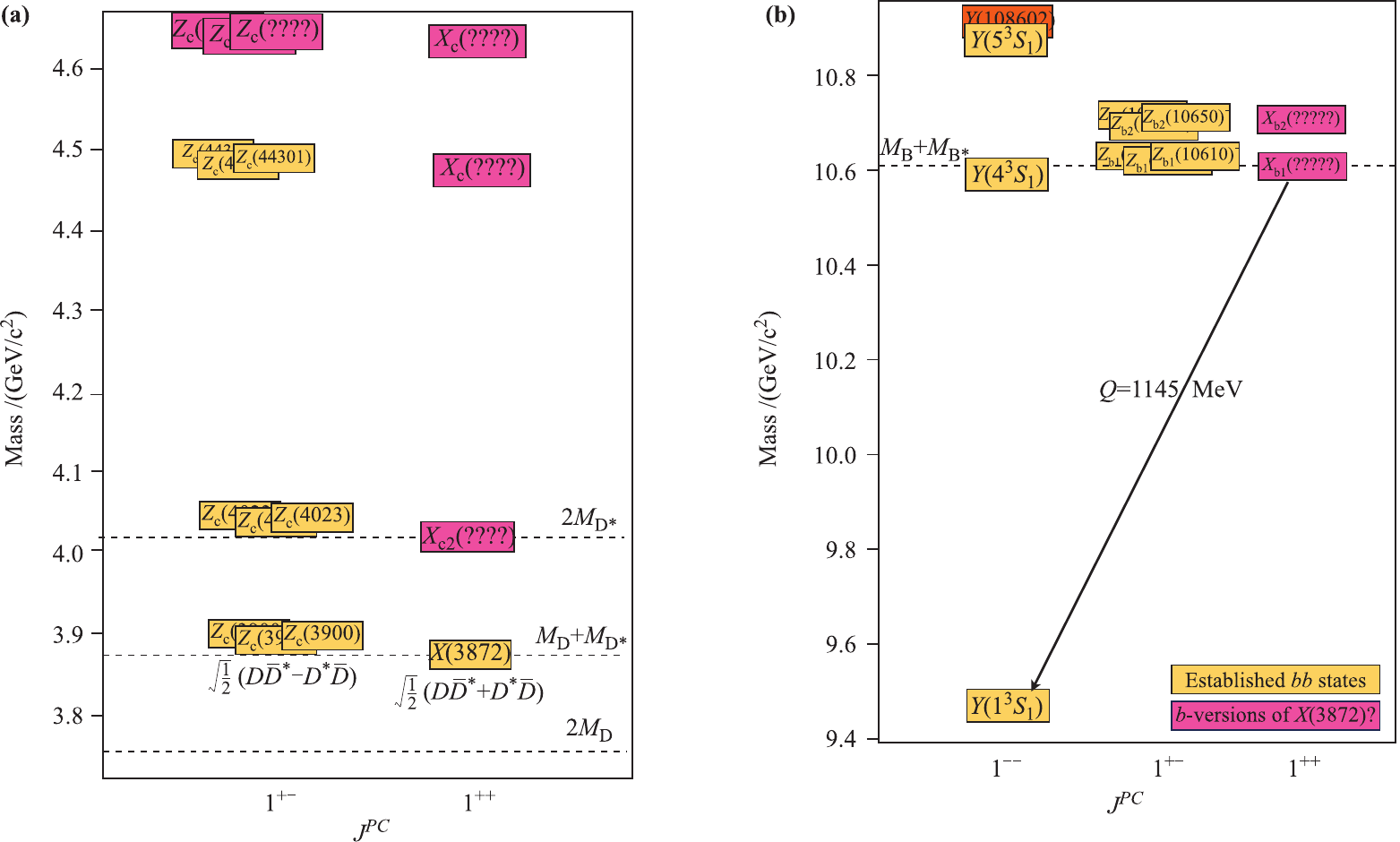}\vspace{1mm}}
 \parbox[c]{160mm}{\baselineskip
10.5pt\renewcommand{\baselinestretch}{1.05}\footnotesize \noindent
{\color{ooorangec}\bf Fig.~22}\quad {\bf(a)} Level diagram for the
$X(3872)$, the recently discovered $Z_c(3900)$ \& $Z_c(4020)$
isotopic triplets, and the $Z(4430)$ isospin triplet.~The
salmon-colored boxes indicate other states that are suggested by the
molecule picture.~{\bf(b)} Level diagram for the recently discovered
$Z_b$ states, a conjectured b-sector equivalent of the $X(3872)$ at
the $m_{B}+m_{B^*} $ threshold and an additional isoscalar partner
of the $Z_b(10650)$ at the $2m_{B^*}$ theshold.~The transition
between a $m_{B}+m_{B^*}$ threshold state to the $\Upsilon(1S)$
would have a $Q$-value of $\simeq 1145$~MeV, well above the mass of
the $\rho$ and $\omega$ mesons.

}\vspace{4mm}

\begin{multicols}{2}
\setlength{\parindent}{1em}    


If the $X_{c2}$ exists with a mass that is below the $D^*\bar{D}^*$
threshold and a relatively narrow width, it might have a significant
branching fraction to $\omega\jp$ final states.~If its mass is above
$2m_{D^*}$ and it is relatively narrow, it might be accessible in
$B^-\rt K^- D^*\bar{D}^*$ decays.~BaBar has reported large branching
fractions for  $B^-\rt K^-D^{0*}\bar{D}^{*0}$ ($1.1\pm 0.1$\%) and
$K^- D^{*+}D^{*-}$ ($0.13\pm 0.02$\%) but did not publish any
invariant mass distributions [169].

The $Z_c$ mesons have a minimal four-quark structure of $\ccbar
q_i\bar{q}_j$ ($i=1,2$,~$j=1,2$~\&~$q_1=u,~q_2=d$).~The locations of
the $Z_c(3900)$ and the $Z_c(4020)$ near the $D\bar{D}^*$ and
$D^*\bar{D}^*$ thresholds and the similarity of the
$M_{Z_c(4020)}-M_{Z_c(3900)} = 133.9\pm 4.0$~MeV mass splitting with
the $\langle m_{D^*}-m_{D} \rangle\simeq 141$~MeV mass difference
indicates that configurations where $c\bar{q}_j$ form a $D^{(*)}$
and $\bar{c} q_j$ form a $D^*$ are important, as is the case for
molecular pictures.~The BaBar group has recently identified the
$n_r=2$ excitations of the $D$ and $D^*$ with masses
$M_{D(2S)}=2540\pm 8$~MeV and $M_{D^*(2S)}=2609\pm 4$~MeV [1708].~An
$S$-wave $D\bar{D}^*(2S)$ combination would have have $J^P=1^+$ and
a ``threshold'' mass around 4480~MeV, very close to the Belle-LHCb
average $M_{Z(4430)}= 4477\pm$ 19~MeV.

The charged $Z_1(4050)^+$ and $Z_2(4250)^+$ states that decay to
$\pip\chi_{c1}$ are less easy to associate with a threshold.~We do
not know their $J^P$ -- we do not even know if the two states have
the same $J^P$ values -- but if the $\pip$-$\chi_{c1}$ system is in
an $S$-wave, $J^{P}=1^-$.~We do know that their neutral $I_3=0$
partner, which has not yet been seen, has to have even $C$ parity.
If we assume $J^P=1^-$, a $D\bar{D}_1(2420)$ would give the right
quantum numbers and a nearby mass threshold for the $Z_2(4250)$,
where the $D_1$ is a $J^P=1^+$, $D^*\pi$ resonance with mass
$M=2421.3\pm 0.6$~MeV and width ${\it\Gamma}=27.1\pm 2.7$~MeV.~One
could imagine a situation where the $Y(4260)$ is the $C$ odd
isosinglet and the $Z_2(4250)$ is the $C$ even isotriplet
$D\bar{D}_1$ molecular states $1/\sqrt{2}(D\bar{D}_1\pm
D_1\bar{D})$.~However, the scenario where the $Y(4260)$ is a
$D\bar{D}_1$ molecule [48] was examined by BESIII as part of their
study of $Y(4260)\rt \pi D\bar{D}^*$ [69].~Since the $Y(4260)$ mass
is $\sim 30$~MeV below $m_D + m_{D_1} \simeq 4290$~MeV, a prominant
$D\bar{D}_1$ component of its wave function should show up as a
distinct clustering of $\pi\bar{D}^*$ invariant masses near their
upper kinematic limit.~BESIII reports no evidence for such
clustering and, thus, no indication of a $D\bar{D}_1$ component of
the $Y(4260)$.~On the other hand, there are no interesting nearby
$D^{(*)}\bar{D}^{(*)}$ mass thresholds for the $Z_1(4020)$,
especially for $J^P=1^-$.

In the $b$-quark-sector, the $Z_b$ states are isospin triplets with
$J^P=1^+$ near the $BB^*$ and $B^*\bar{B}^*$ threshold and
suggestive of virtual $S$-wave $B\bar{B}^*$ and $B^*\bar{B}^*$
molecules.~Here we might expect  $C$-even equivalents of the
$X(3872)$ and $X_{c2}$ (labeled as $X_{b1}$ and $X_{b2}$ in the
right panel of Fig.~22), but finding them may be difficult with
existing data.~The CMS group searched for a $b$-quark version of the
$X(3872)$ in the inclusive $\pipi\Upsilon(1S)$ invariant mass
distribution produced in proton-proton collisions at
$\sqrt{s}=8$~TeV, but found no evidence for peaks other than those
due to $\Upsilon(2S)$ and $\Upsilon(3S)$ to $\pipi\Upsilon(1S)$
transitions [171].~However, if, as expected, the $b$-quark-sector
equivalent of the $X(3872)$ has $J^{PC}=1^{++}$, zero isospin, and
is near the $B\bar{B}^*$ mass threshold, the $\pipi\Upsilon(1S)$
decay mode, for which the $\pipi$ would have to originate from
$\rho\rt\pipi$, would violate isospin and be suppressed relative to
decays to the isospin-conserving $\omega\Upsilon(1S)$ final
state.~This is not the case for the $X(3872)$ where the
isospin-allowed  $\omega\jp$ decay mode is kinematically suppressed:
i.e., $Q_c\simeq m_{D^0}+m_{D^{*0}}-m_{\jp}=776$~MeV, which is about
one $\omega$ natural width below its peak mass
$m_{\omega}=783$~MeV.~In the $b$-quark-sector,
$m_{B}+m_{B^*}-m_{\Upsilon(1S)}=1145$~MeV, which is well above
$m_{\omega}$.~Thus, $\omega\Upsilon(1S)$ final states are probably
more relevant than $\pipi\Upsilon(1S)$ for searches for $X(3872)$
counterparts of the $Z_b(10610)$ and $Z_b(10650)$.~This would
require studies of decay final states that contain a $\pi^0$, which
may be difficult to do with existing LHC experiments, but could be
done at BelleII [85].

\vspace*{5mm} \noindent {\color{ooorangec}\it 3.3.2\quad
Tetraquarks? }\vspace{3.5mm}

\noindent Maiani and collaborators note that the
$M_{Z(4430)}-M_{Z_c(3900)}= 589\pm 30$~MeV mass difference is the
same as $m_{\psip}-m_{\jpsi}=589$~MeV, with not such a large
error.~This, together with the preference for the $Z(4430)^+$ to
decay to $\pip\psip$ rather than $\pip\jpsi$, suggests that in the
$Z(4430)$, configurations where the $c$ and $\bar{c}$ form a $\psip$
are important.~This is in line with QCD tetraquark expectations as
discussed in Ref.~[144].

In Ref.~[143], which appeared shortly after the $Z_c(3900)$
discovery was announced, Maiani and collaborators proposed a test
between the tetraquark and molecular model: their model predicted a
second charged $Z_c$ state with a mass that is about 100~MeV lower,
while the molecular picture predicted a second state at higher mass,
close to the $2m_{D^*}$ threshold.~The BESIII group's subsequent
discovery of the $Z(4020)$ confirmed the molecular picture.~However,
the Ref.~[143] prediction of a lower mass partner state was based on
a particular model for diquark-diantiquark spin-spin
interactions.~Perhaps modifications that incorporate measured
properties of the $Z_c(4020)$ can fix this.~The same paper also
predicted that the $D\bar{D}^*$ decay width of the $Z_c(3900)^+$
would be $4$~MeV, and small compared to that for $\pip\jpsi$, which
they estimated to be $\simeq 29$~MeV.~Subsequently the $D\bar{D}^*$
decays have been seen, with a decay with that is $6.2 \pm 2.9$ times
larger than that for $\pip\jpsi$ [69], not six times
smaller$^{11)}$\footnote{\hspace*{-5mm}$^{11)}$~Here I assume that
the $Z_c(3885)$ and $Z_c(3900)$ are the same state.}.

One attractive feature of the tetraquark feature is that it can
explain the large partial width for transitions such as
$Y(4260)\rt\pipi\jpsi$,~$X(3915)\rt
\omega\jpsi$,~$Z(4430)\rt\pip\psip$, etc., in a natural way.~The
corresponding transitions in the charmonium system are
OZI-suppressed, and the partial widths are relatively small
\linebreak $\le 100$~keV.~In molecular models, although
OZI-suppression can be evaded, the $Q$ and $\bar{Q}$ reside in
different constituent mesons with small spatial overlap.~In addition
they have uncorrelated colors and spins.~In contrast, tetraquarks
are compact.~The the $Q$ and $\bar{Q}$ necessarily have a large
overlap and tightly correlated color and spin and the presence of
the additional light quarks eliminates OZI suppression.~A tetraquark
interpretation of the charged bottomoniumlike $Z_b(10610)$ and
$Z_b(10650)$ states is advocated by Ali and collaborators [172,
173].

\vspace*{5mm} \noindent {\color{ooorangec}\it 3.3.3\quad
QCD-hybrids?}\vspace{3.5mm}

\noindent Since $Q\bar{Q}$-gluon hybrid mesons cannot account for
any of the charged $Z$ states, their role in explaining the $XYZ$
meson puzzles described here has to be limited.~As mentioned above,
there has been a number of papers that identify the $1^{--}$ $Y$
states as charmonium hybrids.~In $1^{--}$ hybrids, the $Q$ and
$\bar{Q}$ are in a relative $P$-wave, which suppresses their
${\it\Gamma}_{\ee}$ partial widths.~The experimental 90\% CL upper
limit ${\it\Gamma}_{\ee}(Y(4260))<580$~eV [45] is smaller than
corresponding width for the nearby $3^3S_1$ state:
${\it\Gamma}_{\ee}(\psi(4040))= 860 \pm 7$~eV  and about the same as
that for the $4^3S_1$ state: ${\it\Gamma}_{\ee}(\psi(4415))=580 \pm
70$~eV [13].~For similar reasons, decays to $S$-wave plus $P$-wave
open charmed mesons are supposed to be dominant.~As mentioned above
in the discussion related to the study of $Z_c(3885)\rt D\bar{D}^*$
in $Y(4260)\rt \pi D\bar{D}^*$ decays, there is no strong evidence
for  subthreshold $Y(4260)\rt D\bar{D}_1(2420)$ decays in these
final states [69].~Thus, although there is a lot of theoretical
enthusiasm for identifying the $1^{--}$ $Y$ states as $\ccbar$-gluon
hybrids, there is not much experimental evidence to support this
assignment.~Perhaps theoretical analyses of the recent BESIII
observations of strong decay widths for $Y(4260)\rt\pip Z_c(3900)^-$
\& $\pip Z_c(4020)^-$ [49, 52] and the radiative transition
$Y(4260)\rt \gamma X(3872)$ [115] will clarify the situation.

\noindent {\color{ooorangec}\it 3.3.4\quad
Hadrocharmonium?}\vspace{3.5mm}

\noindent Voloshin points out that the $\simeq 23$~MeV mass
difference between the $Z_c(3900)$ (measured in the $\pip\jpsi$
channel) and the $D^{+}\bar{D}^{*0}$ (or $D^{*+}\bar{D}^0$)
threshold is not that small [147].~The momenta of the
``constituent'' $D^{(*)}$ mesons in the $Z_c$ restframe would be
substantial, $\sim
200$~MeV/c$^{12)}$\footnote{\hspace*{-5mm}$^{12)}$~With my average
of all $Z_c(3900)$ mass measurements, $\langle M_{Z_C(3900)}\rangle
= 3890\pm 3$~MeV, and the difference between the $Z_c(3900)$ mass
$\frac{}{}~~~~$and the $D\bar{}^*$ mass threshold is lower, at
$\simeq 14$~MeV, and the constituent $D^{(*)}$ momentum in the $Z_c$
restframe decreases to $\simeq 170$ $\frac{}{}~~~~$MeV.}.~Since
$J^{P}=1^+$, the mesons would be in an $S$-wave with no centrifugal
barrier to hold them together, in which case, the total width $29\pm
10$~MeV$^{13)}$\footnote{\hspace*{-5mm}$^{13)}$~This is my average
of the measurements reported in Refs.~[49] and [69].} seems too
small.~Also, in a $D\bar{D}^*$ bound state, the spins of the $c$ and
the $\bar{c}$ quarks would be uncorrelated and, thus, the $c\bar{c}$
system should consist of equal amounts of $S=0$ and $S=1$.~This
suggests that the decays $Z_c(3900)\rt \pi\jpsi$ and $\pi h_c$
should occur at a similar rate, contrary to the observation reported
in Ref.~[51].~As discussed above, BESIII measurements show that the
$Z_c(3900)\rt \pi h_c$ rate is suppressed relative to that for the
$\pi\jpsi$ mode by at least a factor of five.~While the
hadrocharmonium model fixes the above problems, it, like the
tetraquark model, predicts a lower mass partner state while BESIII
found the $Z_c(4020)$ at higher mass.~Also, the hadrocharmonium
prediction for the $D\bar{D}^*$ decay width is similar to the 4~MeV
value predicted by the tetraquark model, and is in
 strong contradiction with experimental observatiions.

\vspace*{5mm} \noindent {\color{ooorangec}\it 3.3.5\quad A unified
model?}\vspace{3.5mm}

\noindent None of the models discussed above gives a compelling
picture of all the observed data.~Thus, in our current situation, we
are forced to a attribute some of the observed states to one picture
and others to a different scenario.~In some cases, like the
$X(3872)$, mixtures of different models have been invoked.~In all
cases, these pictures accept the standard potential-model based
quark model ideas and apply molecule or tetraquark, etc., ideas to
those states where the conventional approach fails.~Thus, molecular
models work for near-threshold ``exotic'' states, tetraquarks for
scalar mesons, $\ccbar$-gluon hybrids for the $1^{--}$ $Y$ states,
etc.~Since all these approaches have problems, the situation is not
very satisfying.

An exception to these piecemeal approaches is an ambitious project
by Friedmann that eschews potential model and non-relativistic
quantum mechanics ideas altogether and, instead, attempts to build
the spectrum of hadrons directly from QCD principles alone, using
quark and diquark constituents on an equal footing [174].~With this
approach, Friedmann is able to categorize all of the established
hadrons as designated by the PDG at the time of her publication,
including many of the ``exotic'' ones listed above,  in an elegant
concise way.~A striking feature of this scheme is that all of the
states can be incorporated in her scheme without invoking any radial
excitations.~In conventional models, radial excitations are
essential features but, with them, the model runs into lots of
problems, especially with the categorization of the baryon
spectrum.~For example, the $J^P=1/2^+$, $N(1440)$ ``Roper''
resonance, which defies any compelling classification in the quark
model, is naturally accommodated in Freidmann's scheme as a
$(qq)_A(qq)_A\bar{q}$ state, where the two asymmetric, spin-zero
diquarks ($(qq)_A$) are in a relative $P$-wave.~The $\psip$, which
is the quintessential radially excited meson in the quark model, is
identified as a $(qq)_A(\bar{q}\bar{q})_A$ pair in a $P$-wave.~Of
course this is just a clasification scheme and it remains to be seen
if a tetraquark picture of the $\psip$ can reproduce the success of
the charmonium model description of the $\psip$ properties such as,
for example, the ratio of
${\it\Gamma}_{\ee}(\psip)/{\it\Gamma}_{\ee}(\jpsi)$.~Nevertheless,
no one could accuse Friedmann of being a ``prisoner of conventional
thinking.''$^{14)}$\footnote{\hspace*{-5mm}$^{14)}$~http://scitation.aip.org/content/aip/magazine/physicstoday/news/10.1063/PT.5.3012}

\vspace*{7mm} {\color{ooorangec}\hrule}\vspace{2mm} \noindent
{\color{ooorangec}\large
\usefont{T1}{fradmcn}{m}{n}\xbt 4\quad Summary  }\vspace{3.5mm}

\noindent The QCD exotic states that are much preferred by
theorists, such as pentaquarks, the $H$-dibaryon, and meson hybrids
with exotic $J^{PC}$ values continue to elude confirmation even in
experiments with increasingly high levels of sensitivity.~On the
other hand, a candidate $p\bar{p}$ bound state and a rich
spectroscopy of quarkoniumlike states that do not fit into the
remaining unassigned levels for $\ccbar$ charmonium and $\bbbar$
bottomonium states has emerged.~No compelling theoretical picture
has yet been found that provides a comprehensive description of what
is seen, but, since at least some of these states are near
$D^{(*)}\bar{D}^*$ or $B^{(*)}\bar{B}^*$ thresholds and couple to
$S$-wave combinations of these states, molecule-like configurations
necessarily have to be important components of their wave functions
[112].~This has inspired a new field of ``flavor chemistry'' that is
attracting considerable attention both by the experimental and
theoretical hadron physics communities [29, 30].~With the increased
emphasis of the BESIII and LHC experiments on this subject and the
imminent operation of BelleII [85], the high luminosity
reincarnation of Belle, and the future operation of the PANDA
experiment at FAIR [175], we can expect a continuous flow of
interesting experimental results including, probably, many new
surprises and, hopefully, new, groundbreaking insights into the
long-distance behavior of QCD.\\

\noindent {\baselineskip
10.5pt\renewcommand{\baselinestretch}{1.05}\footnotesize
{\color{ooorangec}\bf Acknowledgements}~~~While preparing this
report I benefited from communications from Eric~Braaten,
Sookyung~Choi, Tamar~Friedmann, Pyungwon~Ko, Tomasz~Skwarnicki,
Sachiko~Takeuchi, Kai~Yi, Qiang~Zhao and colleagues on the Belle \&
BESIII experiments.~This work was supported in part by the Korean
National Research Foundation Grant No.~2011-0029457 and the
Institute for Basic Science (Korea) Project Code IBS-R016-D1.

}

\vspace{7mm}

{\color{ooorangec}\hrule}\vspace{2mm} \noindent
{\color{ooorangec}\large
\usefont{T1}{fradmcn}{m}{n}\bf  References and notes }\vspace{-2mm}

\parskip=0mm \baselineskip 15pt\renewcommand{\baselinestretch}{1.25} \footnotesize

\begin{OOExercises}
\item R.~L.~Jaffe, $Q^{2}\bar{Q}^{2}$ resonances in the baryon-antibaryon system, {\it Phys.~Rev.~D}, 1978, 17(5): 1444
\item T.~Nakano, et al.~(LEPS Collaboration), Evidence for a narrow $S=+1$ baryon resonance in photoproduction from the neutron, {\it
    Phys.~Rev.~Lett.}, 2003, 91: 012002
\item D.~Diakonov, V.~Petrov, and M.~Polyakov, Exotic anti-decuplet of baryons: Prediction from chiral solitons, {\it Z.~Phys.~A}, 1997, 359(3): 305
\item A review of the events during this period together with references to the experimental work is provided in: R.~A.~Schumacher,
    The rise and fall of pentaquarks in experiments, arXiv: nucl-ex/0512042, 2005.
\item See, for example, B.~McKinnon, et al.~(CLAS Collaboration), Search for the $\Theta ^{+}$ pentaquark in the reaction $\gamma
    $d$\to $pK$^{-}$K$^{+}$n, {\it Phys.~Rev.~Lett.}, 2006, 96: 212001

\item K.~Shirotori, T.~N.~Takahashi, S.~Adachi, M.~Agnello, S.~Ajimura, et al., Search for the $\Theta ^{+}$ pentaquark via the $\piup^-$p$\to
    $K$^-$X reaction at 1.92 GeV/c, {\it Phys.~Rev.~Lett.}, 2012, 109(13): 132002 (and references cited therein)
\item W.-M.~Yao, et al.~(Particle Data Group), Review of particle physics, {\it J.~Phys.~G}, 2006, 33(1): 1.~see,
    in particular, the ``Pentaquark Update'' by G.~Trilling on page 1019.
\item R.~L.~Jaffe, Perhaps a stable dihyperon, {\it Phys.~Rev.~Lett}., 1977, 38: 195
\item H.~Takahashi, J.~K.~Ahn, H.~Akikawa, S.~Aoki, K.~Arai, et al., Observation of a ${}_\Lambda^{} {_\Lambda^6 } $He double hypernucleus, {\it
    Phys.~Rev.~Lett.}, 2001, 87(21): 212502
\item S.~R.~Beane, et al.~(NPLQCD Collaboration), Evidence for a bound H dibaryon from lattice QCD, {\it Phys.~Rev.~Lett.}, 2011, 106: 162001

\item T.~Inoue, et al.~(HALQCD Collaboration), Bound H dibaryon in flavor {\it SU}(3) limit of lattice QCD, {\it Phys.~Rev.~Lett.}, 2011, 106: 162002
\item B.~H.~Kim, {\it et al.~}(Belle Collaboration), Search for an H-dibaryon with a mass near 2$m_\Lambda $ in $\Upsilon $(1S) and
    $\Upsilon $(2S) decays, {\it Phys.~Rev.~Lett.}, 2013, 110: 222002
\item J.~Beringer, et al.~(Particle Data Group), Review of particle physics, {\it Phys.~Rev.~D}, 2012, 86(1): 010001
\item Reported by T.~E.~Barnes, HADRON05 summary: Heavy quark hadrons and theory, arXiv: hep-ph/0510365, 2005
\item R.~L.~Jaffe, Exotica, arXiv: hep-ph/0409065v2, 2004

\item J.~Z.~Bai, et al.~(BES Collaboration), Observation of a near-threshold enhancement in the p$\rm\bar{p}$ mass spectrum from
    radiative J/$\psi \to \gamma $p$\rm\bar{p}$ decays, {\it Phys.~Rev.~Lett.}, 2003, 91: 022001
\item E.~Fermi and C.~N.~Yang, Are mesons elementary particles? {\it Phys.~Rev.}, 1949, 76(12): 1739
\item S.~Godfrey and S.~L.~Olsen, The exotic {\it XYZ} charmonium-like mesons, {\it Annu.~Rev.~Nucl.~Part.~Sci.}, 2008, 58(1): 51
\item B.~S.~Zou and H.~C.~Chiang, One-pion-exchange final-state interaction and the p$\rm\bar{p}$ near threshold enhancement in J/$\psi\to
    \gammaup$p$\rm\bar{p}$ decays, {\it Phys.~Rev.~D}, 2004, 69(3): 034004
\item A.~Sibirtsev, J.~Haidenbauer, S.~Krewald, U.~G.~Mei{\ss}ner, and A.\ Thomas, Near threshold enhancement of the p$\rm\bar{p}$ mass
    spectrum in J/$\psi $ decay, {\it Phys.~Rev.~D}, 2005, 71(5): 054010

\item G.~J.~Ding and M.~L.~Yan, Proton--antiproton annihilation in baryonium, {\it Phys.~Rev.~C}, 2005, 72(1): 034014
\item M.~Ablikim, et al.~(BES Collaboration), Observation of a resonance X(1835) in J/$\psi \to \gamma \pi ^{+}\pi^{-}\eta'
    $, {\it Phys.~Rev.~Lett.}, 2005, 95: 262001
\item M.~Ablikim, et al.~(BES Collaboration), Spin-parity analysis of p$\rm\bar{p}$ mass threshold structure in J/$\psi $ and $\psi
    $(3686) radiative decays, {\it Phys.~Rev.~Lett.}, 2012, 108: 112003
\item M.~Ablikim, et al.~(BES Collaboration), Confirmation of the X(1835) and observation of the resonances X(2120) and X(2370) in
    J/$\psi \to \gamma \pi ^{+}\pi ^{-}\eta ' $, {\it Phys.~Rev.\ Lett.}, 2011, 106: 072002
\item M.~Ablikim, et al.~(BES Collaboration), Study of J/$\psi $ decaying into $\omega $p$\rm\bar{p}$, {\it Eur.~Phys.~J.~C},
    2008, 53: 15, arXiv: 0710.5369 [hep-ex]

\item S.~B.~Athar, {et al.~}(CLEO Collaboration), Radiative decays of the $\Upsilon $(1S) to a pair of charged hadrons, {\it Phys.~Rev.
    D}, 2006, 73: 032001
\item M.-Z.~Wang, et al.~(Belle Collaboration), Observation of B$^{+}\to $p\={p}$\pi ^+$, B$^0\to $p$\rm\bar{p}$ K$^0$, and B$^{+}\to
    $p$\rm\bar{p}$ K$^{\ast +}$, {\it Phys.~Rev.~Lett.}, 2004, 92: 131801
\item See, for example, M.~Ablikim, et al.~(BESIII Collaboration), Observation of a structure at 1.84 GeV/c$^{2}$ in the 3($\pi
    ^{+}\pi ^{-})$ mass spectrum in J/$\psi \to \gamma $3($\pi ^{+}\pi ^{-})$ decays, {\it Phys.~Rev.~D}, 2013, 88: 091502(R)
\item For recent reviews see N.~Brambilla, S.~Eidelman, B.~K.\ Heltsley, R.~Vogt, G.~T.~Bodwin, et al., Heavy quarkonium: Progress,
    puzzles, and opportunities, {\it Eur.~Phys.~J.~C}, 2011, 71(2): 1534
\item G.~T.~Bodwin, E.~Braaten, E.~Eichten, S.~L.~Olsen, T.~K.~Pedlar, and J.~Russ, Quarkonium at the frontiers of high energy physics: A
    snowmass white paper, arXiv: 1307.7425 [hep-ph], 2013

\item K.~Abe, et al$.~$(Belle Collaboration), Observation of a charmoniumlike state produced in association with a J/$\psi $ in
    e$^{+}$e$^{-}$ annihilation at $\sqrt{s}\approx $10.6 GeV, {\it Phys.~Rev.\ Lett.}, 2007, 98: 082001
\item P.~Pakhlov, et al.~(Belle Collaboration), Production of new charmoniumlike states in e$^{+}$e$^{-}\to $J/$\psi $D$^{\ast
    }$\={D}$^{\ast }$ at $\sqrt{s}\approx $10.6 GeV, {\it Phys.~Rev.\ Lett.}, 2008, 100: 202001
\item K.-T.~Chao, Interpretations for the observed in the double charm production at B factories, {\it Phys.~Lett.~B}, 2008, 661(5): 348
\item J.-Z.~Bai, et al.~(BES Collaboration), Measurements of the cross section for e$^{+}$e$^{-}\to $ hadrons at center-of-mass energies
    from 2 to 5 GeV, {\it Phys.~Rev.~Lett.}, 2002, 88: 101802
\item M.~Ablikim, et al.~(BES Collaboration), Determination of the $\psi $(3770), $\psi $(4040), $\psi $(4160) and $\psi $(4415) resonance parameters,
    {\it Phys.~Lett.~B}, 2008, 660(4): 315

\item S.~Godfrey and N.~Isgur, Mesons in a relativized quark model with chromodynamics, {\it Phys.\ Rev.~D}, 1985, 32(1): 189
\item B.~Aubert, et al.~(BaBar Collaboration), Observation of a broad structure in the $\pi ^{+}\pi ^{-}$J/$\psi $ mass spectrum around
    4.26 GeV/c$^{2}$, {\it Phys.~Rev.~Lett.}, 2005, 95: 142001
\item B.~Aubert, et al.~(BaBar Collaboration), Evidence of a broad structure at an invariant mass of 4.32 GeV/c$^{2 }$ in the reaction
    e$^{+}$e$^{-}\to \pi ^{+}\pi ^{-}\psi $(2S) measured at BABAR, {\it Phys.~Rev.~Lett.}, 2007, 98: 212001
\item C.~Z.~Yuan, et al.~(Belle Collaboration), Measurement of the e$^{+}$e$^{-}\to \pi ^{+}\pi ^{-}$J/$\psi $ cross section via
    initial-state radiation at Belle, {\it Phys.~Rev.~Lett.}, 2007, 99: 182004
\item X.~L.~Wang, et al.~(Belle Collaboration), Observation of two resonant structures in e$^{+}$e$^{-}\to \pi ^{+}\pi ^{-}\psi $(2S)
    via initial-state radiation at Belle, {\it Phys.~Rev.~Lett.}, 2007, 99: 142002

\item G.~Pakhlova, et al.~(Belle Collaboration), Measurement of the near-threshold e$^{+}$e$^{-}\to $D$^{(*)\pm }$D$^{(*)\mp }$ cross
    section using initial-state radiation, {\it Phys.~Rev.~Lett.}, 2007, 98: 092001
\item G.~Pakhlova, et al.~(Belle Collaboration), Observation of the $\psi $(4415)$\to $D\={D}$^*_2$(2460) decay using
    initial-state radiation, {\it Phys.~Rev.~Lett.}, 2008, 100: 062001
\item G.~Pakhlova, et al.~(Belle Collaboration), Observation of a near-threshold enhancement in the e$^{+}$e$^{-}\to \Lambda
    _{c}^{+}\Lambda _{c}^{-}$ cross section using initial-state radiation, {\it Phys.~Rev.~Lett.}, 2008, 101: 172001
\item G.~Pakhlova, et al.~(Belle Collaboration), Measurement of the e$^{+}$e$^{-}\to $D$^0$D*$^{-}\pi ^{+}$ cross section using initial-state radiation,
    {\it Phys.~Rev.~D}, 2009, 80: 091101(R)
\item X.~H.~Mo, G.~Li, C.~Z.~Yuan, K.~L.~He, H.~M.~Hu, J.~H.~Hu, P.\ Wang, and Z.~Y.~Wang, Determining the upper limit of ${\it\Gamma}_{ee}$ for
    the Y(4260), {\it Phys.~Lett.~B}, 2006, 640(4): 182

\item See, for example, S.-L.~Zhu, The possible interpretations of Y(4260), {\it Phys.~Lett.~B}, 2005, 625(3--4): 212
\item E.~Kou and O.~Pene, Suppressed decay into open charm for the being a hybrid, {\it Phys.~Lett.~B}, 2005, 631(4): 164
\item Q.~Wang, C.~Hanhart, and Q.~Zhao, Decoding the riddle of Y(4260) and Z$_c$(3900), {\it Phys.~Rev.~Lett.}, 2013, 111(13): 132003
\item M.~Ablikim, et al.~(BESIII Collaboration), Observation of a charged charmoniumlike structure in e$^{+}$e$^{-}\to \pi ^{+}\pi
    ^{-}$J/$\psi $ at $\sqrt{s}$=4.26 GeV, {\it Phys.~Rev.~Lett.}, 2013, 110: 252001
\item Z.~Q.~Liu, et al.~(Belle Collaboration), Study of e$^{+}$e$^{-}\to \pi ^{+}\pi ^{-}$J/$\psi $ and observation of a
    charged charmoniumlike state at Belle, {\it Phys.~Rev.~Lett.}, 2013, 110: 252002

\item M.~Ablikim, et al.~(BESIII Collaboration), Observation of a charged charmoniumlike structure Z$_{c}$(4020) and search for the
    Z$_{c}$(3900) in e$^{+}$e$^{-}\to \pi ^{+}\pi ^{-}$hc, {\it Phys.\ Rev.~Lett.}, 2013, 111: 242001
\item M.~Ablikim, et al.~(BESIII Collaboration), Observation of a charged charmoniumlike structure in e$^{+}$e$^{-}\to
    $(D$^{*}$\={D}$^{*})^{\pm }\pi ^{\mp }$ at $\sqrt{s}$ = 4.26 GeV, {\it Phys.~Rev.~Lett.}, 2014, 112: 132001, arXiv: 1308.2760 [hep-ex]
\item R.~Mizuk, et al.~(Belle Collaboration), Observation of two resonance-like structures in the $\pi ^{+}\chi _{c1}$ mass distribution
    in exclusive \={B}$^0\to $K$^{-}\pi ^{+}\chi _{c1}$ decays, {\it Phys.\ Rev.~D}, 2008, 78: 072004
\item S.-K.~Choi, et al.~(Belle Collaboration), Observation of a resonancelike structure in the $\pi ^{\pm}\psi ' $ mass
    distribution in exclusive B$\to $K$\pi ^{\pm}\psi ' $ decays, {\it Phys.~Rev.~Lett.}, 2008, 100: 142001
\item B.~Aubert, et al.~(BaBar Collaboration), Search for the Z(4430)$^{-}$ at {\it BABAR}, {\it Phys.~Rev.~D}, 2009, 79: 112001

\item R.~Mizuk, et al.~(Belle Collaboration), Dalitz analysis of B$\to $K$\pi ^{+}\psi ' $ decays and the Z(4430)$^{+}$, {\it
    Phys.\ Rev.~D}, 2009, 80: 031104(R)
\item K.~Chilikin, et al.~(Belle Collaboration), Experimental constraints on the spin and parity of the Z(4430)$^{+}$, {\it
    Phys.\ Rev.~D}, 2013, 88: 074026
\item R.~Aaij, et al.~(LHCb Collaboration), Observation of the resonant character of the Z(4430)$^{-}$ state, {\it Phys.~Rev.~Lett.}, 2014, 112: 222002
\item See, for example, P.~Pakhlov, Charged charmonium-like states as rescattering effects in decays, {\it Phys.~Lett.~B}, 2011, 702(2--3): 139
\item P.~Pakhlov and T.~Uglov, Charged charmonium-like Z$^{+}$(4430) from rescattering in conventional B decays, arXiv: 1408.5295
[hep-ph], 2014

\item B.~Aubert, et al.~(BaBar Collaboration), Measurements of the absolute branching fractions of B$^\pm\to$K$^\pm$X$_{\rm c\bar{c}}$,
    {\it Phys.~Rev.~Lett.}, 2006, 96: 052002
\item K.~Chilikin, et al.~(Belle Collaboration), Observation of a new charged charmonium-like state in B$\to $ J/$\psi $ K $\pi $ decays,
    arXiv: 1408.6457, 2014 (submitted for publication in {\it Phys.~Rev.\ D})
\item For discussions of factorization in heavy quark decays see, for example, M.~Beneke, G.~Buchelle, M.~Neubert, and C.~Sachrajda, QCD
    factorization for B $\to \pi\pi$  decays: Strong phases and CP violation in the heavy quark limit, {\it Phys.~Rev.~Lett.}, 1999, 83(10): 1914
\item M.~Beneke, G.~Buchelle, M.~Neubert, and C.~Sachrajda, QCD factorization for exclusive non-leptonic -meson decays: General
    arguments and the case of heavy-light final states, {\it Nucl.~Phys.\ B}, 2000, 591(1--2): 313
\item M.~Beneke, G.~Buchelle, M.~Neubert, and C.~Sachrajda, QCD factorization in B $\to  \pi $K, $\pi  \pi $ decays and
    extraction of Wolfenstein parameters, arXiv: hep-ph/0104110, 2001

\item J.~P.~Lees, et al.~(BaBar Collaboration), Search for the Z$_1$(4050)$^{+}$ and Z$_2$(4250)$^{+}$ states in \={B}$^{0}\to \chi
    _{c1}$K$^{-}\pi^+ $ and B$^{+}\to \chi _{c1}$K$^0_{\rm S}\pi ^{+}$, {\it Phys.~Rev.\ D}, 2012, 85: 052003
\item C.~Zhang, Studies on BEPC upgrade from pretzel to double-ring, {\it Sci.~China G}, 2010, 53(11): 2084
\item M.~Ablikim, et al.~(BESIII Collaboration), Design and construction of the BESIII detector, {\it Nucl.~Instrum.~Methods A},
    2010, 614: 345, arXiv: 0911.4960 [physics.ins-det]
\item M.~Ablikim, et al.~(BESIII Collaboration), Observation of a charged (D\={D}*)$^{\pm }$mass peak in e$^{+}$e$^{-}\to \pi $D\={D}*
    at $\sqrt{s}$=4.26 GeV, {\it Phys.~Rev.~Lett.}, 2014, 112: 022001
\item M.~Ablikim, et al.~(BESIII Collaboration), Observation of e$^{+}$e$^{-}\to \pi ^{0}\pi ^{0}h_{c }$ and a neutral charmoniumlike
    structure Z$_{c}$(4020)$^{0 }$, 2014 (in preparation)

\item S.~K.~Choi, et al.~(Belle Collaboration), Observation of a near-threshold $\omega $J/$\psi $ mass enhancement in exclusive
    B$\to $K$\omega $J/$\psi $ decays, {\it Phys.~Rev.~Lett.}, 2005, 94: 182002
\item P.~del Amo Sanchez, et al.~(BaBar Collaboration), Evidence for the decay X(3872)$\to $J/$\psi \omega $, {\it Phys.~Rev.~D}, 2010, 82: 011101(R)
\item B.~Aubert, et al.~(BaBar Collaboration), Observation of Y(3940)$\to $J/$\psi \omega $ in B$\to $J/$\psi \omega $K at {\it BABAR}, {\it
    Phys.~Rev.~Lett.}, 2008, 101: 082001
\item S.~Uehara, et al.~(Belle Collaboration), Observation of a charmoniumlike enhancement in the $\gamma \gamma \to \omega $J/$\psi
    $ process, {\it Phys.~Rev.~Lett.}, 2010, 104: 092001
\item J.~P.~Lees, et al.~(BaBar Collaboration), Study of X(3915)$\to $J/$\psi \omega $ in two-photon collisions, {\it Phys.~Rev.~D}, 2012, 86: 072002

\item T.~Barnes, S.~Godfrey, and E.~S.~Swanson, Higher charmonia, {\it Phys.~Rev.~D}, 2005, 72(5): 054026
\item F.~K.~Guo, and U.~G.~Meissner, Where is the $\chi_{c0}$(2P)? {\it Phys.~Rev.~D}, 2012, 86(9): 091501
\item S.~Uehara, et al.~(Belle Collaboration), Observation of a $\chi _{c2}' $ candidate in $\gamma \gamma \to $D\={D} production at
    Belle, {\it Phys.~Rev.~Lett.}, 2006, 96: 082003
\item B.~Aubert, et al.~(BaBar Collaboration), Observation of the $\chi_{c2} $(2P) meson in the reaction $\gamma \gamma \to $D\={D} at
    {\it BABAR}, {\it Phys.~Rev.~D}, 2010, 81: 092003
\item J.~Brodzicka, et al.~(Belle Collaboration), Observation of a new D$_{\rm sJ}$ meson in B$^{+}\to $\={D}$^0$D$^0$K$^{+}$ decays, {\it
    Phys.~Rev.~Lett.}, 2008, 100: 092001

\item B.~Aubert, et al.~(BaBar Collaboration), Study of resonances in exclusive B decays to D\={ }(*)D(*)K, {\it Phys.~Rev.~D}, 2008, 77: 011102(R)
\item This is the weighted average of results reported in Refs.~[74] and [75].
\item Y.~Jiang, G.~L.~Wang, T.~H.~Wang, and W.~L.~Ju, Why X(3915) is so narrow as a $\chi _{c0}$(2P) state, arXiv: 1310.2317 [hep-ph], 2013
\item Y.~C.~Yang, Z.~Xia, and J.~Ping, Are the X(4160) and X(3915) charmonium states? {\it Phys.~Rev.~D}, 2010, 81(9): 094003
\item T.~Abe, et al.~(BelleII Collaboration), Belle II technical design report, arXiv: 1011.0352 [hep-ex], 2010

\item R.~Molina and E.~Oset, Y(3940), Z(3930), and the X(4160) as dynamically generated resonances from the vector-vector interaction,
    {\it Phys.~Rev.~D}, 2009, 80(11): 114013
\item T.~Aaltonen, et al.~(CDF Collaboration), Evidence for a narrow near-threshold structure in the $J$/$\psi \phi $ mass
    spectrum in B$^{+}\to $J/$\psi \phi $K$^{+}$ decays, {\it Phys.\ Rev.~Lett.}, 2009, 102: 242002
\item T.~Aaltonen, et al.~(CDF Collaboration), Observation of the $Y$(4140) structure in the $J$/$\psi\phi $ mass
    spectrum in B$^{\pm }\to $ J/$\psi\phi $K$^{\pm }$ decays, arXiv: 1101.6058, 2011
\item R.~Aaij, et al.~(LHCb Collaboration), Search for the $X$(4140) state in B$^{+}\to $J/$\psi \phi $K$^{+ }$decays, {\it
    Phys.~Rev.~D}, 2012, 85: 091103(R)
\item S.~Chatrchyan, et al.~(CMS Collaboration), Observation of a peaking structure in the J/$\psi \phi $ mass spectrum from B$^{\pm}\to $
    J/$\psi $K$^{\pm}$decays, {\it Phys.~Lett.~B}, 2014, 734: 261

\item J.~P.\ Lees, et al.~(BaBar Collaboration), Study of B$^{\pm,0} \to $ J/$\psi $ K$^{+}$K$^{- }$K$^{\pm,0}$ and search for
    B$^{0} \to $ J/$\psi \phi $ at BABAR, arXiv: 1407.7244, 2014 (submitted for publication in {\it Phys.~Rev.~D})
\item X.~Liu, Z.~G.~Luo, and S.~L.~Zhu, Novel charmonium-like structures in the invariant mass spectra, {\it Phys.~Lett.~B}, 2011, 699(5): 341
\item K.~Yi, Experimental review of structures in the J/$\psi \phi $ mass spectrum, {\it Int.~J.~Mod.~Phys.~A}, 2013, 28(18): 1330020
\item C.~P.~Shen, et al.~(Belle Collaboration), Evidence for a new resonance and search for the $Y$(4140) in the $\gamma \gamma \to
    \phi $J/$\psi $ process, {\it Phys.~Rev.~Lett.}, 2010, 104: 112004
\item S.~K.~Choi, et al.~(Belle Collaboration), Observation of a narrow charmoniumlike state in exclusive B$^{\pm }$$\to $K$^{\pm }\pi
    ^{+}\pi ^{-}$J/$\psi $ Decays, {\it Phys.~Rev.~Lett.}, 2003, 91: 262001

\item D.~Acosta, et al.~(CDF II Collaboration), Observation of the narrow state X(3872)$\to $J/$\psi \pi ^{+}\pi ^{-}$ in p$\rm\bar{p}$
    collisions at $\sqrt{s}$=1.96 TeV, {\it Phys.~Rev.~Lett.}, 2004, 93: 072001
\item V.~M.~Abazov, et al.~(D0 Collaboration), Observation and properties of the X(3872) Decaying to J/$\psi \pi ^{+}\pi ^{-}$ in
    p$\rm\bar{p}$  Collisions at $\sqrt{s}$=1.96 TeV, {\it Phys.~Rev.~Lett.}, 2004, 93: 162002
\item B.~Aubert, et al.~(BaBar Collaboration), Study of the B$^{-}$$\to $J/$\psi $K$^{-}\pi ^{+}\pi ^{- }$ decay and measurement of the
    B$^{-}$$\to $X(3872)K$^{-}$ branching fraction, {\it Phys.~Rev.~D}, 2005, 71: 071103
\item R.~Aaij, et al.~(LHCb Collaboration), Observation of $X($3872) production in {\it pp} collisions at $\sqrt{s} =7$ TeV, {\it
    Eur.~Phys.~J.~C}, 2012, 72: 1972
\item S.~Chatrchyan, et al.~(CMS Collaboration), Measurement of the $X$(3872) production cross section via decays to J/$\psi\  \pi
    \pi $ in pp collisions at $\sqrt{s}$ = 7 TeV, {\it J.~High Energy Phys.}, 2013, 154: 1304; see also arXiv: 1302.3968 [hep-ex]

\item S.~Eidelman, Talk at the $\phi$ to $\psi$ symposium, Rome, Sep.~9--12, 2013
\item S.-K.~Choi, et al.~(Belle Collaboration), Bounds on the width, mass difference and other properties of X(3872)$\to \pi ^{+}\pi
    ^{-}$J/$\psi $ decays, {\it Phys.~Rev.~D}, 2011, 84: 052004
\item R.~Aaij, et al.~(LHCb Collaboration), Determination of the X(3872) meson quantum numbers, {\it Phys.~Rev.~Lett.}, 2013, 110: 222001
\item A.~Abulencia, et al.~(CDF Collaboration), Analysis of the quantum numbers J$^{\rm PC}$ of the $X$(3872) particle, {\it Phys.~Rev.\ Lett.}, 2007, 98: 132002
\item D.~Acosta, et al.~(CDF Collaboration), Measurement of the dipion mass spectrum in X(3872)$\to $J/$\psi \pi ^{+}\pi ^{-}$ decays, {\it
    Phys.~Rev.~Lett.}, 2006, 96: 102002

\item B.~Aubert, et al.~(BaBar Collaboration), Search for a charged partner of the X(3872) in the B meson decay B$\to $X$^{-}$K,
    X$^{-}\to $J/$\psi \pi ^{-}\pi ^{0}$, {\it Phys.~Rev.~D}, 2005, 71: 031501
\item K.~Abe, et al.~(Belle Collaboration), Evidence for X(3872) $\to \gamma $J/$\psi $ and the sub-threshold decay X(3872) $\to \omega$ J/$\psi
    $, arXiv: 0505037 [hep-ex], 2005
\item T.~Aushev, et al.~(Belle Collaboration), Study of the B$\to $X(3872)($\to $D*$^0$\={D}$^0$)K decay, {\it Phys.~Rev.~D}, 2010, 81: 031103(R)
\item G.~Gokhroo, et al.~(Belle Collaboration), Observation of a near-threshold D$^0$\={D}$^0$$\pi $$^0$ enhancement in B$\to $D$^0$\={D}$^0$$\pi
    $$^0$K decay, {\it Phys.~Rev.~Lett.}, 2006, 97: 162002
\item B.~Aubert, {\it et al.} (BABAR Collaboration), Study of resonances in exclusive $B $ decays to D$^{(\ast )}$\={D}$^{(\ast )}$K,
    {\it Phys.~Rev.~D}, 2008, 77: 011102(R)

\item E.~Braaten and M.~Lu, Line shapes of the X(3872), {\it Phys.\ Rev.~D}, 2007, 76(9): 094028
\item E.~Braaten and H.~W.~Hammmer, Universality in few-body systems with large scattering length, {\it Phys.~Rep.}, 2006, 428(5--6): 259
\item S.~Coito, G.~Rupp, and E.~van Beveren, X(3872) is not a true molecule, {\it Eur.~Phys.~J.~C}, 2013, 73(3): 2351
\item N.~A.~T\"{o}rnqvist, Isospin breaking of the narrow charmonium state of Belle at 3872 MeV as a deuson, {\it Phys.~Lett.~B}, 2004, 590(3--4): 209
\item M.~Ablikim, et al.~(BESIII Collaboration), Observation of e$^{+}$e$^{-}\to \gamma $X(3872) at BESIII, {\it Phys.~Rev.~Lett.}, 2014, 112: 092001

\item N.~A.~T\"{o}rnqvist, From the deuteron to deusons, an analysis of deuteronlike meson-meson bound states, {\it Z.~Phys.~C}, 1994, 61(3): 525
\item N.~A.~T\"{o}rnqvist, Comment on the narrow charmonium state of Belle at 3871.8 MeV as a deuson, arXiv: hep-ph/0308277, 2003
\item See, for example, F.~E.~Close, and P.~R.~Page, The threshold resonance, {\it Phys.~Lett.~B}, 2003, 578(1--2): 119
\item C.~Y.\ Wong, Molecular states of heavy quark mesons, {\it Phys.~Rev.~C}, 2004, 69(5): 055202
\item S.~Pakvasa and M.~Suzuki, On the hidden charm state at 3872 MeV, {\it Phys.~Lett.~B}, 2004, 579(1--2): 67

\item E.~Braaten and M.\ Kusunoki, Production of the X(3870) at the $\Upsilon$(4S) by the coalescence of charm mesons, {\it Phys.~Rev.~D},
    2004, 69(11): 114012
\item E.~S.~Swanson, Short range structure in the X(3872), {\it Phys.\ Lett.~B}, 2004, 588(3--4): 189
\item M.~B.~Voloshin, Heavy quark spin selection rule and the properties of the X(3872), {\it Phys.~Lett.~B}, 2004, 604(1--2): 69
\item S.\ Fleming, M.~Kusunoki, T.~Mehan, and U.~van Kolck, Pion interactions in the X(3872), {\it Phys.~Rev.~D}, 2007, 76(3): 034006
\item E.\ Braaten and M.~Lu, Line shapes of the X(3872), {\it Phys.~Rev.~D}, 2007, 76(9): 094028

\item S.~L.~Zhu, New hadron states, {\it Int.~J.~Mod.~Phys.~E}, 2009, 17(02): 283
\item D.~Gamermann and E.~Oset, Isospin breaking effects in the X(3872) resonance, {\it Phys.~Rev.~D}, 2009, 80(1): 014003
\item D.~Gamermann and E.~Oset, Couplings in coupled channels versus wave functions: Application to the X(3872) resonance, {\it Phys.~Rev.~D},
    2010, 81(1): 014029
\item O.~Zhang, C.~Meng, and H.~Q.~Zheng, Ambiversion of X(3872), {\it Phys.~Lett.~B}, 2009, 680(5): 453
\item M.~B.~Voloshin and L.~B.~Okun, Hadron molecules and charmonium atom, {\it JETP Lett.}, 1976, 23: 333

\item M.~Bander, G.~L.\ Shaw, P.~Thomas, and S.~Meshkov, Exotic mesons and e$^{+}$e$^{- }$ annihilation, {\it Phys.~Rev.~Lett.}, 1976, 36(13): 695
\item A.~De Rujula, H.~Georgi, and S.~L.~Glashow, Molecular charmonium: A new spectroscopy? {\it Phys.~Rev.~Lett.}, 1977, 38(7): 317
\item A.~V.~Manohar and M.~B.~Wise, Exotic states in QCD, {\it Nucl.\ Phys.~B}, 1993, 339(1): 17
\item CDF note 7159, http://www-cdf.fnal.gov/physics/new/ bottom/051020.blessed-X3872/XLife/xlonglivedWWW.ps
\item S.~Chatrchyan, et al.~(CMS Collaboration), Measurement of the X(3872) production cross section via decays to J/$\psi  \pi
    \pi $ in pp collisions at $\sqrt{s}$ = 7 TeV, {\it J.~High Energy Phys.}, 2013, 1304: 154

\item C.~Bignamini, B.~Grinstein, F.~Piccinini, A.~D.~Polosa, and C.\ Sabelli, Is the X(3872) production cross section at $\sqrt{s}$ =1.96
    TeV compatible with a hadron molecule interpretation? {\it Phys.\ Rev.~Lett.}, 2009, 103(16): 162001
\item B.~Aubert, et al.~(BaBar Collaboration), Evidence for X(3872)$\to \psi $(2S)$\gamma $ in B$^{\pm }\to $X(3872)K$^{\pm }$
    decays and a study of B$\to $c\={c}$\gamma $K, {\it Phys.~Rev.\ Lett.}, 2009, 102: 132001
\item V.~Bhardwaj, et al.~(Belle Collaboration), Observation of X(3872)$\to $J/$\psi \gamma $ and search for X(3872)$\to \psi'
    \gamma $ in B decays, {\it Phys.~Rev.~Lett.}, 2011, 107: 091803
\item F.~Aceti, R.~Molina, and E.~Oset, X(3872)$\to $J/$\psi \gamma$ decay in the D\={D}* molecular picture, {\it Phys.~Rev.~D}, 2012, 86(11): 113007
\item E.~S.~Swanson, Diagnostic decays of the X(3872), {\it Phys.\ Lett.~B}, 2004, 598(3--4): 197

\item R.~Aaij, et al.~(LHCb Collaboration), Evidence for the decay X(3872) $\to\psi $(2S)$\gamma $,
    {\it Nucl.~Phys.~B}, 2014, 886: 665
\item L.~Maiani, F.\ Piccinini, A.~D.~Polosa, and V.~Riguer, Diquark-antidiquark states with hidden
    or open charm and the nature of X(3872), {\it Phys.~Rev.\ D}, 2005, 71(1): 014028
\item L.~Maiani, V.~Riguer, R.~Faccini, F.~Piccinini, A.~Pilloni and A.~D.~Polosa, J$^{\rm PC}$=1$^{++}$ charged resonance in the $\Upsilon
    $(4260)$\to \pi ^{+}\pi ^{-}$J/$\psi $ decay? {\it Phys.~Rev.~D}, 2013, 87: 111102(R)
\item L.~Maiani, F.~Piccinini, A.~D.~Polosa, and V.~Riguer, Z(4430) and a new paradigm for spin interactions in tetraquarks, {\it
    Phys.~Rev.~D}, 2014, 89(11): 114010
\item D.~Horn and H.~Mandula, Model of mesons with constituent gluons, {\it Phys.~Rev.~D}, 1978, 17(3): 898

\item L.~Liu, et al.~(Hadron Spectrum Collaboration), Excited and exotic charmonium spectroscopy from lattice QCD, {\it J.~High Energy
    Phys.}, 2012, 07: 126
\item M.~B.~Voloshin, Z$_{c}$(3900)-what is inside? arXiv: 1304.0380, 2013
\item M.~Takizawa and S.~Takeuchi, X(3872) as a hybrid state of the charmonium and the hadronic molecule, {\it Prog.~Theor.~Exp.~Phys}.,
    2013, 9: 093D01, arXiv: 1206.4877
\item K.~K.~Seth, The quintessential exotic X(3872), {\it Prog.~Part.\ Nucl.~Phys.}, 2012, 67(2): 390
\item W.~S.~Hou, Searching for the bottom counterparts of X(3872) and Y(4260) via $\pi^{+}\pi^{-}\Upsilon$, {\it Phys.~Rev.~D}, 2006, 74(1): 017504

\item K.-F.~Chen, et al.~(Belle Collaboration), Observation of an enhancement in e$^{+}$e$^{-}\to \Upsilon $(1S)$\pi ^{+}\pi ^{-}$,
    $\Upsilon $(2S)$\pi ^+\pi^ -$, and $\Upsilon $(3S)$\pi ^{+}\pi ^{-}$ production near $\sqrt{s}$=10.89 GeV, {\it Phys.~Rev.~D}, 2010, 82: 091106(R)
\item K.-F.~Chen, et al.~(Belle Collaboration), Observation of anomalous $\Upsilon $(1S)$\pi ^{+}\pi ^{-}$ and $\Upsilon $(2S)$\pi
    ^{+}\pi ^{-}$ production near the $\Upsilon $(5S) Resonance, {\it Phys.~Rev.~Lett.}, 2008, 100: 112001
\item A.~Ali, C.~Hambrock, I.~Ahmed, and M.~J.~Aslam, A case for hidden b\={b} tetraquarks based on $\rm e^+e^-\to b\bar{b}$ cross section
    between $\sqrt{s}=10.54$ and 11.20 GeV,     {\it Phys.~Lett.~B}, 2010, 684(1): 28
\item I.~Adachi, et al.~(Belle Collaboration), First observation of the P-wave spin-singlet bottomonium states h$_{\rm b}$(1P) and h$_{\rm b}$(2P), {\it
    Phys.~Rev.~Lett.}, 2012, 108: 032001
\item A.~Bondar, et al.~(Belle Collaboration), Observation of Two Charged Bottomoniumlike Resonances in $\Upsilon $(5S) Decays, {\it
    Phys.~Rev.~Lett.}, 2012, 108: 122001

\item A.~Bondar, Talk at the $\phi$ to $\psi$ symposium, Rome, Sep.~9--12, 2013
\item P.~Krokovny, et al.~(Belle Collaboration), First observation of the $\rm Z_b^0$(10610) in a Dalitz analysis of $\Upsilon $(10860)$\to
    \Upsilon $($n$S)$\pi ^{0}\pi ^{0}$, {\it Phys.~Rev.~D}, 2013, 88: 052016
\item A.~Garmash, et al.~(Belle Collaboration),Amplitude analysis of e$^{+}$e$^{- }\to \Upsilon $($n$S)$\pi ^{+}\pi ^{- }$ at $\sqrt{s}$ =
    10.865 GeV, arXiv: 1403.0992 [hep-ex] (submitted for publication in {\it Phys.~Rev.~D})
\item I.~Adachi, et al.~(Belle Collaboration), Observation of two charged bottomonium-like resonances, arXiv: 1105.4583, 2011
\item I.~Adachi, et al.~(Belle Collaboration), Study of three-body Y(10860) decays, arXiv: 1209.6450, 2012

\item S.~Eidelman, B.~K.~Heltsley, J.~J.~Hernandez-Rey, S.~Navas, and C.~Patrignani, Developments in heavy quarkonium spectroscopy, arXiv:
    1205.4189 [hep-ex], 2012
\item T.~Aaltonen, et al.~(CDF Collaboration), Precision measurement of the X(3872) mass in J/$\psi \pi ^{+}\pi ^{-}$ decays, {\it
    Phys.\ Rev.~Lett.}, 2009, 103: 152001
\item B.~Aubert, et al.~(BaBar Collaboration), Study of the exclusive initial-state-radiation production of the D\={D} system, {\it
    Phys.\ Rev.~D}, 2007, 76: 111105(R)
\item G.~Pakhlova, et al.~(Belle Collaboration), Measurement of the near-threshold e$^{+}$e$^{-}\to $D\={D} cross section using
    initial-state radiation, {\it Phys.~Rev.~D}, 2008, 77: 011103(R)
\item B.~Aubert, et al.~(BaBar Collaboration), Study of the $\pi ^{+}\pi ^{-}$J/$\psi $ mass spectrum via initial-state radiation at
    BABAR, arXiv: 0808.1543v2 [hep-ex], 2008

\item Q.~He, et al.~(CLEO Collaboration), Confirmation of the Y(4260) resonance production in initial state radiation, {\it Phys.~Rev.~D},
    2006, 74: 091104(R)
\item T.~E.~Coan, et al.~(CLEO Collaboration), Charmonium decays of Y(4260), $\psi $(4160), and $\psi $(4040), {\it Phys.~Rev.~Lett.}, 2006, 96: 162003
\item G.~Pakhlova, et al, Belle Collaboration, Observation of a near-threshold enhancement in the e$^{+}$e$^{-}\to
    \rm\Lambda^+_c \Lambda ^{-}_c$ cross section using initial-state radiation, {\it Phys.~Rev.~Lett.}, 2008, 101: 172001
\item P.~del Amo Sanchez, et al.~(BaBar Collaboration), Measurement of the B$\to $\={D}$^{(*)}$D$^{(*)}$K branching fractions, {\it Phys.~Rev.\ D},
    2011, 83: 032004
\item P.~del Amo Sanchez, et al.~(BaBar Collaboration), Observation of new resonances decaying to D$\pi $ and D*$\pi $ in inclusive
    e$^{+}$e$^{-}$ collisions near $\sqrt{s}=10.58$ GeV, {\it Phys.~Rev.~D}, 2010, 82: 111101(R)

\item S.~Chatrchyan, et al.~(CMS Collaboration), Search for a new bottomonium state decaying to $\Upsilon$(1S)$\pi ^{+}\pi ^{-}$ in pp collisions
    at $\sqrt{s}=8$ TeV, {\it Phys.~Lett.~B}, 2013, 727: 57
\item A.~Ali, C.~Hambrock, and W.~Wang, Tetraquark interpretation of the charged bottomonium-like states Z$_{b}^{\pm }$(10610) and Z$_{b}^{\pm
    }$(10650) and implications, {\it Phys.~Rev.~D}, 2013, 85(5): 054011
\item A.~Ali, C.~Hambrock, and M.~J.~Aslam, Tetraquark interpretation of the BELLE data on the anomalous $\Upsilon$(1S) $\pi
    ^{+}\pi ^{-}$ and $\Upsilon$(2S)$\pi ^{+}\pi ^{-}$ production near the $\Upsilon$(5S) resonance, {\it Phys.~Rev.~Lett.}, 2010, 104(16): 162001
\item T.~Friedmann, No radial excitations in low energy QCD (I): Diquarks and classification of mesons, {\it Eur.~Phys.~J.~C}, 2013, 73(2): 2298
\item W.~Erni, et al.~(PANDA Collaboration), Physics performance report for PANDA: Strong interaction studies with antiprotons,
    arXiv: 0903.3905 [hep-ex], 2009

\end{OOExercises}
\end{multicols}
\end{document}